\documentclass{emulateapj}

\usepackage{natbib}
\usepackage{subfigure}
\usepackage{graphicx}
\begin{document}

%  Sept. 1993
%  CLM
%  Use ``\input mymac.tex'' in a tex file to use these definitions.
%
%\renewcommand{\deg}{\mbox{$^{\circ}$}}
%
\def \fesclya {\mbox{$f_{esc}({\rm Ly}\alpha)$}}
\def \dlow {\mbox{$400 {\rm ~l~mm}^{-1}$}}
\def \dhigh {\mbox{$600 {\rm ~l~mm}^{-1}$}}
\def \i {\mbox{${\rm IRAS}$}}
% PAPER SPECIFIC
% EVAN S.
\newcommand{\be}{\begin{equation}} \newcommand{\ba}{\begin{eqnarray}}
\newcommand{\ee}{\end{equation}} \newcommand{\ea}{\end{eqnarray}}
\def\etal{{\it et al.\thinspace}}
\def\-{{\em{---}}}
\def \mA {\mbox{${\rm m \AA} $} }
\def \rr {\mbox{${\rm RR}$} }
\def \rarb {\mbox{${\rm R_AR_B}$} }
\def \rara {\mbox{${\rm R_AR_A}$} }
\def \dd {\mbox{${\rm DD}$} }
\def \dada {\mbox{${\rm D_AD_A}$} }
\def \dadb {\mbox{${\rm D_AD_B}$} }
\def \dr {\mbox{${\rm DR}$} }
\def \darb {\mbox{${\rm D_AR_B}$} }
\def \dara {\mbox{${\rm D_AR_A}$} }
\def \dbra {\mbox{${\rm D_BR_A}$} }
\def \hMpc      {h^{-1}{\rm\ Mpc}}
\def \hkpc      {h^{-1}{\rm\ kpc}}
\def \h         {\hbox{$\, h$} }
\def \hinv      {\hbox{$\, h^{-1}$} }
\def \hinvseven    {\hbox{$\, h_{70}^{-1}$} }
\def\ewr{\mbox {EW$_r$}}
\def\ewo{\mbox {EW$_o$}}
\def\H7{\mbox {$h_{0.7}$}}
\def\naI{\mbox {\ion{Na}{1}}}
\def\kI{\mbox {\ion{K}{1}}}
\def\mgI{\mbox {\ion{Mg}{1}}}
\def\arI{\mbox {\ion{Ar}{1}}}
\def\nII{\mbox {\ion{N}{2}}}
\def\feI{\mbox {\ion{Fe}{1}}}
\def\ovi{\mbox {\ion{O}{6}}}
\def\nv{\mbox {\ion{N}{5}}}
\def\znII{\mbox {\sc Zn~II~}}
\def\crII{\mbox {\sc Cr~II~}}
\def\alI{\mbox {\sc Al~I~}}
\def\alII{\mbox {\sc Al~II~}}
\def\alIII{\mbox {\sc Al~III~}}
\def\mgII{\mbox {\ion{Mg}{2}}}
\def\mnII{\mbox {\ion{Mn}{2}}}
\def\niII{\mbox {\ion{Ni}{2}}}
\def\feII{\mbox {\ion{Fe}{2}}}
\def\feIII{\mbox {\ion{Fe}{3}}}
\def\cIV{\mbox {\ion{C}{4}}}
\def\sV{\mbox {\ion{S}{5}}}
\def\sII{\mbox {\ion{S}{2}}}
\def\siIV{\mbox {\ion{Si}{4}}}
\def\siIII{\mbox {\ion{Si}{3}}}
\def\siII{\mbox {\ion{Si}{2}}}
\def\siI{\mbox {\ion{Si}{1}}}
\def\cII{\mbox {\ion{C}{2}}}
\def\cIII{\mbox {\ion{C}{3}}}
\def\llambda{\mbox {$\lambda$}}
\def\mstar{\mbox {$M_*$}}
\def\hlen{\mbox {$h_{0.7}^{-1}$}}
\def\lstarlya{\mbox {$L^*_{Ly\alpha}$}}
\def\IZw18{I~Zw~18}
\def\m82{M82}
\def\Ab{Abell~}
\def\gi{\mbox {\rm g-i}}
\def\ug{\mbox {\rm u-g}}
\def\br{\mbox {\rm b-r}}
\def\eqn{equation}
\def\vesc{\mbox {$v_{\rm esc}$}}
% HE PAPER
\def\heha{\mbox {He~I~$\lambda 5876$ / H$\alpha$}}
\def\xhe{\mbox {$\chi({\rm He}) / \chi({\rm H})$} }
\def\heii{\mbox {${\rm He}^+$}}
\def\he{\mbox {\rm He}}
\def\hii{\mbox {${\rm H}^+$}}
\def\h{\mbox {\rm H}}
\def\mab{\mbox {$\rm m_{AB}$}}
\def\ssp{\baselineskip=13pt plus 1pt minus 1pt}
\def\tsp{\baselineskip=5pt plus 1pt minus 1pt}
%
% ASTRO SYMBOLS (revised to work in/out mathmode).
%
\def\deg{\mbox {$^{\circ}$}}
\def\msun{\mbox {${\rm ~M_\odot}$}}
\def\zsun{\mbox {${\rm ~Z_{\odot}}$}}
\def\lsun{\mbox {${~\rm L_\odot}$}}
\def\msunyr{\mbox {$~{\rm M_\odot}$~yr$^{-1}$}}
\def\angs{\mbox {~\AA}}
\def\lya{\mbox {Ly$\alpha$}}
\def\lyb{\mbox {Ly$\beta$}}
\def\Ha{\mbox {H$\alpha$}}
\def\Hb{\mbox {H$\beta$}}
\def\Hg{\mbox {H$\gamma$}}
\def\Pa{\mbox {Pa$\alpha$}}
\def\tion{\mbox {$T_{\rm ion}$~}}
\def\ch{\mbox {$\bigtriangleup$}}
\def\grad{\mbox {$\bigtriangledown$}}
\def\lstar{\mbox {$L^*$}}
\def\line{\mbox {~$\lambda$}}
\def\lines{\mbox {~$\lambda\lambda$~}}
\def\h0{\mbox {~H$_0$}}
\def\q0{\mbox {~q$_0$}}
%
% **** LINE RATIOS ****
%
\def\auroral{[OIII]~$\lambda4363$~}
\def\auroral{[OIII]~$\lambda4363$~}
\def\ohsun{\mbox {(O/H)$_{\odot}$~}}

\def\o3hb{[OIII]$\lambda5007$~/~H$\beta$~}
\def\O1ha{[OI]$\lambda6300$~/~H$\alpha$~}
\def\Ru{[OII]$\lambda\lambda3727$~/~[OIII]$\lambda5007$~}
\def\s2ha{[SII]$\lambda\lambda6717,31$~/~H$\alpha$~}
\def\2z2{HeII~$\lambda4686$~}
\def\z7{[NII]~$\lambda6583$ }
\def\N2{[NII]~$\lambda6583$~/~H$\alpha$~}
\def\16z2{[SII]~$\lambda\lambda6717, 6731$ }
\def\HgI{HgI~$\lambda4358$~}
\def\Sdensity{[SII]~$\lambda6717 / \lambda6731$}
\def\Temp{[OIII]~$\lambda\lambda4959 + 5007 ~{\rm to}~ \lambda4363$~}
%
% **** end ****
%
\def\n{NGC~}
\def\asec{\ifmmode {'' }\else $''~$\fi}  % arc sec
\def\amin{\ifmmode {' }\else $'~$\fi}    % arc min
\def\arcsper{\ifmmode \rlap.{'' }\else $\rlap{.}'' $\fi} % '' %Arcsec period
\def\arcmper{\ifmmode \rlap.{' }\else $\rlap{.}' $\fi} % '  %Arcmin period
\def\sles{\lower2pt\hbox{$\buildrel {\scriptstyle <}
   \over {\scriptstyle\sim}$}} % approximately less than
\def\sgreat{\lower2pt\hbox{$\buildrel {\scriptstyle >}
    \over {\scriptstyle\sim}$}} % approximately greater than
%These are smaller.
%\def\gapp{$_>\atop{^\sim}$}  % approximately greater than
%\def\lapp{$_<\atop{^\sim}$}  % approximately less than
\def\gapp{\mbox {$_>\atop{^\sim}$}}  % approximately greater than
\def\lapp{\mbox {$_<\atop{^\sim}$}}  % approximately less than
%
% UNITS
\def\kms{\mbox {~km~s$^{-1}$}}
\def\ergsec{~ergs~s$^{-1}$~}
\def\sb{~ergs~s$^{-1}$~cm$^{-2}$~arcsec$^{-2}$}
\def\flux{~ergs~s$^{-1}$~cm$^{-2}$}
\def\flam{~ergs~s$^{-1}$~cm$^{-2}$ \AA$^{-1}$}
\def\cm3{~cm$^{-3}$}
\def\col{\mbox {~cm$^{-2}$}}
\def\mpc3{~Mpc$^{3}$}
\def\mpc-3{~Mpc$^{-3}$}
\def\rate{~sec$~{-1}$}
\def\um{~${\mu}$m}
% ABBREVIATIONS
\def\fig{{Figure}}
\def\figs{{Figures}}
\def\tbl{{Table}~}
\def\sec{{Sec.}~}
\def\x{{X-ray}~}
\def\xs{{X-rays}~}
\def\X{{X-Ray}~}

%
% REFERENCES
\def\et{{\rm et\thinspace al.}\ }   % et al.
\def\ets{{\rm et\thinspace al.'s}\ }   % et al.'s
\def\reff{\par\noindent\parskip=1pt\hangindent=3pc\hangafter=1}
\def\apj{ApJ}
\def\apjs{ApJS}
\def\pasp{PASP}
\def\aj{AJ}
\def\mn{MNRAS}
\def\nat{Nature}
\def\aa{A\&A}
\def\aasup{A\&AS}
\def\baas{BAAS}
\def\annrev{ARA\&R}
\def\aar{A\&AR}
\def\pasj{PASJ}
%

% USE THIS FOR TABLE REFERENCES
%
\def\beginrefs{
         {\normalsize}
         {\noindent}
         \small
        \baselineskip=11pt
        \parindent=0pt
        \frenchspacing
        \parskip=1pt plus 1pt
%        \interlinepenalty=1000\tolerance=400
        \everypar={\hangindent=0.42in}}

\def\rev{}

\title{The \lya\ Line Profiles of Ultraluminous Infrared Galaxies: Fast Winds and Lyman
Continuum Leakage}

\author{Crystal L. Martin\altaffilmark{1}, 
{Mark Dijkstra\altaffilmark{2},}
{Alaina Henry\altaffilmark{3},}
{Kurt T. Soto\altaffilmark{4},}
{Charles W. Danforth\altaffilmark{5},}
{Joseph Wong\altaffilmark{1}}}

\altaffiltext{1}{Department of Physics, University of California, 
Santa Barbara, CA, 93106, USA, cmartin@physics.ucsb.edu}

\altaffiltext{2}{Institute of Theoretical Astrophysics, University
of Oslo, Postboks 1029, 0858 Oslo, Norway}

\altaffiltext{3}{Astrophysics Science Division, Goddard Space Flight
Center, Code 665, Greenbelt, MD 20771, USA}

\altaffiltext{4}{Institute for Astronomy, Department of Physics, ETH Zurich,
CH-8093 Zurich, Switzerland}

\altaffiltext{5}{CASA, Department of Astrophysical and
Planetary Sciences, University of Colorado, 389-UCB,
Boulder, CO, 80309, USA}

\begin{abstract}
We present new Hubble Space Telescope Cosmic Origins Spectrograph 
far-ultraviolet (far-UV) spectroscopy and Keck Echellete optical spectroscopy
of 11 ultraluminous infrared galaxies (ULIRGs), a rare population
of local galaxies experiencing  massive gas inflows, extreme
starbursts, and prominent outflows. We detect \lya\ emission
from 8 ULIRGs and the companion to \i09583+4714. In contrast to the
P~Cygni profiles often seen in galaxy spectra, the \lya\ profiles 
exhibit prominent, blueshifted emission out to Doppler shifts 
exceeding -1000\kms\ in three HII-dominated and two
AGN-dominated ULIRGs. To better understand the role of resonance scattering 
in shaping the \lya\ line profiles, we directly compare them to
non-resonant emission lines in optical spectra. We find that the line 
wings are already present in the intrinsic nebular spectra, and scattering 
merely enhances the wings relative to the line core. The \lya\ attenuation (as measured
in the COS aperture) ranges from that of the far-UV continuum to over 
100 times more. A simple radiative transfer model suggests the \lya\ photons
escape through cavities which have low column densities of neutral hydrogen 
and become optically thin to the Lyman continuum in the most advanced mergers. 
We show that the properties of the highly blueshifted line wings on the \lya\ 
and optical emission-line profiles are consistent with emission from clumps of gas 
condensing out of a fast, hot wind. The luminosity of the \lya\ emission increases 
non-linearly with the ULIRG bolometric luminosity and represents about 0.1 to 1\% of 
the radiative cooling from the hot winds in the HII-dominated ULIRGs.

%Simple radiative transfer
%models suggest relatively low column densities and/or non-unity 
%covering fractions of neutral gas are needed to produce the \lya\ line
%profiles. 
%
%The \lya\ flux ranges from 0.01 to 0.3 of that predicted
%from the \Ha\ flux and Case~B recombination theory, 

%and the 
%implied escape fraction of \lya\ photons increases with the velocity 
%width of the \lya\ line profile and decreases with the color excess
%measured from the Balmer decrement confirming expectations from
%radiative transfer theory. 

%is similar to other samples
%of galaxies, the \Ha\ luminosity of ULIRGs greatly underestimates 
%the total production rate of ionizing radiation by massive stars
%due to optical depth effects.  

\end{abstract}

\section{Introduction}  \label{sec:intro}

Hydrogen recombination in star-forming galaxies produces prodigious 
amounts of line radiation including 8 \lya\ photons for every \Ha\ photon. 
The emergent \lya\ and \Ha\ line profiles typically look very different
because galaxies are nearly always optically thick to the \lya\
photons. Even when the neutral gas column becomes optically thin
to Lyman continuum radiation, i.e.,  $N(HI) < 1 / \sigma_{0} \approx 1.59 \times 10^{17}$\col\ 
where $\sigma_{0}$ represents the hydrogen photoionization 
cross section at the Lyman edge, the \lya\ optical depth at line center 
remains very large,
\begin{eqnarray}
\tau_{0}(\lya) = 8.02 \times 10^{3} 
\left (
 \frac{15 {\rm ~km~s}^{-1}}{b} 
  \right ) \nonumber \\
\left ( 
 \frac{N(HI)}{1.59 \times 10^{17} \col}
  \right ),
\label{eqn:lya_tau0}  \end{eqnarray}
{\rev
using the atomic data in \cite{Morton:2003p498} and 
a Doppler parameter $b \equiv \sqrt{2kT/m_H}$.}
%\citep{Draine:2011}. 
The \lya\ photons random walk 
out of the galaxy and undergo diffusion in frequency space 
with each scattering. The increased path length makes the line
photons more susceptible than the underlying continuum to 
absorption by dust.
The luminosity and frequency of the emergent \lya\ photons are therefore 
sensitive to the dust content, phase structure, and kinematics of interstellar gas
\citep{Dijkstra:2014}.

We can better understand these properties of the interstellar medium 
by studying \lya\ emission from nearby galaxies for which many
physical properties have been measured, and, as a result, also learn
how to best derive physical properties of high-redshift galaxies
from \lya\ properties when few other diagnostics are possible.
Given the large number of factors influencing the radiative transfer 
of \lya\ photons \citep{Gronke:2014}, empirical studies are needed for galaxies
sampling the full gamut of dust properties and gas kinematics.

Prior to the Galaxy Evolution Explorer (GALEX) spectroscopic survey,  
limited access to the space ultraviolet restricted observations of  
\lya\ 1215.67 \AA\ to less than two dozen galaxies at low redshift 
\citep{Giavalisco:1996,Kunth:1998}. A systematic search of GALEX data
produced 66 \lya - emitting galaxies at $0.2 < z < 0.35$
\citep{Deharveng:2008}; and a study of nine deep fields extended the 
redshift range of GALEX \lya\ emitters \citep{Cowie:2010}.
Multiwavelength follow-up demonstrated that these low-redshift, 
\lya-emitters, have undergone very recent star formation \citep{Cowie:2011}.

These studies have introduced the concept of a \lya\ escape fraction, $f_{esc}(\lya)$, 
which compares the flux detected in the \lya\ line to that predicted from the extinction 
corrected \Ha\ flux under Case~B recombination conditions at low density. \cite{Atek:2009p38}  
demonstrated the systematic decline in $f_{esc}(\lya)$  with increasing dust extinction. 
The attenuation of the resonance line in many galaxies exceeds the attenuation of the 
underlying far-UV continuum. This result suggests resonance scattering increases the 
pathlength of the \lya\ photons and therefore the probability of absorption by
dust grains \citep{Atek:2014}. Resonance scattering  also explains the larger
angular size of the galaxies in \lya\ relative to the population of massive stars \citep{Hayes:2013}. 

%, so the discovery of a few galaxies with $f_{esc}(\lya) >  f_{esc}(\lambda 1215)$
%stimulated  a number of models for the geometry of the dust and gas \citep{Hansen:2006,
%Scarlata:2009,Laursen:2013,Duval:2014}. 

%but spans a wide range of values in part because of
%its sensitivity to the ISM kinematics \citep{Atek:2008}.  

%extremely compact central object (radius of order 100~pc) 
%

The recent addition of the Cosmic Origins Spectrograph (COS, \cite{Green:2012} to the 
Hubble Space Telescope (HST) substantially improved ultraviolet sensitivity. This
new opportunity has allowed the exploration of the \lya\ properties for a wider range of galaxy
types \citep{Heckman:2011,Wofford:2013,Leitherer:2013,Jaskot:2014}. 
{\rev
Particularly relevant COS observations for understanding \lya\ emission from high-redshift galaxies include the 
pioneering studies of low-redshift Lyman Break Analogs (LBAs, \cite{Heckman:2011}) and of lower mass galaxies with 
extreme specific star formation rates (Green Peas, \cite{Jaskot:2014}).
New COS observations of the most luminous, dusty starbursts (introduced and discussed here) 
and a forthcoming analysis of a larger sample of Green Pea galaxies (Henry et al.,  in prep.)
build on this work.
}

{\rev
The properties of some LBAs and Green Pea galaxies make them candidates for Lyman continuum
leakage.  \cite{Heckman:2001} showed that sightlines optically thin to the Lyman continuum 
would not generate completely black troughs in low-ionization state absorption lines. The
discovery of significant residual intensity in the \ion{C}{2} $\lambda 1334.5$ feature
of both the LBAs with the most dominant central objects (DCOs, \cite{Heckman:2011}) 
and a few Green Pea galaxies \citep{Jaskot:2014} therefore generated much interest, and
\cite{Borthakur:2014} subsequently measured an escape fraction of a few percent from
the direct detection of the Lyman continuum from the DCO J0921+4509. The observations
of both populations also suggested the \lya\ emission properties were modified in
systems with escaping Lyman continuum.  In particular, \cite{Heckman:2011} noted a
correlation between the presence of blueshifted \lya\ absorption and a high (inferred)
escape for the  Lyman continuum.
}

%Roughly half of the eight LBAs observed with COS show the canonical 
%P~Cygni \lya\ profiles typical of high-redshift galaxies \citep{Shapley:2003}, but 
%the three with dominant central objects (DCOs) have a significant amount of blueshifted 
%\lya\ emission.  

None of these studies have observed the
most luminous sites of star formation in the local universe, deterred perhaps by their
enormous infrared excess.  Dusty starbursts, however, comprise an increasing fraction of the 
cosmic star formation rate (SFR) with lookback time \citep{LeFloch:2005}; and, somewhat surprisingly, 
sub-mm selected galaxies at high redshift actually show \lya\ in emission more frequently than do 
Lyman-break selected galaxies \citep{Chapman:2005}. Observations of the \lya\ emission from local
Ultraluminous Infrared Galaxies (ULIRGs) are largely missing from the literature yet would
provides important insight into the physics of \lya\ radiative transfer and feedback in intense 
star-forming regions at a much earlier time.

%Considering the ubiquity of galactic outflows in spectra
%of high redshift galaxies, their winds may generate holes through 
%the interstellar gas via a similar process, so these studies
%of local galaxies bear directly on the question of how winds
%affect the escape of ionizing radiation from galaxies.

In this paper, we present new COS observations of 11 galaxies selected
on the basis of their high far-infrared luminosity, $L_{IR} > 10^{12}\lsun$,
but representing a range of stages in the transformation of gas-rich spiral
galaxies to field ellipticals via  major mergers \citep{Genzel:2001}.
As a class, these objects emit a tiny fraction of their luminosity in the far-UV, but
their enormous luminosity makes them detectable with COS. We expect 
substantial variations among their absorption and emission properties in \lya\
due to both the wide variation in the UV/FIR flux ratio among ULIRGs \citep{Murray:2007p211}
and the evolution of the interstellar medium (ISM) during the merger.
Because of their warm dust temperatures and high luminosities during
the latest stages of mergers, a subset of ULIRGs  have been recognized as quasars
partly hidden by a large central concentration of dust
\citep{Sanders:1988a,Sanders:1988b}.

In Section~\ref{sec:observations}, we describe the selection
of the ULIRGs, the new HST observations, and the supporting
optical spectroscopy. We characterize the \lya\ radiation from 
ULIRGs in Section~\ref{sec:results} emphasizing the unusual shape 
of the line profile. Our interpretation of the \lya\ profile is enhanced 
by a quantitative comparison to recombination and forbidden lines 
in optical spectra  in Section~\ref{sec:lya_optical}. We
discuss the physical origin of the broad emission in starburst galaxies
in Section~\ref{sec:discussion} and present simple radiative transfer
modeling of the \lya\ line profiles in Section~\ref{sec:rad_transfer}. 
The results point towards the highly blueshifted emission coming from a 
physically distinct component of the outflow related to the hot wind fluid,
and we summarize the evidence for this results and its implications
in Section~\ref{sec:summary}.

%draws further support from a comparison of the velocities measured in
%emission to that of low ionization state (low-ionization-state) absorption in 
%Section~\ref{sec:absorption_lines} and the scaling of the \lya\ luminosities
%with  the far-IR luminosities which is presented in Section~\ref{sec:ll}.

%After showing that the \lya\ escape fraction and equivalent
%width are typical of low-redshift galaxies, we draw attention to a highly 
%unusual feature of the emission, namely the prominent blue wing on the 
%emergent line profile. 

We adopt a cosmology with $H_0 = 70$\kms Mpc$^{-1}$, $\Omega_0 = 0.3$,
and $\Omega_{\Lambda} = 0.7$ throughout. At the median galaxy redshift
of $z = 0.11$, one second of arc subtends 2.0~kpc.

%The low ionization state (low-ionization-state) absorption, for
%example, is well known to be blueshifted by several hundred
%\kms\ in this galaxy class \citep{Martin:2005}. The nebular emission 
%lines often show broad wings ($\sigma > 150$\kms) which have been shown to 
%be excited by shocks traveling at velocities of a few hundred 
%\cite{Soto:2012b}. This violent feedback results from an extreme
%starburst fueled by the rapid influx of gas from the outer disks of 
%the merging galaxies \cite{Soto:2010}.

\section{Observations} \label{sec:observations}

We have previously used the Echellete Spectrograph and Imager (ESI,\cite{Sheinis:2002})
on the Keck~II telescope to obtain high-resolution optical spectra of
many of the  2~Jy ULIRGs \citep{Strauss:1992} imaged in K and R bands
by \cite{Murphy:1996}. We found outflows of low-ionization-state gas
via  interstellar \naI\ and \kI\ absorption lines that are blueshifted
several hundred \kms\ with respect to the systemic redshift \citep{Martin:2005,Martin:2006}. We also
resolved broad wings on many optical emission lines and demonstrated that shocks
traveling at several hundred \kms\ excite this component of the emission
\citep{Soto:2012a,Soto:2012b}. To gain further insight into these gas flows,
we have obtained new G130M spectra of ULIRGs from the above sample with the 
Cosmic Origins Spectrograph (COS) on the Hubble Space Telescope (HST,
GO programs 12533 and 13407).

\subsection{The COS ULIRGs Sample}

\subsubsection{UV Selection} \label{sec:sample_selection}

We selected 11 of the UV-brightest ULIRGs at redshifts from 0.084 to 0.151 
for the COS observations.  Like all ULIRGs, our targets have extremely 
high ratios of far-infrared to far-UV flux. Most of the radiation from 
their stars and AGN is absorbed by dust, and the energy re-radiated as a 
thermal spectrum with maximum intensity between 60\um\ and  100$\mu m$.  
Their bolometric luminosities are just 0.30~dex brighter than their 
far-infrared luminosities which we list in Table~\ref{tab:cos_galaxies}
for reference. Very little 
of their bolometric luminosity comes out in the ultraviolet. By choosing the 
UV-brightest ULIRGs, however, our sample is biased towards systems with
lower than average (for ULIRGs) color excess, and this selection effect does
weaken the average strength of interstellar \naI\ absorption 
\citep{Martin:2005,Chen:2010p445}.

The bolometric luminosities of the COS ULIRGs span a relatively narrow range
from $\log (L_{IR}/\lsun) = 12.07$ to  12.55 but are correlated with the 
merger stage. Following \cite{Veilleux:2002}, we estimate the merger state from 
the separation of the nuclei and prominence of the tidal tails as described
in the notes to Column~8 of Table~\ref{tab:cos_galaxies}. The most luminous object, 
\i11598-0112, has progressed to stage~IVb of the merger sequence. The lowest 
luminosity targets (\i09583+4714 and \i08030+5243) are wide binaries still in 
the pre-merger stage (IIIa) with two distinct galaxies still recognizable. The
far-infrared luminosity of \i09583+4714 is dominated by one of the two galaxies, which
we call \i09583+4714; but it is the companion 14.5~kpc away which emits
most of the far-UV flux. We pointed COS at this companion which 
we will refer to as \i09583+4714-B.

The properties of the COS ULIRG sample are also consistent with the 
emergence of an AGN at late times.  
Most of the COS ULIRGs  have the cool far-infrared color typical
of starburst dominated ULIRGs. However, the relative flux densities at 25\um\ 
and 60\um, which are listed in column~4 of Table~\ref{tab:cos_galaxies}, identify two
objects --  \i01003-2238 and \i12071-0444 -- with the higher dust 
temperatures indicative of a more compact source. The upper limit 
on the color for \i11598-0112, $F_{\nu}(25\mu m)/ F_{\nu}(60\mu m) \le 0.22$,
also allows a classification as a warm ULIRG. These warm ULIRGs are less 
common than cool ULIRGs and mark the transition from cool ULIRGs to optically 
selected quasars \citep{Sanders:1988b}. 
The three warm ULIRGs in our sample show compact morphologies indicative of a late
stage merger. For example, \i01003-2238, is the most compact merger in the sample 
(stage V). Both \i12071-0444 and \i11598-0112 are classified morphologically as 
(stage~IVb) compact mergers. 

\subsubsection{The AGN Contribution to $L_{Bol}$}  \label{sec:sample_agn}

%The notes to Table~\ref{tab:cos_galaxies} provide
%definitions of these \cite{Veilleux:2002} interaction classes for the curious
%reader, but the main point is that 

In recent years, much progress has been made towards understanding the AGN contribution 
to the bolometric luminosity of ULIRGs. In particular, the ULIRG evolution study of 
\cite{Veilleux:2009}, used six independent methods to estimate the AGN fraction and 
found consistent luminosity fractions to within 10-15\%. We can use these results
to estimate the AGN contribution to the bolometric luminosity of each COS ULIRG.

\begin{figure}[t]
% \hbox{\hfill \includegraphics[height=9cm,angle=-90,trim=0 50 0 0]{figures/agn_i11598-0112.ps}}
 \hbox{\hfill \includegraphics[height=9cm,angle=-90,trim=0 50 0 0]{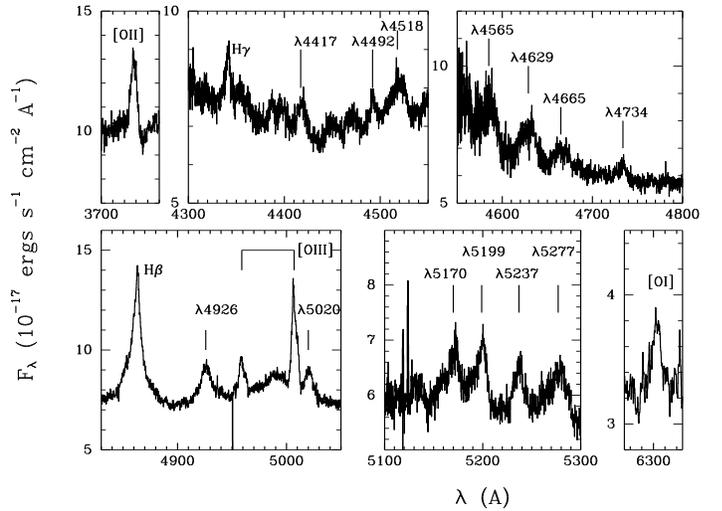}}
   \caption{\footnotesize 
     Optical spectra of the Seyfert~1 ULIRG \i11598-0112 zoomed in on features
     that identify this object as a narrow-line Seyfert 1 galaxy.
     In addition to broad Balmer lines and more
     narrow forbidden lines, the optical spectrum shows strong \ion{Fe}{2} emission complexes.
   }
   \label{fig:esi_11598}    \end{figure}

For example, Figure~\ref{fig:esi_11598} shows the optical spectrum of
\i11598-0112. The narrow forbidden lines and the broad Balmer emission lines 
identify this ULIRG as a Seyfert~1 galaxy \citep{Osterbrock:2006}. The presence of strong 
\feII\ emission in the optical spectrum defines this object as a Narrow Line
Seyfert~1 (NLS1) \citep{Leighly:2004}. The ULIRG evolution study showed that AGN provide roughly 
75\% of $L_{IR}$ in Seyfert~1 ULIRGS, so the luminosity of \i11598-0112 is likely 
dominated by the AGN.

%IRAS 11598
%
%While this AGN was missed in the excitation diagrams where it lies
%in the HII region locus \citep{Soto:2012b}, 

The emission-line ratios measured from the optical spectra of \i01003-2238 and 
\i12071-0444 place them in the Seyfert~2 region of the excitation diagrams \citep{Soto:2012b}. 
The Seyfert~2, \i01003-2238, is the most advanced merger (stage V) in the sample;
and, since this is consistent with the scenario where the AGN emerges towards the 
end of the merger, we consider this object AGN dominated.  
The emission-line nebula in \i12071-0444 is larger than the typical narrow-line 
region of an AGN, however,  and \citep{Soto:2012a} previously attributed the
excitation to shocks related to a galactic wind rather than an AGN. Still, 
he high far-infrared luminosity of \i12071-0444, its warm color,
and late (stage IV) merger class are all consistent with a significant AGN contribution.
The \cite{Veilleux:2009} results find that AGN power as much as 50\% of the total luminosity 
from Seyfert~2 ULIRGs, we infer that the AGN luminosity could be as large as the
starburst luminosity in both \i01003-2238 and \i12071-0444.
For most of the COS ULIRGs, their HII-dominated optical spectra and cool far-infrared color 
strongly suggest the starburt dominates the bolometric luminosity. For this class of ULIRG,
\cite{Veilleux:2009} estimate that AGN contribute no more than 15\% to 35\% of the bolometric 
luminosity.

%The escape rate of UV photons is likely directly connected to 
%the cavities excavated by stellar winds and supernova. Higher
%escape rates may be associated with viewing orientations along
%the outflow axis of a wind and/or later stages in the merger progression.

%FIGURE 
% lbrt

\subsection{HST Imaging \& COS Target Acquisition}

%At the median sample redshift of $z = 0.11$, 
%this shift is just 0\farcs15 on average;
%
%the standard COS target acquisition procedures automatically centered the 2\farcs5 diameter  
%(PSA) on the brightest UV knot. 

For ULIRGs with single-nuclei, we centered the brightest UV knot in the 2\farcs5 diameter 
primary science aperture (PSA) using the dispersed-light acquisition procedure recommended 
by the COS Instrument Handbook for Cycle 19. In those ULIRGs that have double nuclei, however, 
the UV emission may be dominated by star formation in either galaxy or even compact knots in the
tidal tails \citep{Surace:2000}. For all but the compact ULIRGs associated
with the later merger stages,  we obtained short exposures with HST through
far-UV, near-UV, and optical filters and found the coordinates for the UV-brightest 
knots prior to acquiring each target with COS.  Table~\ref{tab:cos_imaging} lists 
the parameters of these imaging observations.  

The standard pipeline processing for each imaging configuration proved sufficient
for target acquisition purposes. To construct images suitable for photometry, however,
we applied a zero-level bias correction to the raw images and ran LACosmic
to remove cosmic rays before combining frames with AstroDrizzle. After these
steps, significant trails can still be seen on the detector from the incomplete transfer 
of charge toward the amplifier. 

\begin{figure*}[h]
%  \hbox{\hfill \includegraphics[height=20cm,angle=90,trim=50 -100 100 0]{figure2.ps}
  \hbox{\hfill \includegraphics[height=20cm,angle=180,trim=0 100 50 25]{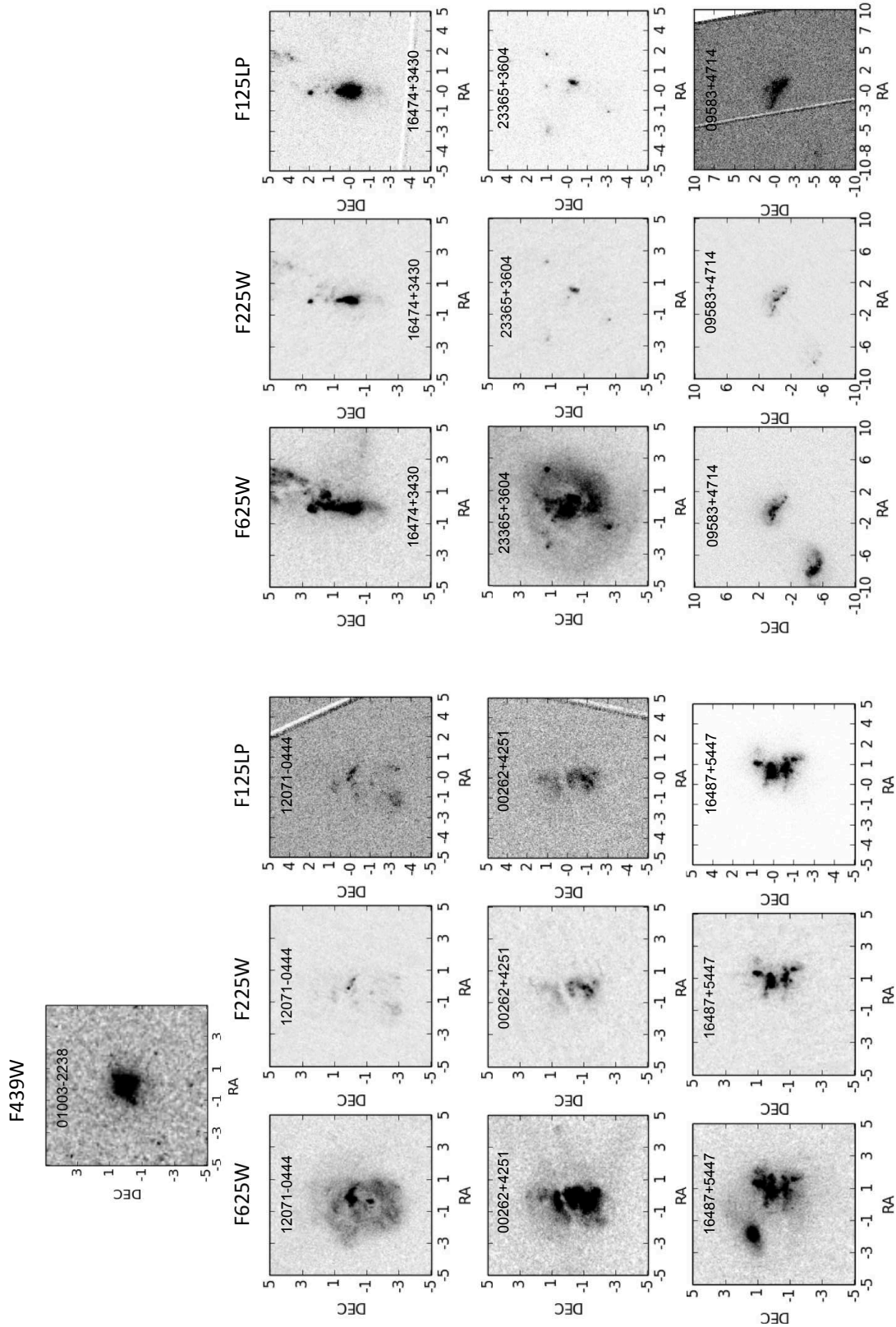}
  \hfill}
   \caption{\footnotesize 
     HST images of the ULIRGs in the optical (WFC3/F625W), near-UV (WFC3/F225W), and far-UV (ACS/SBC/F125LP)
     bands plus an archival image of  \i01003-2238  in WFPC2/F439W. 
     Targets are ordered from the highest to lowest \lya\ luminosity. 
     The images are registered with axes labeled in arcseconds relative to 
     the coordinates of the COS pointing given in Table~\ref{tab:cos_observations};   
     north is up and east is the the left.      The diameter of the COS aperture is
     2\farcs5, but the flux is heavily weighted by the central 0\farcs8 diameter
     region due to vignetting.
     The galaxies are compact in the ultraviolet bands relative to their optical images.
     The UV knots are typically offset by a few tenths of an arcsecond from the peak 
     optical light  consistent with the \cite{Goldader:2002} study.
%F814W
   }
   \label{fig:images} \end{figure*}

Figure~\ref{fig:images} shows these optical, near-UV, and far-UV images.
The ACS SBC/F125LP and WFC3 UVIS/F225W images generally detect the same knots, but 
many of the morphological features present in the optical F625W images remain 
completely undetected in the ultraviolet. Even the optical images  can be blind 
to the most active regions which are veiled by thick layers of dust, e.g., $A_V 
\sim 50$ \citep{Genzel:1998}. 

The far-UV images illustrate the distribution of light in the COS Primary
Science Aperture (PSA) for our spectroscopic observation. The sizes of the 
ULIRGs range from 0\farcs17 to 1\farcs07 FWHM,  smaller than the full
diameter of the PSA but similar to the unvignetted portion. For example,
the COS PSA subtends 5.03~kpc at the median target redshift; but the
diameter of the unvignetted region is just 1.61~kpc. It is unknown
whether a low surface-brightness halo of scattered \lya\ emission
extends beyond the PSA for some targets.

% DO WE NEED TEH OPTICAL VS UV OFFSETS?
%These knots are typically offset by a few tenths of an arcsecond from the peak optical light. 
%These shifts are consistent with expectations based on the previous UV imaging of ULIRGs by 
%\cite{Goldader:2002} 
%who found an average displacement of a few hundred parsecs between the brightest UV knot and the 
%peak near-IR emission. The relatively small size of these spatial offsets suggests that the
%highly obscured region (i.e., tens of magnitudes of optical extinction) in ULIRGs is 
%largely confined to the nucleus.
%
%This comparison also shows that the detection of any UV line or continuum
%emission from ULIRGs requires holes in the distribution of 
%neutral gas and dust through which we directly see some of 
%the star-forming regions. 

Our strategy for computing blind offsets used all 3 images. The 
F125LP bandpass provides the best match to our spectroscopic bandpass,
so we used it to identify the brightest knot;
but the field of view is typically too small to include 
a bright point-source. The WFC3/UVIS F225W images provide a larger field-of-view which typically 
includes a star or quasar bright enough for a near-UV imaging acquisition; and the blind offsets
were computed from these images.
The short WFC3/F625W exposures provide easy registration 
with optical images and were used to verify absolute pointing and estimate the red leaks in the UV filters. 
In one case, \i09583+4714, the stars in the F225W image were not bright enough for a blind
offset. Following the same procedure adopted for ULIRGs without pre-imaging, we centered 
the PSA on a bright knot significantly offset from the position of the maximum far-infrared 
intensity, hereafter \i09583+4714-B.

% after verifying that the ULIRGs had the same positions in the  near-UV and far-UV images.  
%
%. Table~\ref{tab:cos_observations} lists coordinates for {\it Position B},
%, as well as the COS ULIRG spectra.

%When no high-resolution, UV imaging was available, we centered the brightest UV
%knot in the 2\farcs5 diameter PSA using the dispersed-light acquisition
%procedure recommended by the COS Instrument handbook for Cycle 19. We moved the
%telescope in a spiral pattern to cover a square grid around the coordinates 
%of the maximum optical (R-band) surface brightness (ACQ/SEARCH  with SCAN-SIZE = 2 and 
%STEP-SIZE = 1\arcsec) recording a spectrum at each stop. The telescope was then moved
%to the flux-weighted mean position of the target. We then improved the centering
%in the cross-dispersion direction by taking a spectrum, collapsing it in the dispersion
%direction, finding the offset in the cross-dispersion direction to the mean of
%the collapsed spectrum, and finally slewing the telescope by this offset. After
%this ACQ/PEAKXD, we moved the telescope along the dispersion direction to obtain
%several ACQ/PEAKD exposures  (STEP-SIZE=1.118; CENTER FLUX WEIGHTING) and find
%the center of the knot along the dispersion direction. The science spectra were then
%obtained through the PSA.

%We followed the instructions in the WFC3 Data 
%Handbook to correct for CTE (charge transfer efficiency).
%{\bf Note, above step not yet completed. Will have to drop unless J Wong completes it.}

% FIGURE
%l b r t

\subsection{COS Far-UV Spectroscopy}

We obtained COS spectra in TIME-TAG mode through the PSA using the medium 
resolution far-UV grating, G130M.  We observed at four focal plane offset 
positions in order to reduce the impact of fixed-pattern noise associated 
with the micro-channel plate.
The spectra covered much of the bandpass from 1137 to 1468\AA.
The \lya\ line fell on Detector B in the observation of the lowest redshift target, 
\i23365+3604, but was redshifted onto Detector A for all other observations. 
In each case, the redshift of the galaxy cleanly separated the \lya\ profile from 
geocoronal emission. Table~\ref{tab:cos_observations} lists the central wavelengths 
(CWLS) of each pointing and additional details about the observations. 

Individual frames were processed with version 2.13.6 of the CalCOS pipeline.
Calcos corrects the data for instrumental effects, assigns a vacuum wavelength
scale, and extracts flux-calibrated spectra.  It applies a heliocentric
correction to the final x1d files for each exposure.  We combined 
the x1d files using software developed by the COS GTO team \citep{Danforth:2010}.
A simple mean pixel combination gave better results than the default variance
weighting for these low S/N data.

%listed there cover much of the bandpass from 1137 (1173) - 1432 (1468)\AA,
%but each spectrum has a gap between detector Segments B and A. For some targets,
%we could simply chose a CWL to avoid loosing important lines in 
%this gap, but for others we needed continuous coverage and requested
%exposures at two CWL's as indicated in column~11. 
%
%All spectra 
%were obtained at four focal plane offset positions in order to reduce the 
%impact of fixed-pattern noise associated with the micro-channel plate.

%The COS spectrosopic aperture affects the interpretation of
%galaxy spectra in several ways. 

%We address the first two concerns here and return to the discussion
%of scattered emission in Section~\ref{sec:results}

\subsubsection{Wavelength Calibration}

Emission from geocoronal airglow fills the COS aperture, so the observed wavelengths of
these lines are independent of target position in the PSA. We measured the wavelengths of 
Lya      1215.67,   
OI       1302.1685, 
OI       1304.8576, 
OI       1306.0286, 
OI]      1355.5977, 
and 
OI]      1358.5123 
but  did include NI 1199.967 line because it is a 
         blend of N I  1199.55, 1200.22, 1200.71. 
The velocity offsets among these lines were always  $\sles\ 40$\kms, which
we adopt as the maximum error in the pipeline dispersion solution.

Any error in centering the galaxy in the COS aperture will generate systematic
errors in the wavelength calibration. To verify the centering of the target in the PSA, 
we identified Milky Way absorption lines in the spectra with continuum detections.
An offset of the target from the 
center of the aperture will systematically shift the wavelengths of these lines.
In the spectra \i01003-2238, \i11598-0112, \i16487+5447, \i00262+4251, and
\i09583+4714, we measured five lines --
Si II   1190.42
Si II   1193.29
Si II   1260.42
Si III  1206.50
C II    1334.53;
% Not Measured:  Si II  1304.37, C II*  1335.71, O I    1302.17, 1304.86 1306.03
and we measured a subset of these in the \i23365+3604 and \i12071-0444 spectra.
The continuum S/N ratio is not adequate to measure any of these lines in
the \i10378+1109, \i17028+5817, and \i08030+5243 spectra. 
Blends with other transitions are the main source of error in the velocity offsets.\footnote{
     The Si II 1260.42 can blend with S II 1259.52. In the \i00262+4251 spectrum,
     for example, the other two lines in the S II 1250.58, 1253.81, 1259.52 triplet are 
     detected and unblended making the S II 1259 blend with Si II 1260.42 very obvious.
     The blend gives the Si II 1260 absorption trough an anomalous blueshift. 
     The CII 1334.53 trough is blended with the weaker CII* 1335.71 line in some spectra, 
     and this blend systematically increases (redshifts) the velocity offset.
     This CII blend lies on top of the \i00262+4251 \lya\ line in that spectrum.}
After rejection of the obvious blends, the average velocity offset of the Milky Way
absorption was within the uncertainties of the wavelength calibration for most targets.

We identified three spectra, however, with Galactic absorption lines that were
Doppler shifted by amounts greater than the uncertainty in the wavelength calibration.
We compared these Doppler shifts to the \cite{Wakker:2004} map of Galactic high
velocity clouds and made the following associations.
Since the only significant velocity offsets we found have the same sign as the Doppler
shifts of high-velocity clouds  in their direction, we concluded that the UV knots were 
well centered in the COS PSA and did not apply any corrections to the pipeline
wavelength scale. Hence when comparing the velocities of UV and optical lines,
we consider features within $\pm 40$\kms\ to be at the same velocity.

For completeness, we also note that
our sightline towards \i00262+4251 passes 3.2\deg\ northwest of M~31, a projected distance of 
just 44~kpc. In the \i00262+4251 spectrum, we detect Si III, Si II, and 
S II at approximately -300\kms, an offset consistent CGM  (see, for example, \cite{Lehner:2014}).

\subsubsection{Spectral Resolution}

The G130M resolution is excellent
when observing point sources, $R \approx  18,000 $; but degrades to  $R \approx 1450$ 
(about 207\kms) when the angular size of the target exceeds the diameter 
of the COS PSA. The effective resolution of each COS ULIRG spectrum depends 
on the angular size of the UV knots but must lie between these limits.

%At 0.0095 \AA\ per pixel, the resolution
%is between 7 and 92 pixels. {\bf 

We estimated the actual resolution of our spectra two ways. 
We first measured the widths of the unblended Milky Way absorption lines
in the 5 spectra with good continuum S/N ratio. Then we estimated the
resolution from the size of each source in the near-UV image.\footnote{We have
        measured the profile of the near-UV surface brightness and then
        used the scaling given by  \cite{Heckman:2011} for COS observations
        of the DCO's, i.e., FWHM of 0\farcs1 to 0\farcs5 yields a resolution
        $R = 13,000$ to 3000 with the medium resolution gratings.}
These measurements had four objects in common, and we found the Galactic
absorption lines always indicated lower resolution by a factor of two to
three relative to the images. We attribute this to the extended nature
of the UV emission which is not captured by the FWHM. Using the Galactic
lines, the COS G130M resolution for the most compact ULIRG, \i01003-2238,
is 69\kms\ while the stage~IIIa merger \i16487+5447 fills the COS aperture
and degrades the resolution to approximately 200\kms.  For \i16474+3430,
\i00262+4251, and \i11598-0112, we measure a spectral resolution of
150\kms, 154\kms, and 159\kms, respectively. Based on the size of
\i09583+4714-B in the near-UV image, we would estimate the resolution
$R < 1830$ or, equivalently,  $FWHM > 164$\kms.

\subsubsection{The Ultraviolet Continuum}  \label{sec:beta}

When the ultraviolet continuum was detected, we  typically
found spectral signatures of young stars.  Broad, 
shallow \ion{O}{6} and \ion{N}{5} absorption troughs
accompanied by redshifted emission, for example, indicated
the presence of stellar winds from very massive stars.
The ultraviolet continuum is solidly detected in six of the COS spectra,
and only the spectrum of \i11598-0112 fails to directly indicate a
population of ionizing stars.

Since the UV spectral slope of a young stellar population is more
sensitive to reddening than to age, we could estimate the steller 
reddening $E_{B-V,I}$ from the observed shape of the UV continuum.
We first corrected each spectrum for Galactic reddening using the 
\cite{Fitzpatrick:1999} extinction curve. Next, we assumed an 
intrinsic UV spectral slope appropriate to a young stellar population, 
i.e.,  $\beta \approx -2.3 $ where $F_{\lambda} \propto \lambda^{\beta}$
\citep{Leitherer:1999}, and reddened this template using a Calzetti extinction law \citep{Calzetti:2000}. 
Using custom software which allowed spectral binning and interactive 
adjustments to the fitting window, we fit the steller reddening $E_{B-V,*}$ and
the UV continuum simultaneously. 

The uncertainty in our estimate of the stellar reddening
was taken to be the larger of (1) the range resulting from 
placement of the continuum or (2) the range obtained by repeating 
the exercise 
with $\beta = -2.1$ and $\beta = -2.5$.  The resulting values of the stellar reddening in 
Table~\ref{tab:cos_measure} range from 0.02 to 0.32. These values are not large compared 
to normal galaxies and demonstrate that the UV light emerges through holes
in the overall dust distribution.

The global continuum fits generally seemed reasonable around individual spectral lines.
Where the pseudo-continuum formed by the blending of lines in the underlying galaxy/AGN 
spectrum was significant, however, we fit a linear continuum to bandpasses on either side 
of the spectral line and normalized the spectrum by the fitted continuum. We illustrate the 
allowed range for the continuum placement near \lya\ in Figure~\ref{fig:lya_luminosity}. 
We used these limits to estimate the uncertainties in our measurements of the \lya\ emission. 

%%% MOVED PARAGRAPH ON INFRARED EXCESS TO SECTION 4
%
%We attribute this
%problem to the presence of stellar photospheric and wind features which are poorly delineated
%at the resolution and S/N ratio of our spectra. 

%SAVE - THESE ARE THE CONTINUUM ADJUSTMENTS
%For \i01003-2238, \i11598-0112, and \i16487+5447, the stellar continuum did not 
%describe the continuum slope adequately at \lya, presumably due to the contribution
%from the AGN in these objects, and we fit a local continuum. 

\subsection{Echellete Optical Spectroscopy}

We previously obtained optical spectra with ESI in echellete configuration on the Keck~II
telescope as described previously \citep{Martin:2005}.  The resolution through the 
1\farcs0 by 20\farcs0 ESI slit, $R \approx 5000$ or 60\kms\ FWHM, is comparable 
to that obtained for the most compact ULIRGs with COS G130M. These slits intersect
the locations of the COS pointings and are shown in \cite{Soto:2012b}. We extracted
an ESI spectrum along a 2\farcs5 length of each slit centered at the position 
of the COS aperture. The rows of the spectrum were weighted by their distance
from the center of this aperture using the vignetting function for the G130M 
grating.\footnote{
          COS ISR-2010-09 provides throughput measurements in only the XD (cross dispersion) 
          direction for G130M. The Instrument Science Report compares the vignetting function for
          G160M in the XD and spectral directions and shows that they are nearly identical;
          hence, we have tacitly assumed the G130M vignetting is also quite similar the XD
          and spectral directions. }

% because the COS throughput falls off sharpley at radii greater than 0\farcs4.
%
%of 2-4 \AA\ in most of these matched spectra

\subsubsection{Emission-Line Diagnostics}

All the ULIRGs show strong Balmer and forbidden line emission in these ESI spectra. 
We fit these emission lines simultaneously using the procedure described in
\cite{Soto:2012a}   which includes a stellar Balmer absorption component.
Table~\ref{tab:esi_measure} gives the 
reddening estimated from the Balmer decrement. This nebular color excess is
similar to normal galaxies because it does not probe the dust in the regions 
completely opaque to optical light. The reddening towards these HII
regions is generally, but not exclusively, larger than our rough estimates
of the stellar reddening from the UV spectral slope in Section~\ref{sec:beta}
as would be expected considering that the younger ages of the stars emitting 
the Lyman continuum limit the distances the stars have traveled from their birth clouds.

\cite{Soto:2012b} previoulsy measured well-known, emission-line diagnostics
from the ESI spectra and placed pairs of emission-line ratios in diagrams
that distinguish excitation by photoionization from massive stars from radiative shocks
\citep{Allen:2008}. \cite{Soto:2012a} found the broad component of the line profile 
is excited exclusively by shocks moving at speeds of 200-400\kms.
In Table~\ref{tab:esi_measure}, we list
the shock velocity estimated from the line ratios ${\rm {[OIII]}}/\Hb$ and ${\rm [OI]}/\Ha$ at
the position of the COS aperture. 

\subsubsection{Doppler-shifted Resonance Absorption}

Measurements of the \naI\ resonance absorption for \i00262+4251, \i08030+5243,
\i16487+5447, and \i23365+3604 were collected from \cite{Martin:2005}. Table~\ref{tab:esi_measure}
shows these equivalent widths and new measurements made in a similar manner
for the other COS ULIRGs and \i09583+4714-B. The ESI spectra of \i10378+1109 
and \i08030+5243 show the strongest \naI\ absorption at 6.7 and 7.4 \AA, respectively.
The \naI\ doublet in \i09583+4714-B and \i17028+5817 is not strong enough to
definitively claim the detection of a non-stellar component.
The interstellar \naI\ resonance absorption, when detected however,
is  blueshifted roughly 200\kms\ at maximum depth, and the blue 
wing becomes undetectable around -700\kms\ as it approaches the continuum
level and blends with a \ion{He}{1} emission line.  An exception is
\i08030+5243, the only example of redshifted \naI\ absorption shown in
\cite{Martin:2005}.

\subsubsection{Aperture Corrections for \Ha\ Luminosity}  \label{sec:aperture}

Most applications of the optical spectral measurements rely on line ratios
rather than absolute fluxes, so the geometrical difference between the
circular COS aperture (2\farcs5 diameter) and the rectangular ESI aperture 
(1\farcs0 by 2\farcs5) is not a concern, especially since the COS throughput
falls rapidly at radii beyond $0\farcs4$ from the center. 

The geometrical distinction between the rectangular ESI slit and circular 
COS PSA needs to be taken into account, however, when we compare 
the strengths of \Ha\ and \lya. We used the  R band images \citep{Murphy:1996} to model the 
\Ha\ aperture correction because the scale lengths of galactic disks in the R band and 
in \Ha\ are very similar \citep{Kennicutt:1989}. Specifically, we vignetted
the R image using the COS function described above, and then measured
the R band flux in the 2\farcs5 diameter COS aperture relative to that
in a 1\farcs0 by 2\farcs5 rectangular ESI aperture. To estimate the \Ha\ 
flux that would be measured through a 2\farcs5 diameter aperture,  
we multipled the exctinction-corrected \Ha\ fluxes by these aperture corrections 
which we also list in column~4 of Table~\ref{tab:esi_measure}. 

{\rev
The spatial extent of \lya\ halos can also introduce systematic 
errors in estimates of \fesclya. For example, as demonstrated in
\cite{Hayes:2013}, resonance scattering enlarges galaxies in \lya\ by a factor of 
1 to 3.6 over their size in \Ha; hence, our aperture may include a larger fraction
of the total \Ha\ emission than of the total \lya\ emission. This type of bias
is unlikely for the COS ULIRG sample, however, because the LARS data empirically
show that a low dust content is a necessary requirement for \lya\ halos to 
extend well beyond the region of \Ha\ emission \citep{Hayes:2013}.
}

%Imaging in the \lya\ and \Ha\ lines would clearly increase the accuracy of the 
%\fesclya\ estimates.

\section{COS Spectroscopy of \lya}  \label{sec:results}

\begin{figure}[h]
%  \hbox{\includegraphics[height=10cm,angle=-90,trim=0 400 0 0]{figures/l.ps}}
  \hbox{\includegraphics[height=10cm,angle=-90,trim=0 450 40 0]{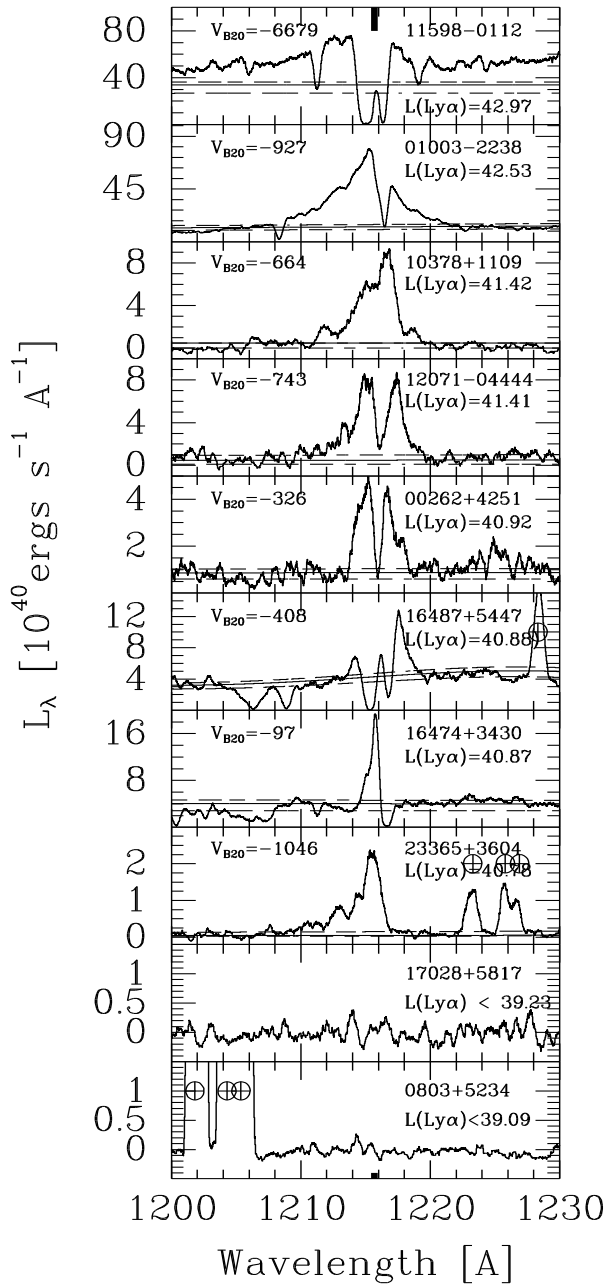}}
   \caption{\footnotesize Rest-frame \lya\ emission in 
     COS G130M spectra of ULIRGs. 
{\rev
Airglow lines are denoted by $\oplus$, and the rest wavelength of \lya is 
indicated by the thick tick marks at the top and bottom.}
     Spectra are ordered from
     the highest to lowest line luminosity (denoted as
     a logarithm in ergs~s$^{-1}$). The Doppler shift of the blue wing,
     $V_{B20}$ is given in \kms.
% (see text for details), is marked by the vertical blue line. 
     The fitted continuum and its uncertainty range are shown
     by the horizontal solid and dashed lines, respectively; 
{\rev note the strong pseudo-continuum from metal lines in the spectrum of the 
mini-BAL  \i11598-0112.}
With the exception of \i23365+3604, the blue line wing reaches higher velocity
as the ULIRG luminosity rises.
   }
   \label{fig:lya_luminosity} \end{figure}

Figure~\ref{fig:lya_luminosity} shows the COS G130M spectra
around \lya.  We detect emission from eight of the ten ULIRGs. 
Considering the large optical depth of the \lya\ transition, 
see Eqn.~\ref{eqn:lya_tau0}, and extreme concentration of dust in ULIRGs, 
it is remarkable that any \lya\ photons escape from ULIRGs. It is 
perhaps easier to understand the absence of \lya\ emission from
\i0803+5234 and \i17028+5817; the upper limits 
on their \lya\ fluxes are far below those detected.

%FIGURE 
%l b r t

\begin{figure}[h]
% lbrt; Right is top for ppt
% \hbox{\hfill \includegraphics[height=10cm,angle=90,trim=25 0 50 -400]{figures/ppt_09583+4714.ps}
% \hbox{\hfill \includegraphics[height=10cm,angle=-90,trim=0 400 0 0]{figures/pcygni.ps}
 \hbox{\hfill \includegraphics[height=10cm,angle=-90,trim=0 400 0 0]{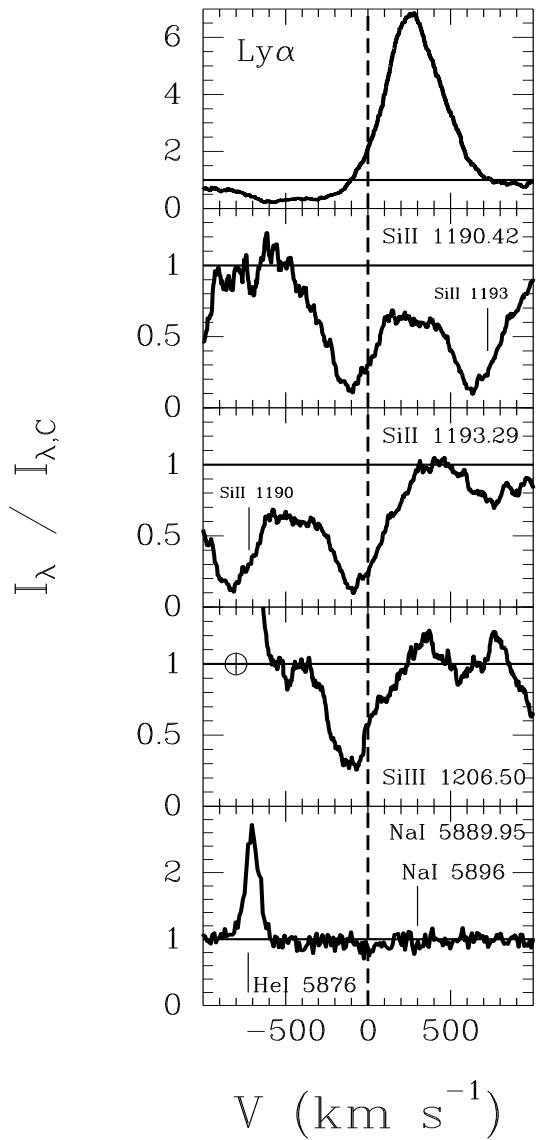}
                \hfill}
   \caption{\footnotesize Continuum normalized spectra of the galaxy \i09583+4714-B.
     The COS G130M spectrum of the \i09583+4714-B galaxy show a P~Cygni 
     \lya\ profile consistent with an outflowing wind or shell.
     The UV absorption lines from low-ionization-state gas
     are blueshifted relative to the systemic velocity set by the 
     optical emission lines. The deepest part of the absorption trough is shifted 
     roughly 100\kms\ whereas the blueshifted wing extends to no more than
     300 to 400\kms. Other lines detected, but not shown,
     include \siII\ 1260.42 which is blended with \sII\ 1259.52,
     \sII\ 1253.81, \arI\ 1066.66, \nII\ 1083.99, and multiple
     \feII\ lines.
     The optical spectrum shows no absorption in the
     \naI\ doublet indicating sodium is mostly ionized.
   }
   \label{fig:pcygni} \end{figure}

To better appreciate the surprising shape of the ULIRG \lya\ profiles, we
show the spectrum of the non-ULIRG, \i09583+4714-B, in Figure~\ref{fig:pcygni}.
The \lya\ emission is redshifted roughly 250\kms\ relative to the optical
emission lines, and the resonance absorption in several \siII\ transitions
and \siIII\  is blueshifted 100\kms\ or more. Blueshifted \lya\ absorption is
visible after smoothing the spectrum. P~Cygni line profiles with redshifted emission 
and blueshifted absorption as seen in this spectrum arise from resonance scattering
in an expanding, neutral medium \citep{Verhamme:2008p89}. 
Six of the 13 \lya\ emitters from the \Ha-selected galaxy sample of \cite{Wofford:2013} 
and all four of the infrared luminous galaxies ($10^{11} < L_{IR} / \lsun < 10^{11.7}$) observed 
by  \cite{Leitherer:2013} show these P~Cygni \lya\ profiles; and they have been modeled
with outflowing shells of neutral gas expanding at velocities $\sles\  200$\kms.

%and are therefore called P~Cygni profiles after the stellar prototype. When
%\lya\ emission is detected from star-forming regions in galaxies \citep{Shapley:2003}, 
%it often has a P~Cygni profile. This shape 

%\footnote{
%    From the reference frame of the shell on the near side of the galaxy,
%    the light emitted by the starburst appears redshifted, and
%    photons emitted at frequencies slightly higher than the
%    resonance will be scattered by outflowing H atoms 
%    producing a broad, blueshifted absorption trough.}
%    
%se profiles are unlike the galaxies without \lya\ emission (13/20), for example, 

The dominance of blueshifted emission among the ULIRG \lya\ profiles
indicates different physical conditions than those which produce
the P~Cygni profiles. The profiles are unlike the local \Ha-selected galaxies 
without P~Cygni profiles; they display \lya\ absorption 
troughs with damping wings fit by \ion{H}{1} column densities in the range of $1.0 \times 
10^{20}$ to $2.3 \times 10^{21}$\col\ \citep{Wofford:2013}. Even among
UV-selected galaxies at $z = 2-3$ where $30\%$ show multiple emission peaks, \cite{Kulas:2012}
find the redshifted peaks are normally stronger than the blueshifted emission component.
Only the line profiles of \i16474+3430 and \i16487+5447 show most of the net emission 
redward of the systemic velocity, and only  \i16474+3430 
may (depending on the continuum placement) have a weak, 
blueshifted absorption component.

We have ordered the ULIRG spectra by decreasing \lya\ luminosity in 
Figure~\ref{fig:lya_luminosity}. The most luminous ULIRG, 
\i11598-0112, is a quasar on account of its dominant AGN and
bolometric luminosity;   blending of its broad \lya\ profile 
with emission lines from \nv, \siII, and \cIII\ produces line
wings much broader than on any other COS ULIRG. If a G160M 
spectrum of this object is obtained in the future, the \cIV\ 
emission could be used as a template profile to deblend this complex as
shown, for example, in  \cite{Leighly:2007}. For now,
the line flux we measure from integrating this broad
emission complex provides an upper limit on the \lya\ 
luminosity due to the contamination from other lines.

{\em Blueshifted emission wings} are prominent on the \lya\ profiles of
\i01003-2238, \i12071-0444, \i10378+1109, and \i23365+3604. 
These wings extend roughly -1000\kms\ from the systemic velocity.
The process that generates these line wings does not appear
to be uniquely associated with an active nucleus. Although
we argued in Section~\ref{sec:sample_selection} that the
starburst and AGN make roughly equal contributions to the
bolometric luminosities of \i01003-2238 and \i12071-0444, 
starbursts power \i10378+1109 and \i23365+3604. 
Radiative transfer modeling has previously linked
blueshifted \lya\ emission to either radial infall
\citep{Dijkstra:2006a,Dijkstra:2006b} or low columns of 
neutral gas \citep{Verhamme:2014,Behrens:2014}.

%FIGURE 

\subsection{Escape Fraction of \lya\ Photons} \label{sec:fesc}

To gain insight into the role of radiative transfer effects
in shaping the \lya\ emission, we compare the strengths of 
the \lya\ and \Ha\ emission. Scattering off neutral hydrogen
atoms increases the pathlength of \lya\ photons through the 
ISM. The \lya\ photons will be more attentuated 
than the \Ha\ radiation due to a combination of the longer distance
traveled due to scattering {\it and} the frequency
dependence of interstellar reddening. We will distinguish 
between these processes using the attenuation of the
far-UV continuum.

%relative intensity of recombination lines emitted by hydrogen atoms 

The intrinsic ratio $F(\lya)/F(\Ha)$ depends only on the electron 
density and temperature for Case~B conditions, i.e., the ionized 
region is optically thick to photons just above the hydrogen edge 
($IP = 13.60$~eV). In the low density limit, we have $n_e << n_{e,crit} = 
1.55 \times 10^4$\cm3 and collisions can be neglected. The 
intrinsic \lya\ flux is then 8.1 times the intrinsic \Ha\ flux
\citep{Draine:2011}. 

Galaxies typically have \lya\ to \Ha\ flux ratios much less 
than the Case~B ratio. It is therefore common to describe 
the suppression of \lya\ photons 
relative to the intrinsic \lya\ flux indicated by the \Ha\ emission.
We define the escape fraction of \lya\ photons as
\begin{eqnarray}
\fesclya\ = \frac{F(\lya)}{8.1 \times F(\Ha)_{cor}}.
\label{eqn:fesclya} \end{eqnarray}

We calculate the intrinsic \Ha\ flux from the observed \Ha\ flux 
and the nebular reddening using the relation appropriate
for a \cite{Cardelli:1989} reddening curve, namely
\begin{eqnarray}
F_{cor}(\Ha) = F_{obs}(\Ha) \times 10^{1.012 E_{B-V}}.
\end{eqnarray}
The color excess for the ionized gas is
\begin{eqnarray}
%E(B-V) = 1 / 0.9692 \log ([F(\Ha) / F(\Hb)]_{obs} / 2.86).
E_{B-V} = \frac{1}{0.9692} \log \left [ \frac{F(\Ha) / F(\Hb)}{2.86} \right ]
\end{eqnarray}
and was computed from the same reddening curve. We
take the intrinsic Balmer decrement from
\cite{Osterbrock:2006} for
Case~B conditions, an electron density of
$n_e = 100$\cm3, and a temperature of  $T_e = 10^4$~K.

%\i17028+5817 and \i08030+5243, Figure~\ref{fig:lya_luminosity} shows the 

\begin{figure}[h]
% \hbox{\hfill \includegraphics[height=10cm,angle=-90,trim = 0 50 0 0]{figures/fesc_ebv.ps}
 \hbox{\hfill \includegraphics[height=10cm,angle=-90,trim = 0 50 0 0]{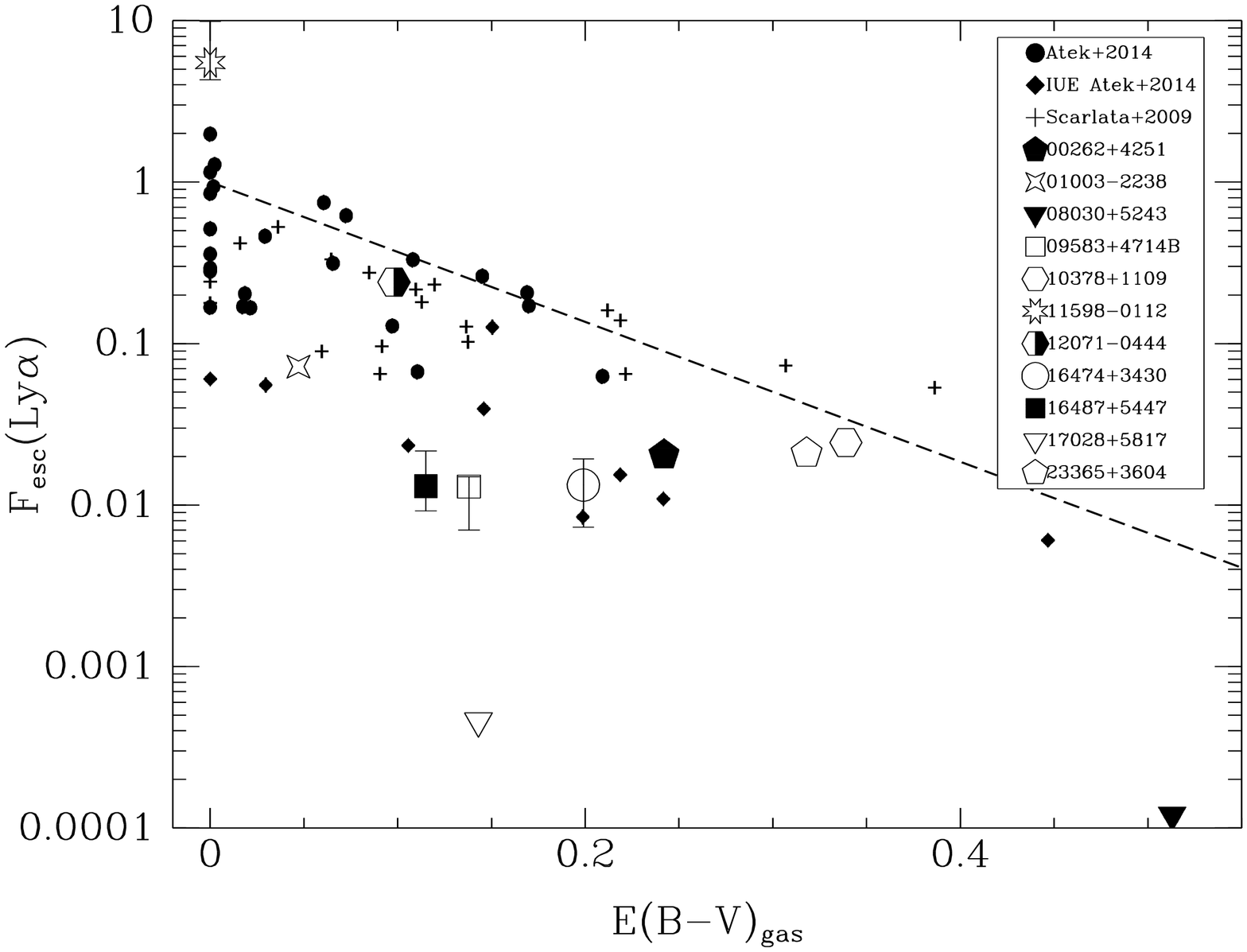}
  \hfill}
   \caption{\footnotesize Escape fraction of \lya\ photons vs.\~(nebular) color excess.
     The inverted triangles are upper limits. 
     The dashed 
%(blue) 
     line illustrates the suppression of the continuum near \lya\
     from dust as given by a \cite{Cardelli:1989} reddening curve and the color 
     excess (measured from the Balmer decrement).  When the \lya\ escape fraction
     of a galaxy lies far below this curve, we attribute the extra \lya\ attenuation
     to the longer path length caused by resonance scattering. 
     The ULIRGs do not stand out from other
     samples of \lya\ 
     emitting galaxies shown here for comparison. See Section~\ref{sec:bal}
 for further discussion 
     of the Seyfert I galaxy, \i11598-0112.
   }
    \label{fig:fesc_ebv}   \end{figure}

The \lya\ escape fraction computed this way ranges from 1-3\% 
among the starburst-dominated spectra with \lya\ emission.
For our two non-detections, we find $ \fesclya\ < 0.5\%$. 
Figure~\ref{fig:fesc_ebv} shows that these escape fractions decline
with increasing color excess. Among the nine \lya\ detections,
the Spearman rank order correlation coefficent is $r_S = 0.42$,
and the correlation deviates from the null hypothesis of no
correlation by just $1.2\sigma$. These escape fractions fall within 
the range measured for local galaxies \citep{Atek:2014} 
as indicated in   Figure~\ref{fig:fesc_ebv}.

%from Table~\ref{tab:cos_measure}

To assess the importance of resonance scattering, we
compare the attenuation of \lya\ to that of the continuum
emission near \lya. The continuum attenuation is shown
as a function of the color excess by the dashed line
in Figure~\ref{fig:fesc_ebv}. 
From this comparison we see that the \lya\ attenuation
in three of the ULIRGs -- \i1207-0444, \i10378+1109,
and \i23365+3604 -- is similar to the continuum attenuation.
In spectra of the  other non-AGN ULIRGs and \i09583+4715-B, 
the \lya\ escape fractions is greatly suppressed relative
to the continuum due to scattering.
The upper limits on the \lya\ flux for the two undetected 
ULIRGs imply the resonance line is attenuated over a factor of 100 
more than the far-UV continuum. 
We conclude that the \lya\ optical depth varies widely among the COS ULIRGs,
scattering has the least affect on the \lya\ emission emerging from
\i1207-0444, {\rev \i23365+3604}, and \i10378+1109. 

%, so the scattering optical depth is lowest in these objects. 

%FIGURE 3

\subsubsection{High \fesclya\ from a Mini-BAL Quasar} \label{sec:bal}

%The strength of \lya\ relative to \Ha\ 
%grows as the gas density increases because collisions
%of hydrogen atoms with protons and electrons are
%effective at promoting electrons from the long-lived
%$2s$ level to the $2p$ level where they rapidly decay
%via emission of a \lya\ photon. 

\begin{figure}[t]
% \hbox{ \includegraphics[height=8cm,angle=-90,trim=50 0 0 250]{figures/h3_11598-0112.ps} }
% \hbox{ \includegraphics[height=8cm,angle=-90,trim=50 0 0 250]{figures/h6_11598-0112.ps} }
 \hbox{ \includegraphics[height=8cm,angle=-90,trim=50 0 0 250]{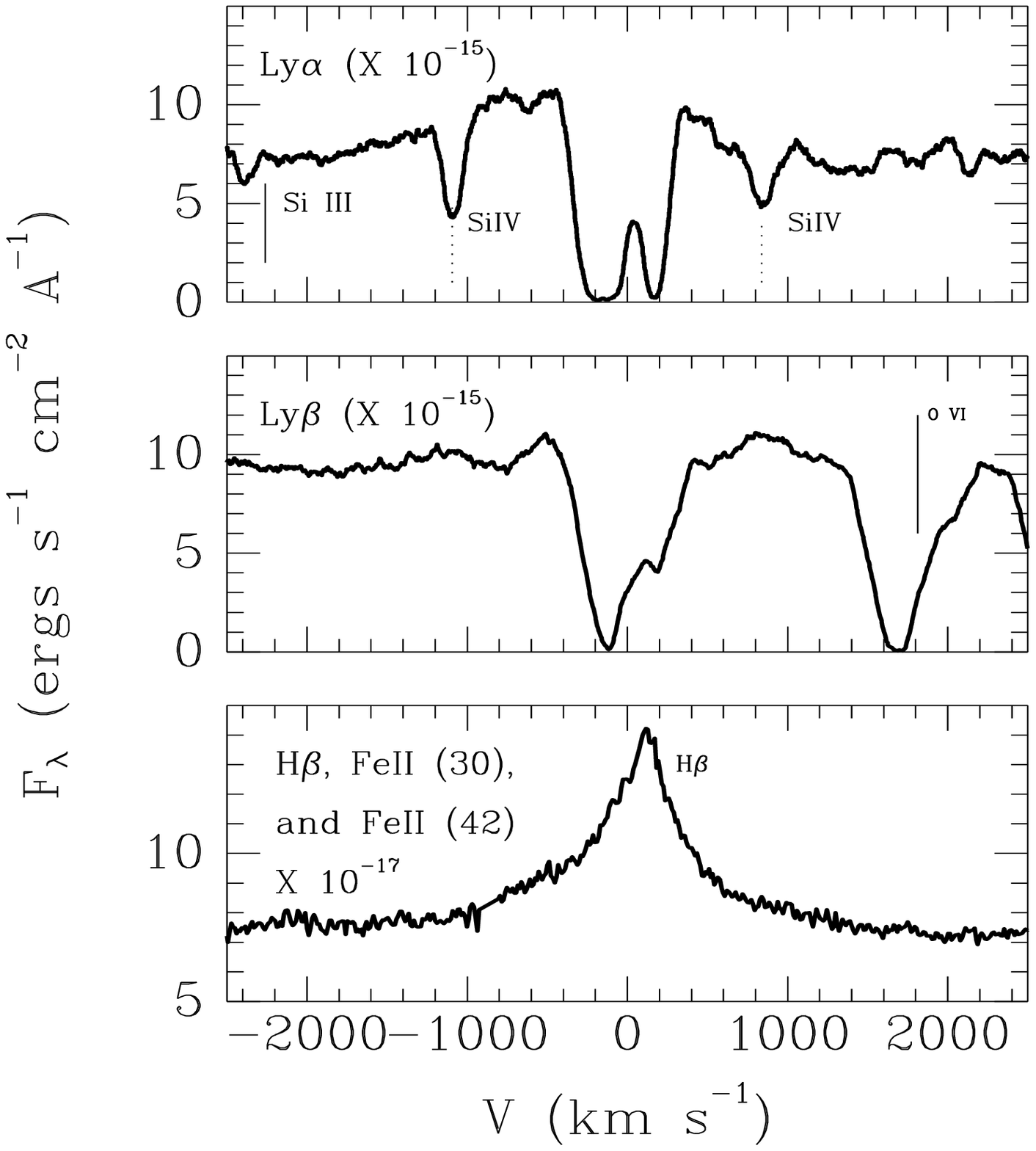} }
 \hbox{ \includegraphics[height=8cm,angle=-90,trim=50 0 0 250]{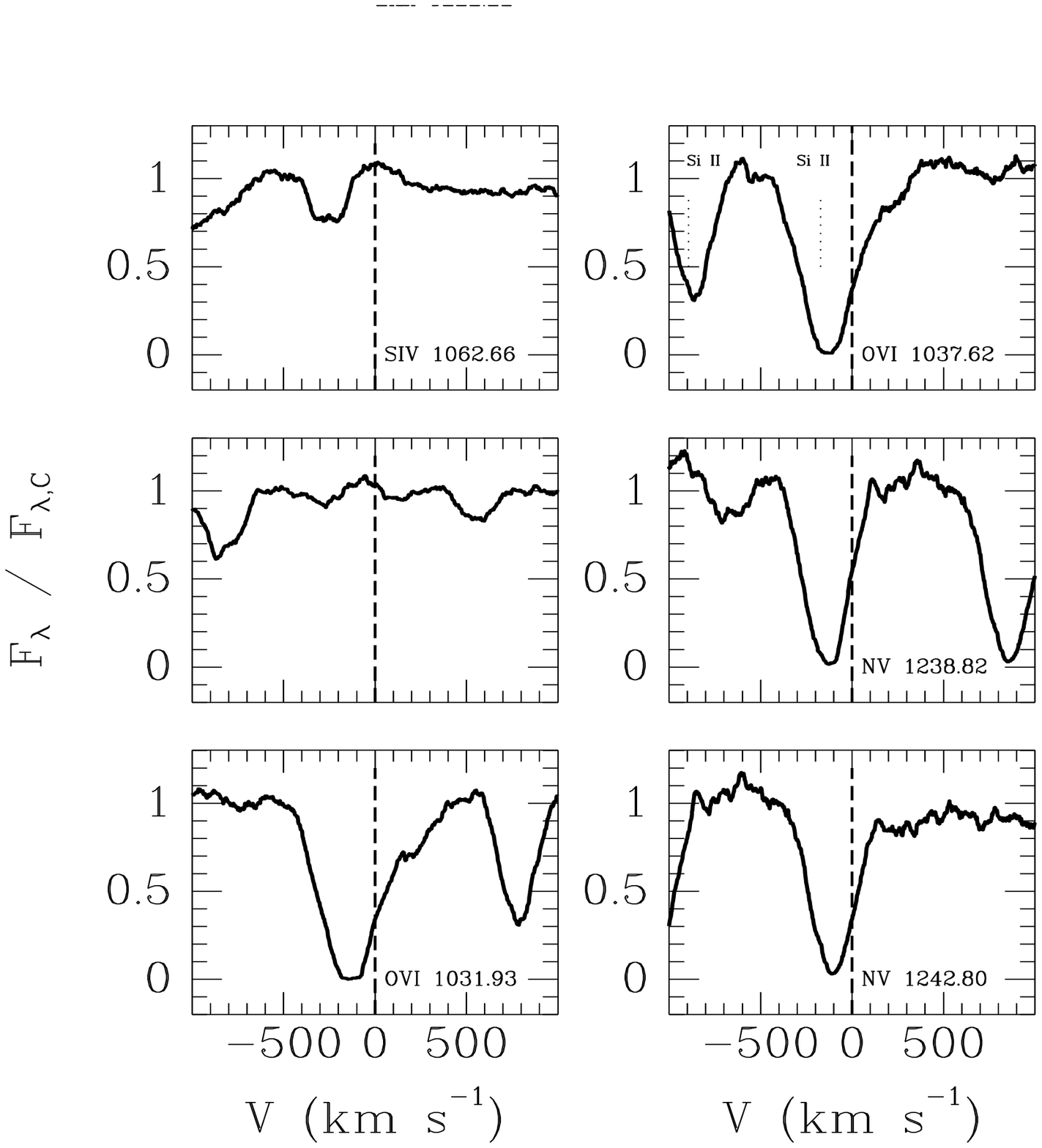} }
   \caption{\footnotesize 
     COS G130M and optical \Hb\ spectra of the narrow-line, Seyfert~1 ULIRG \i11598-0112.
     (a)
     The \lya\ and \lyb\ profiles show strong absorption near the systemic
     velocity. The hydrogen Balmer lines have broad wings reaching nearly 
     2000\kms. While we expect such wings on the \lya\ profile, the
     even broader wings visible in the top panel are believed to
     be heavily  blended with broad emission from metal lines including
     \siII.      See also the [\ion{O}{3}] profile in the Appendix.
     (b) 
     The COS G130M spectrum shows strong absorption from high-ionization gas.
     No absorption is detected from low-ionization-state gas or \siIII. 
     The absorption from high-ionization 
     state gas is blueshifted up to 500\kms.
     The equivalent widths of \ion{O}{6} $\lambda 1032$ and \ion{N}{5} $\lambda 1239$
     are 1.43 \AA\ and 1.29 \AA, respectively.
   }
   \label{fig:cos_11598}    \end{figure}

Our measurements allow unusually high values of \fesclya\
for \i11598-0112{\rev , but} the accuracy of this estimate  merits additional discussion 
because a quasar powers this object. The uniqueness of this object's
far-UV spectrum among the COS ULIRGs might have been anticipated
from the strong optical \feII\ emission lines shown in Figure~\ref{fig:esi_11598}
which \cite{Lipari:1993,Lipari:1994} have previously associated 
with the emergence of nascent broad absorption line quasars in ULIRGs.
Our detection of  broad, high-ionization-state lines in 
Figure~\ref{fig:cos_11598} {\rev solidly identifies} this object as a mini-BAL (broad 
absorption line) quasar.

The assumptions underlying the intrinsic strength of \lya\ in
Equation~\ref{eqn:fesclya} may not hold for \i11598-0112.
For example, the intrinsic Case~B ratio of  $F(\lya) / F(\Ha) $ will 
be larger than 8.1 as $n_e$ approaches the critical density; collisions 
promote electrons from the $2s$ state to $2p$ where they can decay by 
emission of a \lya\ photon. In some quasar spectra, the intrinsic 
ratio is driven in the oppposite direction (i.e., towards
enhanced Balmer emission relative to \lya) by thermalization of 
the energy levels \citep{Kwan:1979}. Thermalization only becomes
important at very high densities however.

The far-UV spectrum illustrated in Figure~\ref{fig:cos_11598} 
helps us place constraints on the electron density. The \i11598-0112 
outflow is only detected in lines from high-ionization-state gas.
We detect blueshifted absorption in the \ovi\ 1032, 1038, 
\ion{S}{4}$\lambda 1062$, and \nv\ 1239, 1243 transitions;
and the widths of these lines place \i11598-0112 in the mini-BAL (broad
absorption line) quasar category. 
To directly measure $n_e$, we need to measure the absorption 
from a collisionally excited level of one of the ions  \citep{Borguet:2013,
Dunn:2012}. Our spectrum covers 
\ion{S}{4}$^*~\lambda 1073$ at comparable S/N ratio to the resonance
transition at $\lambda 1062$~\AA. No absorption is detected from
the excited line, so the electron density lies well below the 
critical density of $3.16 \times 10^{4}$\cm3. To summarize, the density 
is too low for thermalization to determine the level populations of hydrogen but  
is likely high enough for the intrinsic Case~B ratio to be significantly larger than 8.1.

%we naively measure $\fesclya\ \approx  5$ for \i11598-0112
%but this value overestimates the escape fraction. 

{\rev
By assuming the low density limit, we have underestimated the value of the denominator 
in Eqn.~\ref{eqn:fesclya}. As explained above, however, we have 
overestimated the \lya\ flux in the spectrum (i.e., the numerator in Eqn.~\ref{eqn:fesclya})
by including blended metal lines in our \lya\ flux. It follows that we obtain no quantitative 
limit on \fesclya\  for \i11598-0112. Our point is that the fiducial value stands 
out compared to the other ULIRGs, and we attribute this to the dominant AGN in \i11598-0112.
We speculate that this object is viewed through an ionized cone (with neutral walls)
pointed in our direction. 
In radiative transfer calculations,  the escape fraction of \lya\ 
photons from a compact source in an anisotropic gas distributions 
can be boosted by factors of $2 - 4$ relative to the isotropic case when viewed through low-$N(HI)$ holes 
\citep{Behrens:2014}.}

%when dust is included in the clumps, however, boosts in the equivalent width appear to be
%driven primarly by reduced escape fractions of UV photons rather than enhanced \lya\ escape
%\citep{Gronke:2014}.

% THE OLD PICTURE, NOT SO RELEVANT
%
%Scattering of \lya\ photons off the surfaces of neutral clouds in an ionized, dust-free medium  
%can, in theory, preferentially enhance the escape of \lya\ photons relative 
%to the UV-continuum.  
%
% surfaces and encounter little dust attenuation
%\citep{Neufeld:1991,Hansen:2006}.  Recent analyses have shown that this
%requires both extreme density contrasts and very slow outflows
%\citep{Laursen:2013,Duval:2014}. 
%
% EVEN FOR SPHERICAL GEOMETRY, YOU CAN GET BOOSTS
% Gronke:2014} have shown that there can exist

%FIGURE 
% lbrt;  left is the true bottom; right is true top; top is true left; bottom is true right.

%Alternatively, or in addition,  anisotropic escape of \lya\ radiation 
%has been shown to produce higher than expected equivalent widths 
%\citep{Laursen:2013}. In an isotropic but clumpy medium, in contrast,
%boosting the escape of \lya\ photons relative to the continuum requires
%both extreme density contrasts and very slow outflows \citep{Duval:2014}. 
%Hence, one possible explanation for the high escape fraction of
%\lya\ photons (relative to the continuum) in \i11598-0112 
%is resonance scattering off the neutral wall of an ionized
%cone pointing in our direction. 

% through isotropic
%but clumpy medium \citep{Neufeld:1991,Hansen:2006}
%would require extreme density contrasts between very dusty, neutral clouds 
%and an ionized, dust-free intercloud medium 

%% COULD THIS REFLECT GALAXIES BEING EDGE ON?
%The observed growth in the continuum normalized \fesclya\ with the color excess
%of galaxies is consistent with this two-phase geometry \citep{Atek:2014}. 

%Attenuation that is independent of wavelength, as introduced by 
%\cite{scarlata:2009} to describe opaque clouds,

\subsubsection{Correlation of \fesclya\ with the Width of the \lya\ Profiles} \label{sec:line_profile}

%For \i09583+4714-B, 
%     The \lya\ absorption blueward of the systemic velocity, and the emission redward
%     of the systemic velocity, are well-described by the expanding shell (or wind) model.

%\subsubsection{Broad Line Wings}

To facilitate direct comparison of the linewidth with \fesclya,
we need a quantitative characterization of broad wings and/or line asymmetries.
Describing the broad wings and asymmetries of the \lya\ profiles
robustly, however, provides a challenge. The full width of the 
profile at half maximum intensity, for example, is not uniquely 
defined for the double peaked profiles, nor does it flag line profiles
with broad wings or significant asymmetry. In Appendix~A, we define
velocity markers $V_{B}$ and $V_{R}$ as a continuous function of 
the fractional profile area. For any percentile of the profile area, 
the full width of the {\it interpercentile velocity}  (IPV) is well defined. The
widths at the 0.20, 0.50, and 0.80 percentiles are marked on each 
\lya\ profile in Appendix Figures~a-i to illustrate this concept.

%In Appendix~\ref{sec:appendix}, we define measurements 
%of the wing velocity, the line width, and the line asymmetry
%based on profile area rather than height. 

%\subsubsection{Relation of Gas Kinematics to \fesclya\ and Luminosity}

% VB1216 and fesc 0.83 and 2.4 sigma
% VR1216 ($r_s = 0.35$ and null hypothesis rejected at $0.99 \sigma$).

\begin{figure}[h]
% \hbox{\hfill  \includegraphics[height=10cm,angle=-90,trim=0 50 0 0]{figures/fesc_vb.ps}
 \hbox{\hfill  \includegraphics[height=10cm,angle=-90,trim=0 50 0 0]{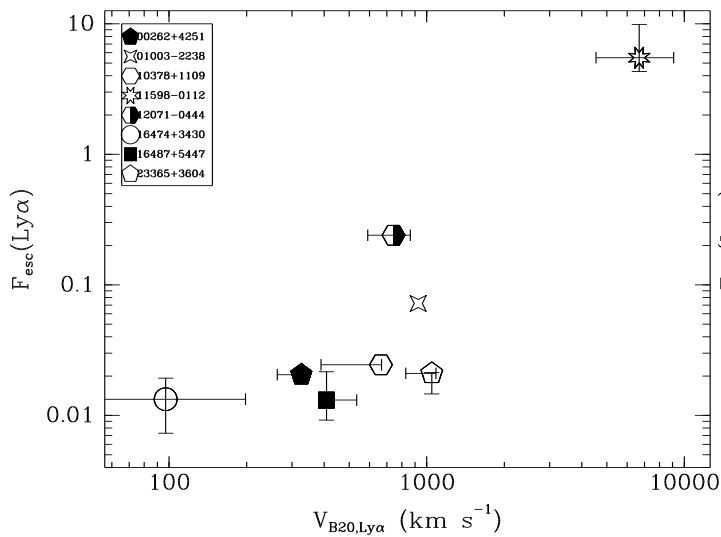}
                   \hfill}
   \caption{\footnotesize 
The \lya\ escape fraction increases with the blueshift of the \lya\ emission wing,
defined here as the velocity beyond which 10\% of the \lya\ emission line flux is measured. 
The Spearman rank order correlation coefficient is $r_s = 0.83$, and the null hypothesis
is rejected at $2.4 \sigma$.
The correlation strength is driven by the large \fesclya\ values inferred for
the ULIRGs with the highest AGN fraction.
   }
   \label{fig:fesc_v}   \end{figure}

The escape fraction increases as the width of the \lya\ 
emission line increases. This correlation is both more significant and
stronger than that for \fesclya\ and color excess. The correlation with line 
width reflects an underlying correlation of \fesclya\ with the Doppler shift
of the blue wing. This correlation is stronger than the one with total linewidth,
and  \fesclya\ does not show a correlation with the velocity of the red wing.
Figure~\ref{fig:fesc_v} compares \fesclya\ to the Doppler shift of the blue wing.

This measured correlation is in 
the opposite direction to that expected for a static slab.
For a static slab, the width of the resonance line would increase
with the number of scatterings. Since the longer pathlengths would
decrease \fesclya, we would expect the \lya\ escape fraction to
decline as the line width increased. 

{\rev
Perhaps the simplest intrepretation of this correlation is that \fesclya\ depends 
strongly on the gas kinematics, not just the column density of gas and dust. 
However, previous empirical evidence for this effect has been limited to studies
of a few galaxies \citep{Kunth:1998,Atek:2008,Wofford:2013}. Among the HII-dominated 
ULIRGs, we find \fesclya\ is correlated more strongly with 
emission-line width than with color excess. }

%We expect the escape of \lya\ photons from galaxies to be enhanced by
%galactic winds because the hot ($10^7 - 10^8$~K) wind fluid is highly ionized. 

%The strength of the correlation of \fesclya\ with
%the equivalent width of the \lya\ emission is similar 
%($r_s = -0.82$ and null hypothesis rejected at $2.3 \sigma$).

%FIGURE 8

\subsubsection{The \lya\ Equivalent Width} \label{sec:ew}

%The equivalent width of the \lya\ emission will necessarily be correlated with \fesclya\
%because the \lya\ flux enters into the numerator of both quantities. 

Often the \Ha\ spectrum is not accessible for \lya\ emitters. In this situation,
the \lya\ equivalent width may be a useful indicator of the attenuation of \lya\ photons.
The equivalent width measures the strength of the emission line relative to the 
continuum.\footnote
     {We define the \lya\ equivalent width by an integral
     over the line such that $
     W(\lya) \equiv \int (F_{\lambda,c}(\lambda) - 
     F_{\lambda}(\lambda)) / F_{\lambda,c}(\lambda) d\lambda$,
     where $F_{\lambda,c}$ is the fitted continuum.
     It represents the continuum bandpass (measured in \AA) 
     with the same flux as the emission line.}
The ULIRGs as a class show moderate equivalent widths similar to the values
measured previously for the local IUE-selected galaxies plotted Figure~\ref{fig:fesc_ebv}.
The median $W(\lya)$  is 11 \AA\ among the COS ULIRGs, and the  strongest equivalent width 
ULIRGs include the AGN.

For continuous star formation, the UV continuum (like the
\Ha\ emission) scales linearly with the SFR. In contrast to the
$F(\lya)/F(\Ha)$ ratio, however, the strength of \lya\ relative to the UV
continuum is quite sensitive to the  age of the stellar population because
the UV continuum can be produced by stars of insufficient mass to generate
a substantial ionizing continuum. Considering the sensitivity of $W(\lya)$ 
to stellar age \citep{Cowie:2011} then, the expected strength of the
correlation between these \lya\ measurements is not clear a priori.

\begin{figure}[h]
% \hbox{\hfill \includegraphics[height=10cm,angle=-90,trim=0 50 0 0]{figures/fesc_ew.ps}
 \hbox{\hfill \includegraphics[height=10cm,angle=-90,trim=0 50 0 0]{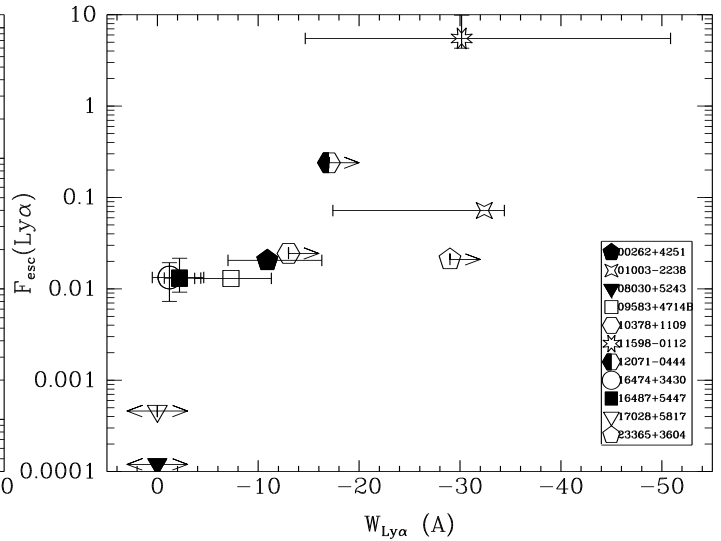}
                \hfill}
   \caption{\footnotesize 
     Correlation of \fesclya\ with equivalent width of \lya\ emission. 
     The ULIRGs with the highest escape fractions
     also have large equivalent widths. The correlation coeffient is $r_s = -0.82$ 
     for the nine detections, and the null hypothesis is rejected at the $2.3\sigma$ level. 
     The lower limits on $|W(\lya)|$ arise from the absence of continuum detections.
     The downward pointing triangles illustrate the upper limits on 
     the two ULIRGs with no \lya\ detection; these galaxies were
     arbitrarily placed at zero equivalent width since the continuum
     is not detected either.  
   }
    \label{fig:lya_ew}   \end{figure}

Figure~\ref{fig:lya_ew} illustrates how $|W(\lya)|$ grows with \fesclya.
Even among the starburst-dominated ULIRGs, there is a nearly one-to-one
correspondence, and we conclude that \fesclya\ and $W(\lya)$ indicate a consistent 
picture of moderate \lya\ escape fractions from starburst dominated ULIRGs.
In the opposite picture where stellar age dictates $W(\lya)$, the ULIRGs 
furthest along the merger sequence would in fact be expected
to have the oldest stellar populations and therefore the weakest 
$W(\lya)$. In contrast, we find the opposite.
The most advanced mergers tend to be AGN dominated ULIRGs,
and these objects show the highest \lya\ equivalent widths.

%The equivalent width has been
%     integrated over the entire line profile, and it can take positive
%     values when the absorption trough dominates over the emission component.

\subsection{Caveats}

%Even at optical wavelengths, we measure modest reddening ($E_{B-V}$ 
%of a just a few tenths) in the presence of enormous infrared excesses;  

In this section, we have shown that \lya\ escape fractions from ULIRGs
are comparable to those in more typical star-forming galaxies. We explain
this surprising result by the fact that the COS spectroscopy is blind to 
the most active and heavily obscured regions in ULIRGs. 
Unlike normal star clusters and starburst galaxies which show
a strong correlation between the far-infrared excess and 
the UV spectral slope \citep{Meurer:1995}, the ULIRGs
have a very blue UV spectral slope for their infrared
excess. No UV radiation would be detected from ULIRGs
if the dust distribution could be approximated by
the foreground screen model which works so well for
local galaxies.  To illustrate this point, we used
the GALEX magnitudes and  far-infrared luminosities from 
Table~\ref{tab:cos_galaxies} and estimated the infrared 
excess following  \cite{Meurer:1995}. We find
the typical infrared excess value in the COS ULIRGs is
$IRX \equiv F_{FIR} / F_{220} \approx 10$. We have only
measured the escape fraction relative to the regions 
transparent to optical light. 

To estimate the escape fraction is in an absolute sense, we
compare the SFR from unobscured regions, as measured from the \Ha\ luminosity (not
corrected for internal extinction), to the SFR 
indicated by the far-infrared luminosity. We estimate that 
unobscured regions account for only 0.3 to 2 percent of the total 
star formation rate among the COS ULIRGs, and the measured escape fractions
refer only to these unobscured regions. The absolute escape fractions are
far smaller than those estimated from UV and optical lines.

{\rev
The COS PSA subtends 5.0~kpc at the median redshift of the 
sample, and measurements of \lya\ flux from the COS 
spectra will underestimate the total \lya\ luminosity. 
A more specific concern relevant for the \fesclya\ results is that 
resonance scattering can produce large \lya\ halos around galaxies
relative to the angular extent of the \Ha\ emission.  Extrapolation
of the \cite{Hayes:2013} results for the LARS galaxies (which found the largest 
differences in spatial extent among galaxies with very low dust abundance) would
suggest little spatial bias for the COS ULIRGs sample. Considering the unusual
conditions in the ISM of ULIRGs, it would be quite interesting
to directly image the spatial extent of their \lya\ emission and test
whether they follow the trends for halo emission defined by more normal galaxies.
Since the far-UV images have peak sensitivity near the 
wavelength of the \lya\ emission, this may be possible using
%from our targets which range from 1376 to 1500 \AA. 
using sophisticated spectral fitting techniques as outlined, for example,
by \cite{Hayes:2009p911}.
}

\subsection{Implications}

Our discovery of blue wings on the \lya\ line profiles raises
the question of whether ULIRGs might have gas flows with
bulk velocities of roughly $\sim 1000$\kms.  The large Doppler shift 
of the \lya\ photons does not necessarily require fast outflows.
The frequency diffusion which accompanies repeated scatterings can 
produce a large Doppler boost. For example, a resonance photon
scattered back and forth repeatedly between opposite sides
of an expanding shell could get a Doppler boost that greatly exceeds
the shell velocity. Distinguishing bulk motion from 
frequency diffusion is the first critical step towards understanding 
the conditions that produce the line wings. 

From our analysis of 
\fesclya\ in Section~\ref{sec:fesc}, we concluded that the
\lya\ photons from \i12071-0444, \i23365+3604, and
\i10378+1109 do not undergo many scatterings because their
attenuation is similar to that of the UV continuum; hence, 
they probably do not receive their large Doppler boost from frequency
diffusion. We also find that \fesclya\ is most strongly correlated
with the Doppler shift of the blue wing on the \lya\ profile, the
\lya\ line width, and the \lya\ equivalent width.

We explore these concepts further in Section~\ref{sec:lya_optical}.
We use the optical line profiles of the ULIRGs as a template
for the intrinsic profile of the emitted \lya\ photons,
i.e., prior to their passage through the galaxy, and discuss
the implications for the \lya\ radiative transfer.

%On the other hand, cattering is clearly important, in contrast, for
%shaping many of the \lya\ near the systemic velocity.

%In contrast, some scattering mechanism must be at work in
%\i11598-0112 to effectively focus the line photons towards
%us, and these scatterings could plausibly boost the line
%wings to Doppler shifts higher than the bulk velocity of
%any gas flows.  

\section{Comparison of H Lyman and Optical Line Profiles} \label{sec:lya_optical}

%It is important to appreciate that the optical spectra of ULIRGs do not 
%probe the most obscured regions of the starburst; this fact can be inferred from 
%
%draws attention to the importance of resonance scattering 
%for shaping the \lya\ line profiles. 
%
%

The optical emission lines in the Keck~II/ESI spectra of the ULIRGs
provide a view limited to the optically transparent holes. This 
perspective is, however, closely related to the regions visible in 
the ultraviolet spectra; many common features can be seen in the
images presented in Figure~\ref{fig:images} for example.
In Figure~\ref{fig:line_shape_a}a,b, we show the line profiles of 
the H Balmer lines and [\ion{O}{3}] $\lambda 5007$ forbidden line 
on the same velocity scale as the COS G130 \lya\ emission. We
begin this section by comparing their line cores and then
demonstrate the relationship between the highly blueshifted gas seen
in the Keck~II/ESI and COS G130M spectra.

\subsection{Profile Core} \label{sec:line_core}

In Figure~\ref{fig:line_shape_a}, the velocity of the \lya\ absorption typically
coincides with the velocity of the maximum optical emission which is near the
systemic velocity. This absorption in the profile core is relatively narrow 
and most prominent in the HII-dominated ULIRGs (\i0026+4251, \i10378+1109, and \i16474+3430)
as well as the three systems with  prominent AGN (\i11598-0112, 
\i01003-2238, and \i12071-04444). The \lya\ absorption follows
our expectation that neutral hydrogen in these galaxies will
scatter \lya\ photons away from the systemic velocity.

One unexpected feature of these line cores is the residual intensity
at the systemic redshift. In the \i16487+5447 spectrum, the central absorption troughs 
lie to either side of the peak \Ha\ intensity. In \i11598-0112, we see
emission in \lya\ in the broad absorption trough near the velocity of the peak optical
emission. We also note that many of the spectra are not completely black 
in the absorption troughs. In a few cases, this residual intensity near 
line center may result from the line-spread function smearing out narrower
lines which are black at line center.  The troughs that are completely 
black (\i11598-0112, \i09583+4714, \i16474+3430, and \i16487+5447) appear 
significantly broader however, and our highest resolution spectrum, \i01003-2238, 
shows a narrow trough which is not black. The residual intensity
requires some sightlines with very low column densities of neutral hydrogen,
a result we will refine quantitatively in Section~\ref{sec:rad_transfer}.

%be an artifact of the resolution in some spectra; 
%\i00262+4151, \i1207-0444
%
%The \i23365+3604 spectrum has a shallow trough which is
%blueshifted about 400\kms. 

Scattering of resonance photons out of the line core will 
broaden the \lya\ line profile. Following absorption  near the
resonance, the \lya\ photons random walk in frequency space until 
they are either absorbed by a dust grain or are Doppler shifted 
to a frequency where they can escape. We therefore expect the 
\lya\ lines to be somewhat broader than the optical lines. 

%Near the core of the line profiles, the greater width of \lya\ is clearly
%attributable to photons scattering out of the absorption troughs and into the
%emission peaks; this makes the \lya\ widths larger than the optical lines at fixed 
%fractional area. 

To investigate the relationship between the observed and intrinsic
\lya\ profiles, we use a non-resonant optical emission-line to approximate
the intrinsic \lya\ profile shape. We have normalized the \lya\ and [\ion{O}{3}]
line profiles by their respective areas and overplotted them in Appendix~A.
To make a quantitative comparison, we use the technique based on 
profile area introduced in Section~\ref{sec:line_profile} and 
further described in Appendix~\ref{sec:appendix}. 
We mark the interpercentile velocity at $p = 0.80$ and take this as a 
measurement of the width of the core of the line profile. In the plot
of IPV as a function of profile area, comparison of the \lya\ and
[\ion{O}{3}]  line widths at $p = 0.80$ illustrates the systematically
broader width of \lya. Further inspection of thsese plots shows that
IPV(\lya) exceeds IPV([\ion{O}{3}]) at all values of the fractional
profile area $p$. It follows that the \lya\ profiles are systematically
broader than the optical line profiles regardless of exactly where
the width is measured.
The only ULIRG with an [\ion{O}{3}] profile broader than \lya\ is 
\i16474+3430 which has been broadened by orbital motion; the 2D version
of the ESI spectrum shows a second nucleus offset 1\farcs84 from the center
of the COS pointing. {\rev We therefore conclude that diffusion of resonance photons 
in frequency space typically explains the broadening of the cores of the \lya\ profiles 
relative to the non-resonant lines.}

% ############################################# HERE 

\subsection{Line Wings}

% moving at several hundred \kms. 

%one hump
%in the [\ion{O}{3}] profile is
%[\ion{O}{3}] profile broader than \lya\ is 
%
%the large absorption trough in this spectrum
%that somewhat artifically  reduces the measured \lya\ linewidth. 

\begin{figure*}[h]
% \hbox{\hfill \includegraphics[height=13cm,angle=-90,trim= 0 0 0 0]{figures/fa.ps}  }
% \hbox{\hfill \includegraphics[height=13cm,angle=-90,trim=0 0 0 0]{figures/fb.ps}  }
 \hbox{\hfill \includegraphics[height=13cm,angle=-90,trim= 0 0 0 0]{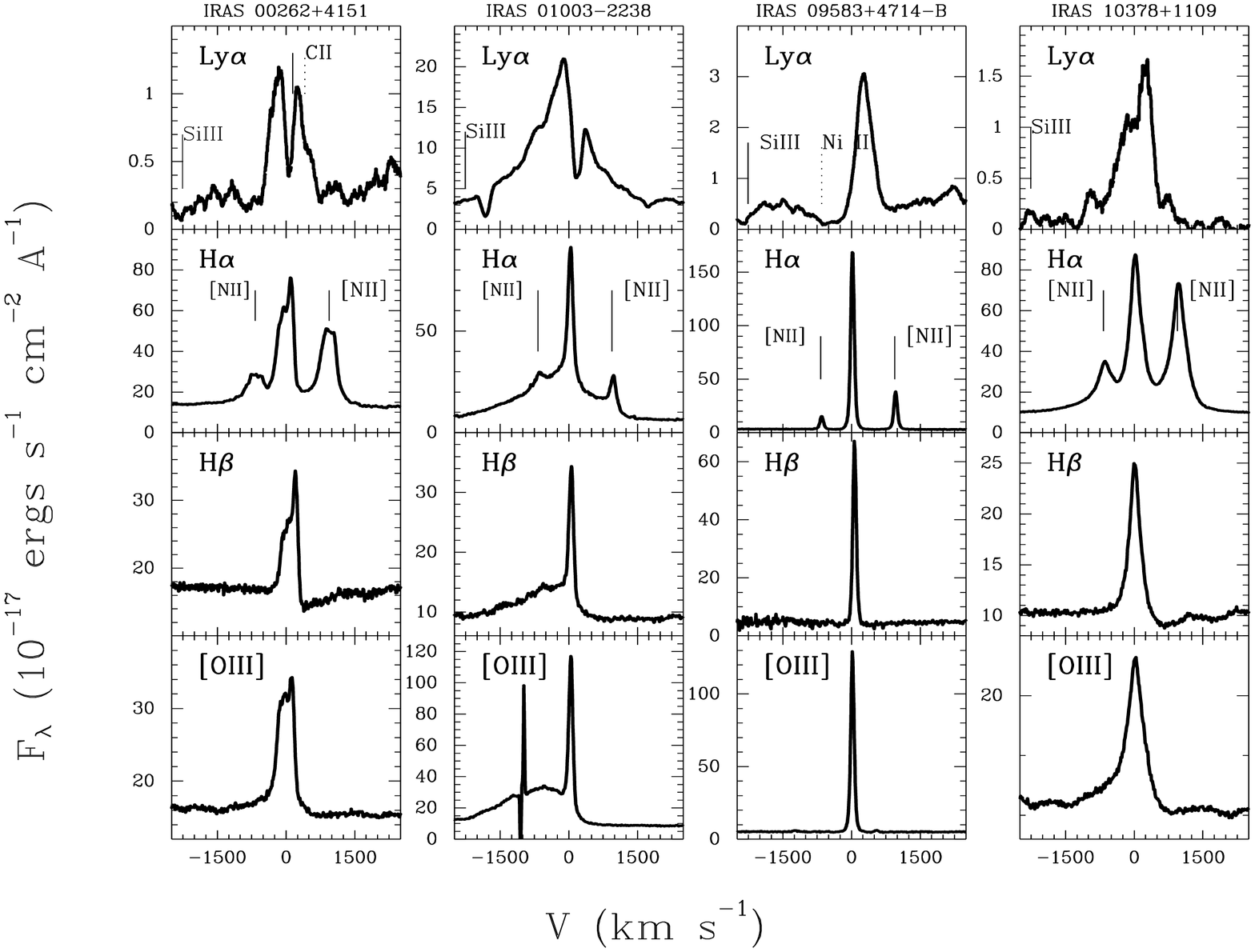}  }
 \hbox{\hfill \includegraphics[height=13cm,angle=-90,trim=0 0 0 0]{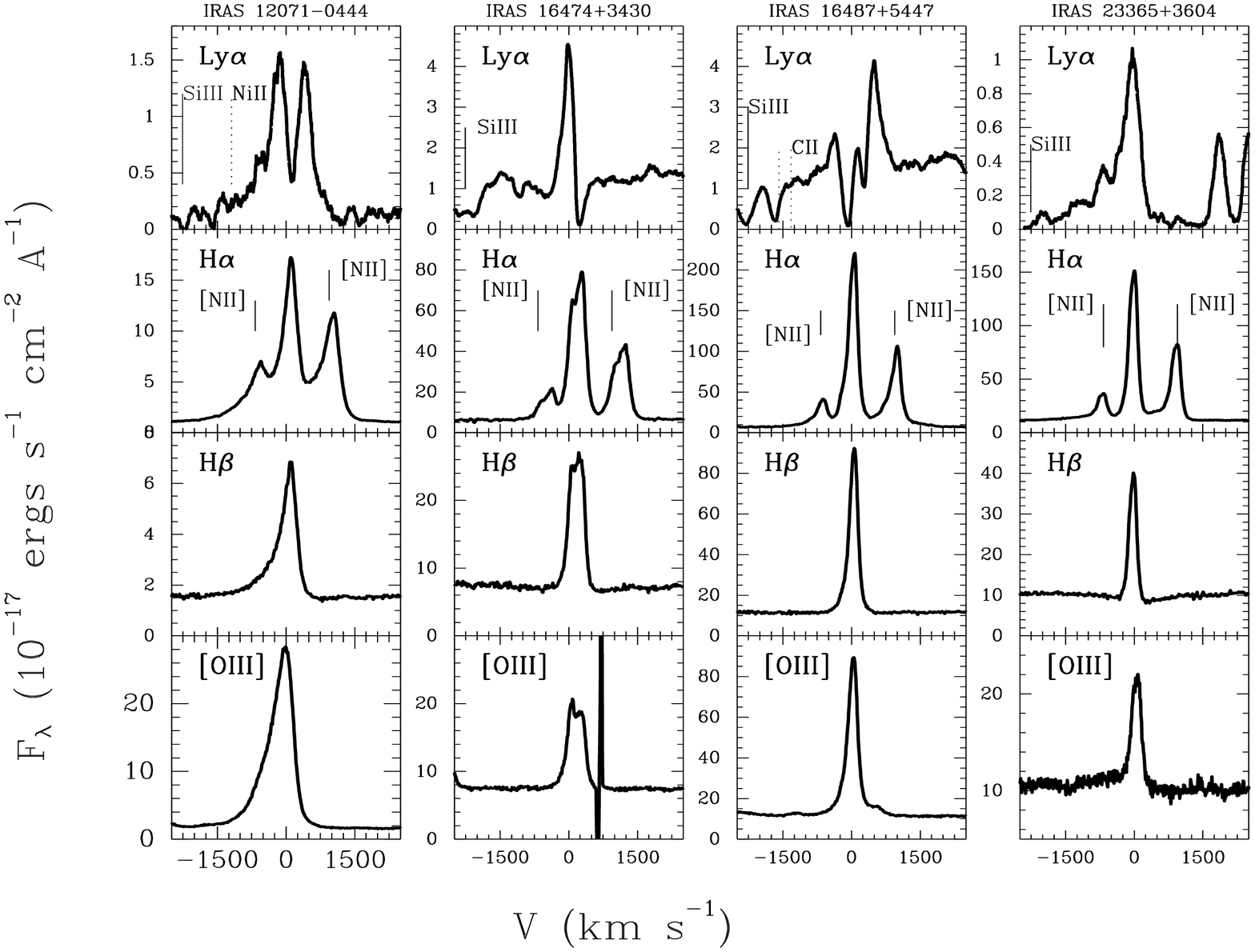}  }
   \caption{\footnotesize 
    Comparison of Lyman series and optical line profiles. The \lya\ absorption troughs
     (top row) line up with the velocities of the maxima in the optical emission lines. 
     Comparison of the blended \Ha\ and [\ion{N}{2}] $\lambda\lambda 6548, 6584$ lines with 
     the \Hb\ line shows the contribution of the broad line wings to the blend. The
     line profiles for [\ion{O}{3}] $\lambda 5007$  demonstrate that broad
     wings are present on the forbidden lines as well as the recombination lines.
     The systemic velocities 
     were obtained from either CO observations or the average of many optical emission 
     lines as summarized in Table~\ref{tab:cos_galaxies}. Blended lines from the
     galaxy are indicated by solid, vertical lines, and Galactic absorption lines
     are indicated by dotted, vertical lines. 
%     Note that we detect \siIII\ $\lambda 1206$ 
%     absorption at high significance towards 
%     \i11598-0112, \i16487+5447, and possibly \i16474+3430.
%The \Ha\ wings blend with forbidden lines of [\ion{N}{2}].
%The wings of the \Hb\ emissionare sometimes superposed on 
%broad stellar absorption troughs \citep{Soto:2010}.
% The extremely broad and prominent wings on the forbidden
%[\ion{O}{3}] and [\ion{N}{2}] emission in the spectrum of \i01003-2238
%draw attention to this object which we classified as an AGN; 
%the broad component is spatially unresolved in this particular ULIRG.
   }
  \label{fig:line_shape_a}   \end{figure*}

In the optical lines shown in Figure~\ref{fig:line_shape_a}, the hydrogen
Balmer lines typically show a broad, blueshifted line wing. The median blueshift of the fitted 
wings is 250\kms\ \citep{Soto:2012a}, but this emission component often extends to 1000\kms\ due to 
its large velocity dispersion. This high velocity gas emits strongly in optical 
forbidden lines of [\ion{O}{3}], [\ion{S}{2}], [\ion{N}{2}], and [\ion{O}{1}].
We focus the discussion on the comparison between the [\ion{O}{3}] 5007 line and \lya\
because the former lies in a spectral region free of blends with other transitions.
In contrast, if we use \Ha\  or \Hb, then the deblending of [\ion{N}{2}] 6548, 84 or, 
respectively, stellar \Hb\ absorption introduces additional noise into the analysis.

The only measurable consequence of this choice is that our template for the optical emission
line profile has slightly stronger wings relative to the line core than it would had we used 
a Balmer line. The shape of these optical  profiles differs slightly due to a change in excitation 
mechanism.  Line ratio diagrams show that the broad component is not excited by
the ionizing continuum from massive stars. Since its spatial extent is also
typically larger than the broad line region of an AGN, \citep{Soto:2012a} have
previously fitted the diagnostic line ratios measured for the broad 
emission component with models for radiative shocks.

%FIGURE 9a

\begin{figure}[h]
% \hbox{\hfill \includegraphics[height=10cm,angle=-90,trim=0 50 0 0]{figures/vb_comp.ps}
 \hbox{\hfill \includegraphics[height=10cm,angle=-90,trim=0 50 0 0]{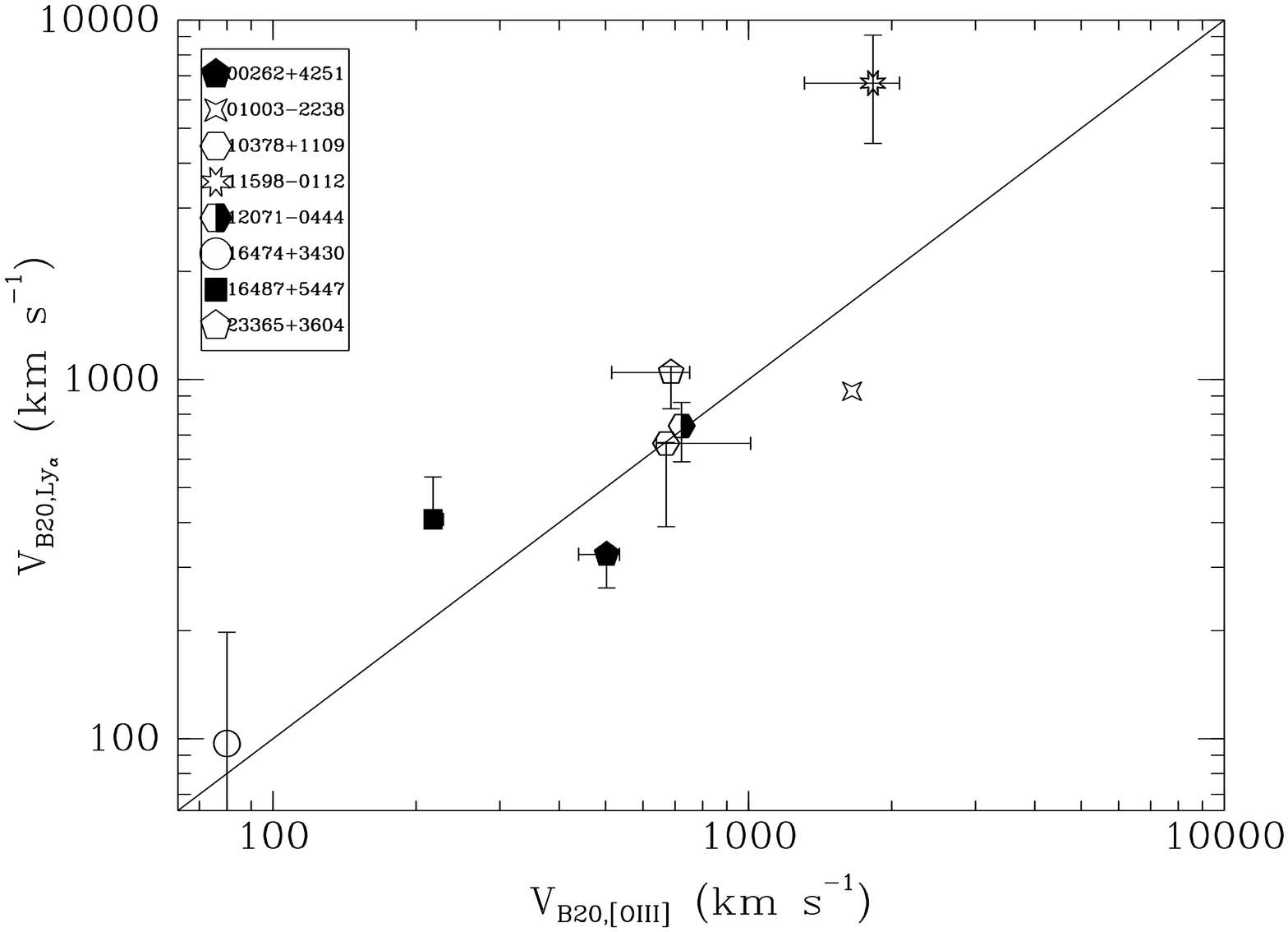}
  \hfill}
   \caption{\footnotesize 
     Velocity of blue wing in \lya\ vs. [\ion{O}{3}] at 20\% of the net profile area. 
     We give the Doppler shift relative to the systemic velocity
     rather than the median velocity of the net emission. The plot omits the two ULIRGS
     without \lya\ detections and \i09583+4714-B which shows a P~Cygni profile (the
     blueshifted line is purely in absorption).  The correlation coefficient
     is $r_S = 0.93$, and the null hypothesis is rejected at the $2.6\sigma$ level.
   }
   \label{fig:vb_comp}   \end{figure}

Figure~\ref{fig:vb_comp} compares the extent of the blue wings 
of the [\ion{O}{3}] and \lya\ emission. For purposes of illustration,
we chose the interpercentile velocity markers at $p = 20\%$ as a 
compromise between being away from the line core yet not too 
sensitive to continuum placement. The velocity of the blue \lya\
wing is correlated with that of the [OIII] wing. We also see that
the measurements of $V_{B20}(\lya)$ and $V_{B20}({\rm[OIII]})$ 
scatter around the locus of equality. Even though the cores of
the line profiles are relatively broader in \lya, the blue wings are
not any more extended than the [\ion{O}{3}] line wings at the same
fractional profile area. This result is noteworthy because the line 
profiles of the non-resonant line are {\it not} affected by scattering.  
The good agreement suggests the {\it intrinsic profile} of the \lya\ 
emission contains blue line wings.

%It suggests resonance scattering does not create the wings; they must
%be present in the intrinsic \lya\ profile.
%Evidently some \lya\ photons were emitted by 
%gas with large Doppler shifts, and the widths of the \lya\ 
%emission profiles should {\it not} be entirely attributed to
%upscattering in frequency space. 

%FIGURE 

\begin{figure}[h]
% \hbox{\hfill \includegraphics[height=10cm,angle=-90,trim=0 50 0 0]{figures/ipv_fesc_o3.ps}
 \hbox{\hfill \includegraphics[height=10cm,angle=-90,trim=0 50 0 0]{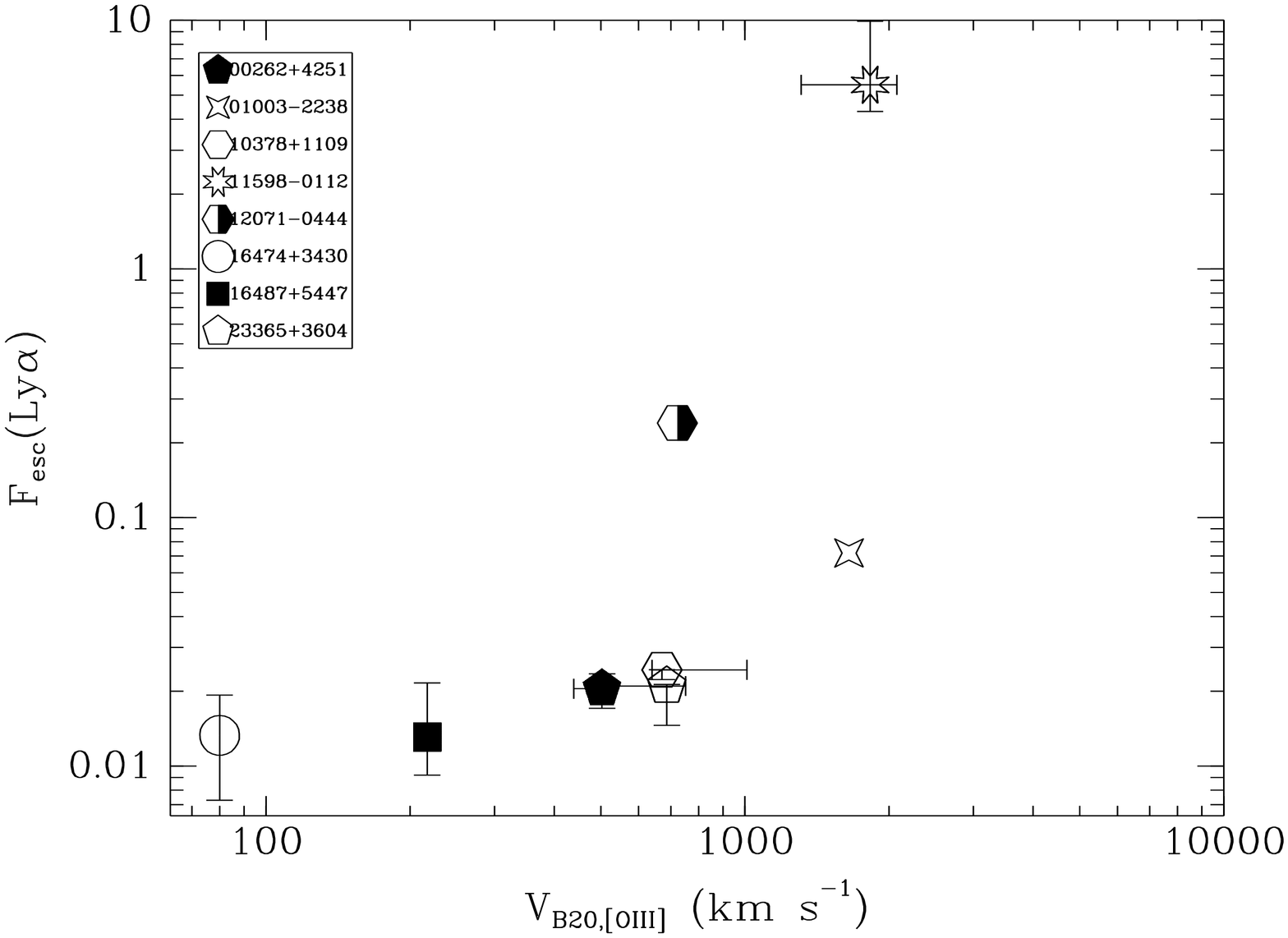}
  \hfill}
   \caption{\footnotesize 
The \lya\ escape fraction increases with the blueshift of the \lya\ emission wing,
defined here as the velocity beyond which 10\% of the \lya\ emission line flux is measured. 
This correlation ($r_s = 0.95$ and null hypothesis rejected at $2.7 \sigma$)
is slightly stronger and more significant that that of \fesclya\ with the velocity
of the \lya\ blue wing.
   }
   \label{fig:ipv_fesc_o3}   \end{figure}

The Doppler shifts of the blue wings therefore arise at least in part 
from bulk motion along the line-of-sight and cannot be attributed solely to
frequency upscattering in a static medium. Since the same flows are
detected in the optical emission lines, the Doppler shift of the optical
lines provides as good an indicator of \fesclya\ as did \lya\ in 
Figure~\ref{fig:fesc_v}. Figure~\ref{fig:ipv_fesc_o3} shows this 
correlation of the \lya\ escape fraction with the velocity of the blue 
wing on the [\ion{O}{3}] profile.

Inspection of the area normalized profiles in the Appendix enables a 
closer comparison of the blue wings in \lya\ and [\ion{O}{3}]. The
more similar the line profiles, the less radiative transfer
shapes the \lya\ profile. We draw particular attention to 
subfigures~d  (\i10378+1109), f (\i12071-0444), and i (\i23365+3604). 
Between velocities of roughly -200\kms\ and -1000\kms, the shapes of
their \lya\ profiles are nearly identical to that of the optical line
profile.   We therefore conclude that radiative transfer effects play
little or no role in shaping the blue wings in these ULIRGs. This
result further strengthens our conclusion that the bulk velocity of 
the \lya\  emitting gas must reach at least 1000\kms in these objects.

%FIGURE 

%
%but in a $V_{R20}(\lya)$ vs. $V_{R20}([O3])$,
%
%That the greater
%width of the \lya\ profile can be attributed primarily to the redshifted 
%\lya\ emission supports the conjecture that the large Doppler shifts of
%the blue wings on the \lya\ profiles largely reflect bulk motion
%rather than radiative transfer effects.

In stark contrast to the blue wings, the velocities of the red wings on 
the \lya\ profiles are not correlated with the optical line profiles. 
Inspection of the area-normalized profiles in Appendix Figures~a-i 
shows that the red wings on the \lya\ line profiles extend to larger Doppler shifts
than do the [\ion{O}{3}] profiles. The greater extent of the red wing in 
the resonant line makes the linewidths systematically larger for \lya.
The red wings on \lya\ also weaken the correlation between the \lya\ and [\ion{O}{3}] 
linewidths.\footnote{
         The linewdiths are not quite as strongly correlated as are the velocities 
         of the blue wings. At $IPV_{20}$ for example, we measure $r_S = 0.85$ with 
         the null hypothesis rejected at the $2.4\sigma$ level.}
Due to the presence of the red wings, the \lya\ profiles are actually less symmetric
than the optical line profiles by some measures.

For example, in the Appendix, we define a line asymmetry measurement across the
line profile (i.e., computed at each interpercentile velocity). The asymmetry
compares the relative strength of the emission redward and blueward of the median
velocity and is positive for a net blueshift. The panels for \i12071-0444 illustrate
this concept for a \lya\ profile with two maxima. Near the core ($p \sgreat\ 0.4$  in
this spectrum), the red velocity marker is further from the median velocity than
is the blue velocity marker at the same IPV; however, in the line wings ($ p \sles\ 0.3$) 
the blueshifted profile has the larger Doppler shift.  Since
the optical lines do not have the redshifted components in their line profiles,
their net asymmetry parameter is positive and biased towards the blue. The
redshifted component of the \lya\ emission is typically not present in the optical
line profiles and therefore provides another profile metric that successful 
models for the \lya\ emission must explain. 

%{\bf Note the greater extent of the redshifted \lya\ wing in 
%\i12071-0444 and \i10378+1109 relative to [\ion{O}{3}].}

%\subsection{Further Requirements for Models}

%The strong \Ha\ profile asymmetry marked by 

%Radiative transfer modeling will
%be needed to address whether this gas lies on the near
%side of the ULIRG or is backscattered by an expanding shell. 
%In the former case, emission near the systemic velocity
%would be scattered to both bluer and redder wavelengths. 

Indeed, the main conclusion from our comparison to non-resonant 
line profiles is that resonance scattering does not create
the high velocity wings.
The absence of red wings on the optical line profiles suggested
this emission arises from an outflow near the starburst region. In 
this scenario, we see emission from blueshifted gas on the near side, but dust 
in the ULIRG blocks the redshifted emission from the far side \citep{Soto:2012a}. 
Measurements of line profiles at wavelengths that propagate through the dust
generally support this interpretation. The CO~1-0 line profile 
of \i23365+3604, for example,  exhibits broad (albeit very weak) 
symmetric line wings \citep{Cicone:2014}. It therefore seems
likely that the blue wings observed on the \lya\ line profiles 
are directly emitted by  outflowing gas, and dust between us and the far
side of this outflow attenuates the red wings on the intrinsic
\lya\ profile. The broad \lya\ wings must be emitted by gas moving 
at velocities of 500 to 1000~\kms.

%

%The \lya\ photons emitted from the far side of an expanding shell or 
%continuous wind may traverse the entire system without further resonance scattering
%due to their redshift relative to the resonance. These \lya\ photons will, however, 
%still be destroyed by dust at a higher rate than will \Ha\ photons as the light 
%traverses the nebula. The redshifted \lya\ line wings therefore place
%interesting constraints on the ouflowing gas.  Using a \cite{Cardelli:1989} 
%reddening curve for illustration, the extinction of the continuum near  \lya\ 
%is $10.9 E_{B-V}$ but drops to $ 2.53 E_{B-V}$ at \Ha. 
%Assuming recombination in the ionized regions produces roughly
%8.1 \lya\ photons for every \Ha\ photon, we infer that \lya\ will be
%brighter than \Ha\ when the color excess $E(B-V) < 0.27$.  In fact
%the two ULIRGs with larger color excess (along the optically transparent 
%sightlines) in Table~\ref{tab:esi_measure}, \i10378+1109 and \i23365+3604, 
%show prominent red wings on their \lya\ profiles.
%SO THE LARGE EBV DID NOT BLACK THE LYA.

\subsection{Comparison to Absorption-Line Kinematics} \label{sec:absorption_lines}

%because non-resonant line
%photons (Balmer photons and forbidden transitions of common metals) also
%show broad, blue wings in many ULIRGs. 
%
%Sodium will be ionized along unshielded sightlines. 

%In this section, we have shown that resonance scattering shapes
%the core of the line profile, but the blueshifted line wings cannot be attributed
%solely to photon diffusion in frequency space. 

%of the \naI\ absorption are summarized in Table~\ref{tab:esi_measure}

The fast winds indicated by the line wings present a paradox because their
Doppler shifts  are significantly larger than the blueshifts of absorption 
lines from low-ionization-state  gas. These blueshifts unambiguously 
mark outflowing gas because the absorbing gas must lie on the near side of the continuum 
source. Among ULIRGs, absorption in the \naI\ 5890, 96 doublet is typically blueshifted
only 200-300\kms\ at the deepest part of the absorption trough \citep{Martin:2005,Martin:2006,Rupke:2005b}. 

\begin{figure}[h]
% \hbox{\hfill  \includegraphics[height=10cm,angle=-90,trim=0 50 0 0]{figures/ebv_nad.ps}
 \hbox{\hfill  \includegraphics[height=10cm,angle=-90,trim=0 50 0 0]{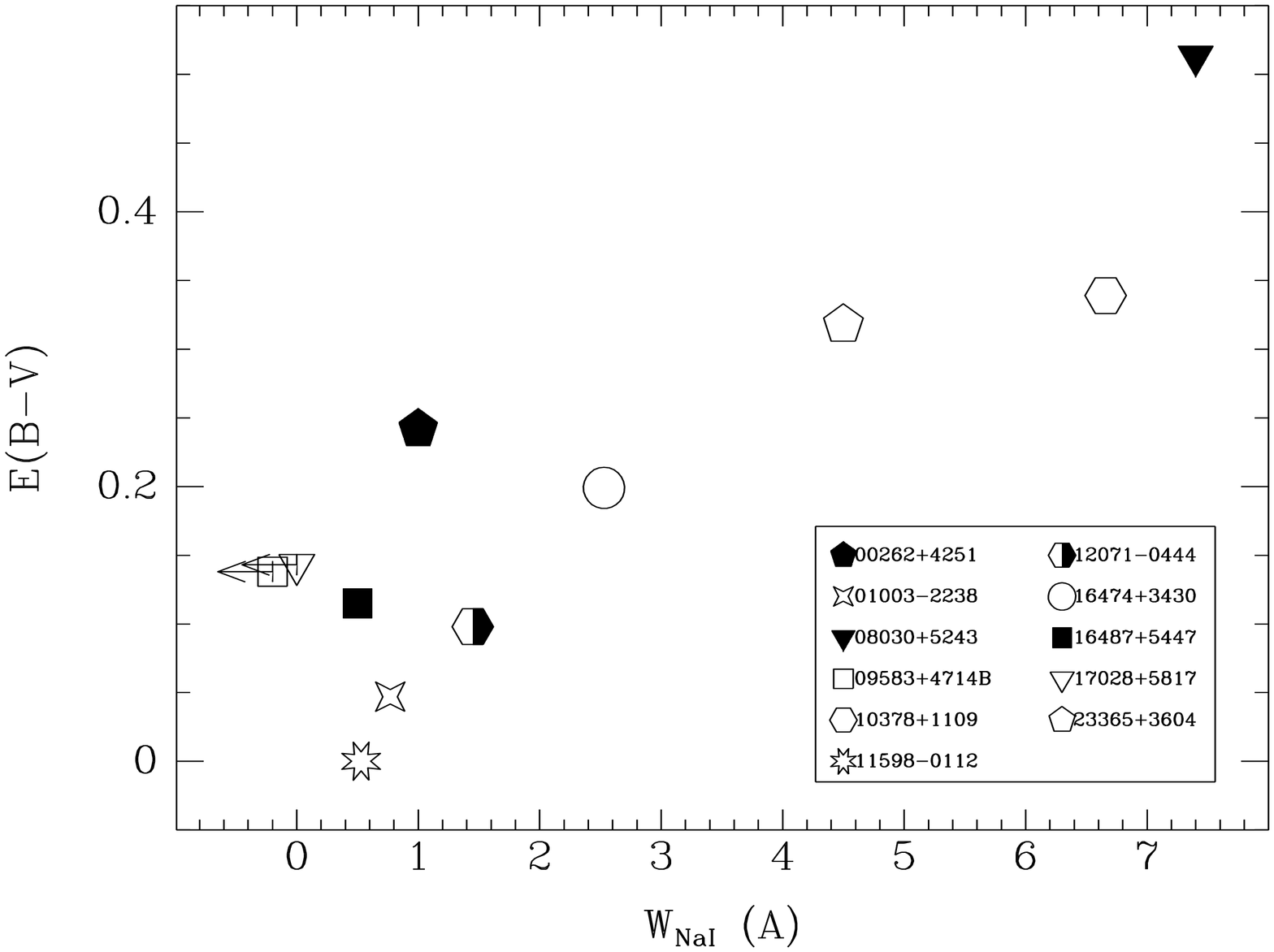}
                \hfill}
   \caption{\footnotesize 
The COS ULIRGs with the largest color excesses show the strongest \naI\ absorption 
(e.g., \i23365+3604, \i10378+1109, and \i0803+5243). 
We interpret this correlation as evidence for an increasing 
neutral sodium fraction with increasing dust column following \cite{Martin:2005,Chen:2010p445}.
Choosing the UV-brightest ULIRGs for the COS observations selected the systems with 
the least reddening and therefore the weakest \naI\ absorption. Weak \naI\ absorption
in many of the COS ULIRGs does not preclude outflows of low-ionization-state gas.
%
%The \lya\ escape fraction is not correlated with the \naI\ strength.
%{\bf Check whether 0958 and 1702 were measured at zero or set there due to missing spectra.}
   }
\label{fig:ebv_nad}   \end{figure}

Many of our COS targets have weaker \naI\ absorption than the average ULIRG because
we observed the UV brightest ones. Figure~\ref{fig:ebv_nad} illustrates the strong 
correlation between the \naI\ equivalent width and the color excess.  We interpret 
this correlation as an ionization effect. Sodium must be shielded from near-UV
photons to remain ionized; it takes just 5.1~eV to remove its outer $3s$ electron.
The UV selection picks the ULIRGs with less reddened sightlines, and
sodium is therefore more likely to be ionized.

\begin{figure*}[t]
% \hbox{\hfill  \includegraphics[height=18cm,angle=-90,trim=0 0 0 0]{figures/vabs_16487+5447.ps}
 \hbox{\hfill  \includegraphics[height=18cm,angle=-90,trim=0 0 0 0]{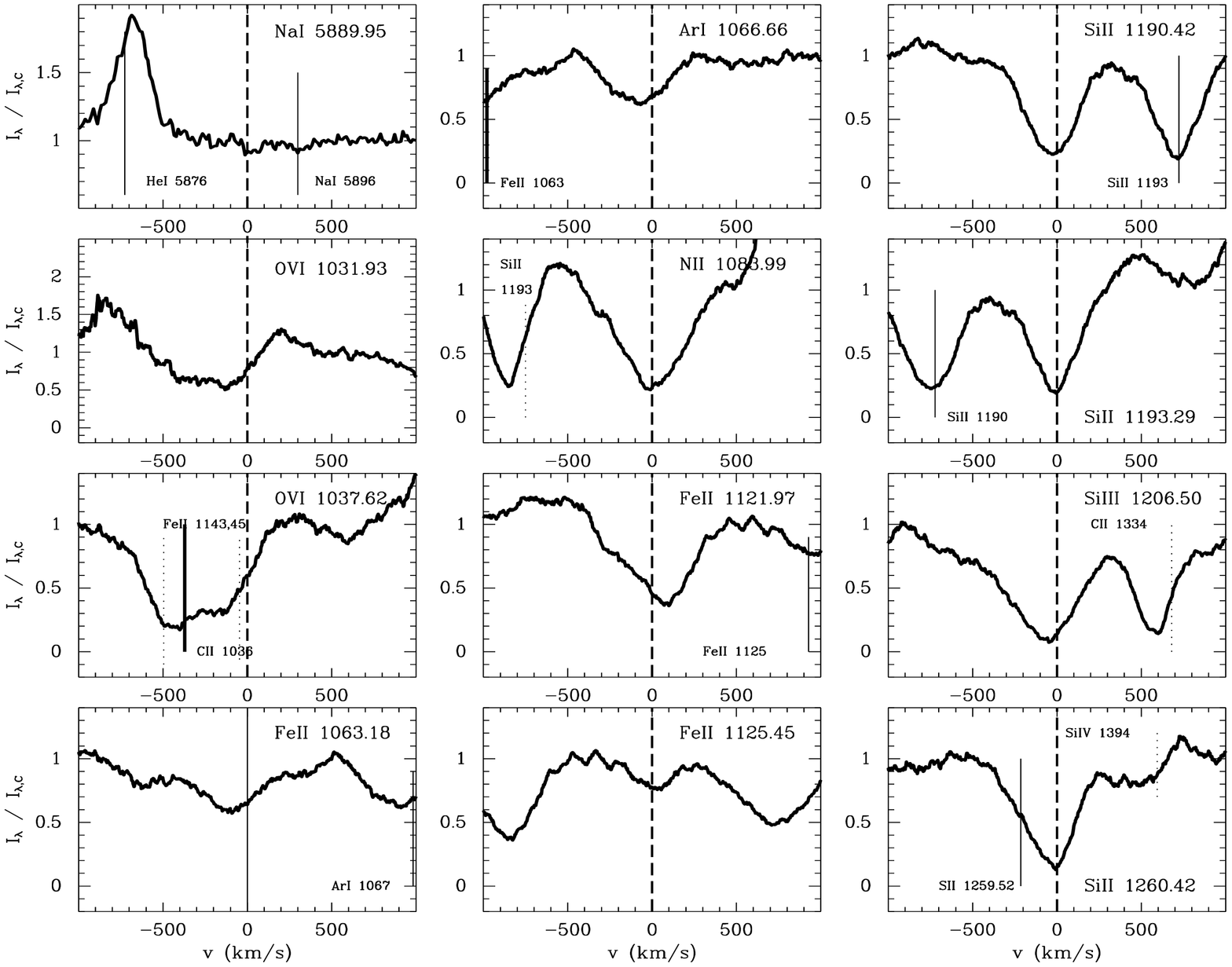}
                \hfill}
\caption{\footnotesize
Absorption lines in the COS G130M and ESI (top left only) spectra of \i16487+5447.  
Note the lack of \naI\ absorption compared to the strong \ion{Si}{2} 1190, 1193, and 1260 
lines. The \ion{Si}{2} lines are not as broad as the \ion{Si}{3} 1206 absorption. The \ovi\ 1032
absorption trough reaches even higher blueshifts of at least -500\kms. (Note the blended \ovi\ 1038 trough
precludes the same measurement.) The \ion{O}{6} absorption may have some contribution from stellar winds as well
as  warm gas in a galactic outflow. Precise measurements of these troughs are
sensitive to the modeling of the line spread function and the continuum placement.
}
\label{fig:vabs_16487+5447} \end{figure*}

The absence of \naI\ absorption does not indicate the absence of an
outflow of low-ionization-state gas.
The ionization state of sodium does not uniquely determine whether hydrogen
is neutral or ionized along these sightlines \citep{Murray:2007p211}. 
The wider range of ionization states accessible with far-UV transitions proves this
point. We saw this in both Figure~\ref{fig:pcygni} and Figure~\ref{fig:cos_11598},
and we demonstrate it for the HII-dominated ULIRG with the best continuum
S/N ratio in Figure~\ref{fig:vabs_16487+5447}. 

%Sodium is much easier to ionize than hydrogen because the first
%ionization potential is just 

The spectrum of \i16487+5447 also demonstrates how the gas kinematics changes
with ionization state. The absorption troughs from low-ionization-state
gas have their deepest point near the systemic velocity. The greater Doppler
shift of their blue wings (relative to the red wings) indicates a contribution
from an unresolved outflow component, but these wings do not reach the 
-500\kms\ blueshift of the \lya\ emission wing. Absorption in the medium ionization state 
gas probed by \ion{Si}{3} probably reaches $-500$\kms. The warmer gas detected via
\ovi\  absorption is  offset blueward of the systemic velocity and extends to 
velocities as large as the blueshifted \lya\ emission from this object. While general
conclusions should not be drawn from a single object, we note that FUSE spectroscopy
of nearby starburst galaxies has previously revealed higher  mean outflow velocities in \ovi\ 
than low-ionization-state lines in the same spectra \citep{Grimes:2009}. We add to 
the emerging picture the result that the Doppler shift of the broad emission-line wings
appears to be more closely related to the Doppler shift of the \ovi\ absorption than
to the absorption from low-ionization-state gas.

%, but the maximum velocity 
%is not precisely measured because the continuum placement
%is highly uncertain in this spectrum at these wavelengths.
%Because of this uncertainty, we also view the possible presence 
%of \ovi\ emission cautiously but do note the presence of a 
%broad bump just redward of  \nv\ as well.

%low-ionization lines including \ion{Fe}{2}, \ion{Ar}{1}, \ion{N}{2}, 
%\ion{Si}{2},  \ion{S}{2}, and \ion{P}{2} in 

%\item
%As the continuum S/N ratio declines among the other HII dominated 
%ULIRGs, a progressively smaller subset of these features are detected.
%The next best example is \i16474+3430, but the blue wing of the \ion{Si}{3}
%trough is suppressed by blending with the wings of a Milky
%Way \ion{C}{2} and \ion{C}{2}$^*$ absorption.
%
%The stronger \ovi\ 1032
%line is not convincingly detected towards \i16474+3430; and we 
%attribute the stronger absorption near \ovi\ 1038 to \ion{C}{2}.

%\item \i01003-2238
%The spectrum of the other AGN, \i01003-2238, shows very
%broad but weak emission at wavelengths just redward of both
%\ovi\ and \nv.

% FIGURE 

\section{Origin of Broad Emission}  \label{sec:discussion}

%a small number of emitting clouds whose 
%
%a lower velocity outflow is detected in absorption (i.e., on the
%near side) of the continuum source. 
%
%We therefore associate the highly 
%blueshifted gas detected in the \lya\ line profile with outflowing gas,
%
%
%{\bf Last few sentences of above paragraph hard to follow. Explain as you did on
%the phone.}
%
%.... outflow conclusion which we will show requires relatively small neutral
%gas columns.

The presence of broad line wings on the optical emission lines
as well as \lya\ suggests gas moving at high velocity emits the
broad, blueshifted \lya\ line wings. The predominance of blueshifted 
line wings over redshifted wings rules out a scenario involving 
large random motions and requires bulk flows to generate the asymmetry.

As recognized previously in the context
of the narrow line region of AGN \citep{Whittle:1985a}, however, the
direction of radial flow cannot be determined unambiguously from the
asymmetry of an emission-line profile alone. Either wing of the intrinsic
line profile can be suppressed in an outflow. Suppression of the red 
wing corresponds to a source of line opacity distributed throughout 
the outflow region. In contrast, line opacity internal to discrete clouds
preferentially blocks radiation from the near side of the outflow 
because the ionized side of each cloud faces the central source; and
the resulting line profile has a red wing.

We follow \cite{Soto:2012b} and argue the bulk flow
is almost certainly outward. Our primary consideration is
the blueshifted absorption seen in resonance lines such as
\naI, \siII, \siIII, and \ovi. The gas detected in absorption
is clearly outflowing because it must lie between us and the 
galaxy. Sources of energy and momentum for the outflows include
massive stars and the central AGN whereas, in contrast, we do not have 
a physical reason to expect gas to be flowing into galaxies at speeds 
around 1000\kms.  In the outflow interpretation, emission from
the fast outflow on the far side of the galaxy produces broad, redshifted 
line wings. We interpret the absence of such wings in our spectra 
as evidence that the receeding outflow lies largely behind
the dusty galaxy and note that the broad, symmetric line wings
on the CO(1-0) profile of \i23365+3604 \citep{Cicone:2014}.

%--------------------------------------------------------
%GAS INFALL
%\citep{Dijkstra:2006a,Dijkstra:2006b} 
%find that the blue peak is significantly
%enhanced.  The total blueshift of the line increases with the total
%column density of neutral hydrogen of the gas and with its infall
%speed. They find blueshifts as large as 2000\kms. 
%The infall velocity during a ULIRG merger is not expected to be anywhere
%near this fast. Infall therefore seems like an unlikely explanation
%for the blue wings on the \lya\ profiles of the ULIRGs. Indeed, infall
%would be further challenged by the blue wings on the optical emission
%line profiles and the lack of redshifted low-ionization-state absorption.
%--------------------------------------------------------

%Although common in ULIRG spectra, the
%prominent wings are absent in the optical spectra of M82, 
%the best studied example of a starburst-driven wind. With
%perhaps a few exceptions, the optical spectra of nearby dwarf 
%galaxies observed in a starburst phase generally do not show these 
%wings even though X-ray and \Ha\ images show outflows \citep{Martin:1998}.

{\rev
Our analysis generates two primary puzzles. First, what 
is the relationship between the fast outflow emitting the 
broad, emission-line wings and the lower Doppler shifts measured for the 
low-ionization-state absorption lines? Second, why do only the most extreme
galactic outflows show these broad, emission-line wings? For example,
the optical spectra of some high-redshift galaxies \citep{Newman:2012_wings,
Schreiber:2014,Genzel:2014} show broad, emission-line wings 
similar to those in ULIRG spectra. We note that one common
property of these high-redshift galaxies and ULIRGs appears to be exceptionally 
high SFR surface densities.
}

%This coincidence suggests that our  ULIRG study provides insight
%into the stellar feedback process in high-redshift winds, 
%and it identifies a wind-related feature that can, at least
%in principle, be studied from galaxies over much of cosmic
%time without the need for high continuum S/N ratio. 

%The presence of these line wings is of particular interest for
%several reasons. 

\subsection{Cool Gas Condensing Out of a Hot Wind} \label{sec:hot_wind}

%Several properties of the fast outflows suggest an
%association of the nebular emission wings, as observed in \lya\ 
%and optical lines, with the hot wind fluid. 

We are suggesting a new picture wherein
the emission wings come from filaments
that condense out of the cooling hot wind. 
The large velocity of the blueshifted nebular emission provides 
the primary motivation for considering a power source associated 
with the hot wind fluid. 
Support for this picture comes from the scaling of the \lya\ luminosity
with the SFR and the structure of simple models for galactic winds
launched from regions of extremely high SFR surface density.

%the cooling time and radius in winds generated by
%starbursts with exceptionally high SFR surface density.

%The highly blueshifted emission might arise from pre-existing
%clouds photoionized by the ionizing radiation produced by the cooling 
%wind, but this alternative picture requires a physical mechanism 
%to accelerate the clouds to the observed velocities without 
%destroying them. In general, however, since the $L_x^{hot}
%\propto SFR^2$ in a  CC85 type wind, we expect any cooler 
%gas which is physically related to the hot phase to show
%a similar scaling dependence, albeit at a much smaller fraction
%of the total luminosity.

\subsubsection{Velocity of Hot Wind Fluid}

When some of the supernova energy associated with star formation is
thermalized, a hot wind is driven out of the starburst region.
\cite{CC85} provided a simple analytic solution for the
Mach number, thermal pressure, and density in a fast, 
adiabatic wind. The wind reaches Mach~1 at the launch 
radius $R_0$ where the mass and energy injection from the 
starburst region terminates. At $r = R_0$, the wind
temperature in the CC85 solution, as shown for example
in Eqn. 6 of \cite{Zhang:2014}, is 
\begin{eqnarray}  \label{eqn:temp}
T(r) = 1.8 \times 10^7 {\rm ~K~} \mu \left (
\frac{\alpha}{\beta} \right ),
\end{eqnarray}
where $\mu$ is the mean molecular weight per gas particle,
and $\alpha$ and $\beta$ are parameters of order unity
which describe the efficiency of thermalizing the 
supernova energy ($\alpha = \dot{E}_{hot} / \dot{E}_{SN}$)
and entraining gas into the wind ($\beta = \dot{M}_{hot} / SFR$).
The specific thermal energy of this gas is sufficient to
accelerate every particle up to a velocity
\begin{eqnarray}
v_{\infty} = 950 \kms\ \sqrt{\mu \frac{\alpha}{\beta} \frac{T}{1.8 \times 10^7 {\rm ~K}}}.
\end{eqnarray}

The similarity between the specific thermal energy of 
hot winds and the specific kinetic energy of the gas
emitting the nebular line wings motivates a conjecture
that the highly blueshifted nebular emission arises
from gas condensing out of a hot wind. Since the
initial velocity of these clouds would be the speed
of the hot wind, this picture explains their high
velocities without introducing the problem of accelerating 
cool clouds to velocities much larger than their internal 
sound speeds; clouds tend to be destroyed by such processes.

\subsubsection{Gas Density \& Instability of a Cooling Hot Wind}

% Emission measure is defined as $EM \equiv \int n_e^2 dV$ where $V$ 
%is the volume of the nebula.

%ULIRGs, and perhaps some high-redshift galaxies,

%decreases as the concentration of massive stars in 
%the starburst region increases. Winds should therefore cool on
%the shortest timescale in 

%The faster cooling in winds
%driven by the most extreme starbursts brings 

\cite{Zhang:2014} also show that the gas density in the 
CC85 solution increases lineary  with the SFR surface density,
hereafter $\dot{\Sigma_*} \equiv SFR / \pi R_0^2$. {\rev Adopting
the radiative cooling function for gas in collisional ionization 
equilibrium (e.g., \cite{Draine:2011} and references therein),}
one finds that the cooling timescale for the hot wind,
\begin{eqnarray}
t_{cool}(R) = \frac{P(n,T)}{(\gamma - 1) \Lambda(T) n_e n_H},
\end{eqnarray}
{\rev declines as the starburst becomes more concentrated, i.e., 
 $t_{cool} \propto \Sigma_*^{-1}$.}  

We can define a cooling radius where the cooling time 
is reduced to the outflow timescale, $t_{out} = R / v(R)$,
and this cooling radius moves inward as $R_{cool} \propto \Sigma_*^{-1}$
for the CC85 solution {\rev and standard cooling function}.
Plugging in typical numbers for ULIRGS, we expect the hot wind to be cooling
very close to the starburst region, in the central few kpc
of ULIRGs. We propose that the ULIRG winds start to cool
at smaller radii than do winds in starbursts like M82 because
the  higher SFR surface density drives winds with higher densities
and smaller cooling radii.

The high gas density at relatively small radii leads to a much 
higher emission measure at the cooling radius than would be 
expected in more typical starburst galaxies. As the wind cools,
we expect a highly unstable situation to develop as the
temperature drops from the free-free portion of the cooling
curve to a minimum (near $T \approx 2 \times 10^7$~K
for solar metallicity gas in CIE). 
Between $10^5 < T (K) < 10^{7.3}$,
the cooling rate per atom begins to increase rapidly
as the gas cools,  $\Lambda(T) \propto T^{-0.7}$  
\citep{Draine:2011}. Resolving this process and the associated
condensations of clouds in numerical simulations would be
enlightening.

%The cooling rate increases as $\Lambda(T) \propto
%T^{1/2}$ towards higher temperture due to the increasing importance 
%of free-free emission. Critically, however, bound-bound 
%transitions also increases the cooling rate 

%{\bf Give full equation if you are going to plug in numbers.}

%{\bf Above: nice to make this point earlier when you first present scaling relation.}
%
%{\bf The paragraphs don't tell a story. Tie it back to the wings.
%Can they be explained by cooling?}

\subsubsection{The Correlation of \lya\ and Bolometric Luminosity} \label{sec:ll}

\begin{figure}[h]
% \hbox{\hfill \includegraphics[height=15cm,angle=-90,clip=true]{figures/ll.ps}
 \hbox{\hfill \includegraphics[height=15cm,angle=-90,clip=true]{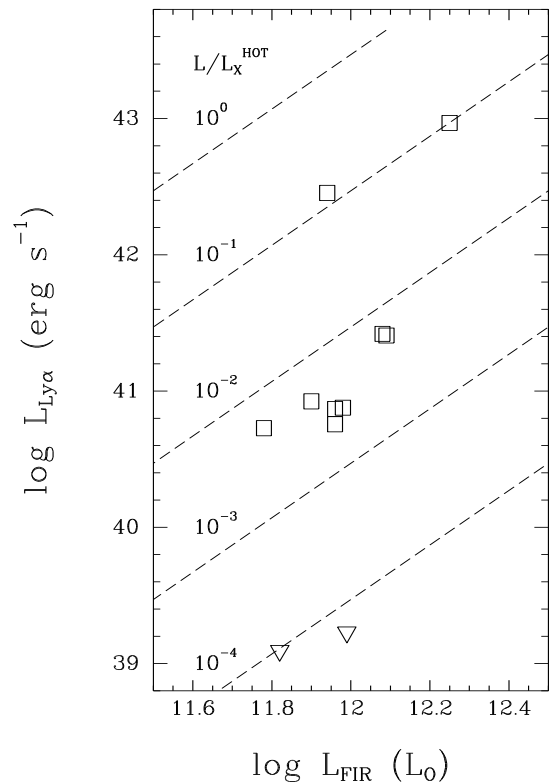}
  \hfill}
   \caption{\footnotesize The ULIRGs show a much larger dynamic range in
     \lya\ luminosity than they span in far-infrared luminosity. Excluding 
     the two non-detections (triangles), we find a strong, positive correlation between
     the \lya\ luminosity and the far-infrared luminosity. The dashed
     lines illustrate a power-law scaling, $L(\lya) \propto L_{IR}^2$.
   }
   \label{fig:lya_lir} \end{figure}

The ULIRG \lya\ luminosities span a much larger dynamic range, 
about four orders of magnitude, than do their far-infrared luminosities. 
Figure~\ref{fig:lya_lir} illustrates that a power-law scaling, 
$L(\lya) \propto L_{FIR}^{2}$, describes the luminosities of
the ULIRGs not dominated by AGN. Including the two AGN, \i11598-0112
and \i01003-2238, the exponent steepens to approximately
four.

%FIGURE 7

A quadratic scaling like this has been predicted,
but not observed, for the X-ray luminosity emitted by
starburst winds. Using the analytic solution for a 
strong wind driven out of a starburst region where
most the supernova energy has been thermalized
\citep{CC85}, \cite{Zhang:2014} calculate 
the 0.5-8.0~keV luminosity of the hot wind fluid
and find $L_X \propto SFR^2$. This scaling requires
constant thermalization and mass-loading efficiencies.
It is interesting that \cite{Zhang:2014} found that
the $L_X - SFR$ relation measured among star-forming galaxies 
is shallower than their prediction. The shallow relation
can be attributed at least in part to the diffuse X-ray 
emission from galaxies being brighter than the hot wind fluid.
The hot wind fluid is extremely challenging to detect directly
due to its low emission measure.  The extended X-ray
emission surrounding starburst galaxies is widely believed
to emanate from intermediate temperature gas at the boundary
between gas phases because it is spatially correlated with \Ha\ 
emission \citep{Strickland:2004}. That the \lya\ luminosity 
increases as roughly the square of the SFR, as traced by $L_{FIR}$,
suggests to us that this luminosity may be physically
related to the cooling of the hot wind.

\cite{Zhang:2014} predict the 0.5-8.0~keV X-ray luminosity 
of the hot wind fluid per unit SFR,
\begin{eqnarray}
L_X^{hot}[0.5-8.0 {\rm ~keV}] = 10^{43} {\rm ~ergs~s}^{-1}
\left (
 \frac{SFR}{100 \msunyr}
  \right )^2  \nonumber \\
\left (
 \frac{200 {\rm ~pc}}{R_0}
  \right ),
\end{eqnarray}
where $R_0$ describes the radius of the starburst region.
We substitute the \cite{Kennicutt:1998araa}
relation between SFR and far-infrared luminosity and
obtain 
\begin{eqnarray}
\frac{L_X^{hot}}{L_{FIR}} \approx 0.0045
\left (
 \frac{SFR}{100 \msunyr}
  \right )
\left (
 \frac{200 {\rm ~pc}}{R_0}
  \right ).
\end{eqnarray}
The dashed lines in Figure~\ref{fig:lya_lir}
compare the \lya\ luminosity to $L_X^{hot}$,
the predicted cooling luminosity.  
Among the starburst-dominated ULIRGs, 
the \lya\  luminosity is  $\approx 0.3$\% of the pedicted cooling
radiation from the hot wind fluid.  The closet analogy may
be the cooling of hot gas in the centers of galaxy clusters where 
calculations for $T \sim 10^7$~K gas  cooling non-uniformly 
show of order 1\% of the losses in strong UV and optical emission lines
\citep{Voit:1994}.

As shown in Section~\ref{sec:line_core}, most of the
\lya\ photons emitted near the systemic redshift are
scattered by neutral hydrogen; but the blueshifted wing
on the \lya\ profiles does not arise from scattering
and may therefore be emitted at larger radii. The
luminosity of just this highly blueshifted component
may be more directly related to the cooling wind than
is the total \lya\ luminosity.

\begin{figure}[t]
% \hbox{\hfill \includegraphics[height=10cm,angle=-90,trim=0 400 0 0]{figures/ll500.ps}
 \hbox{\hfill \includegraphics[height=10cm,angle=-90,trim=0 400 0 0]{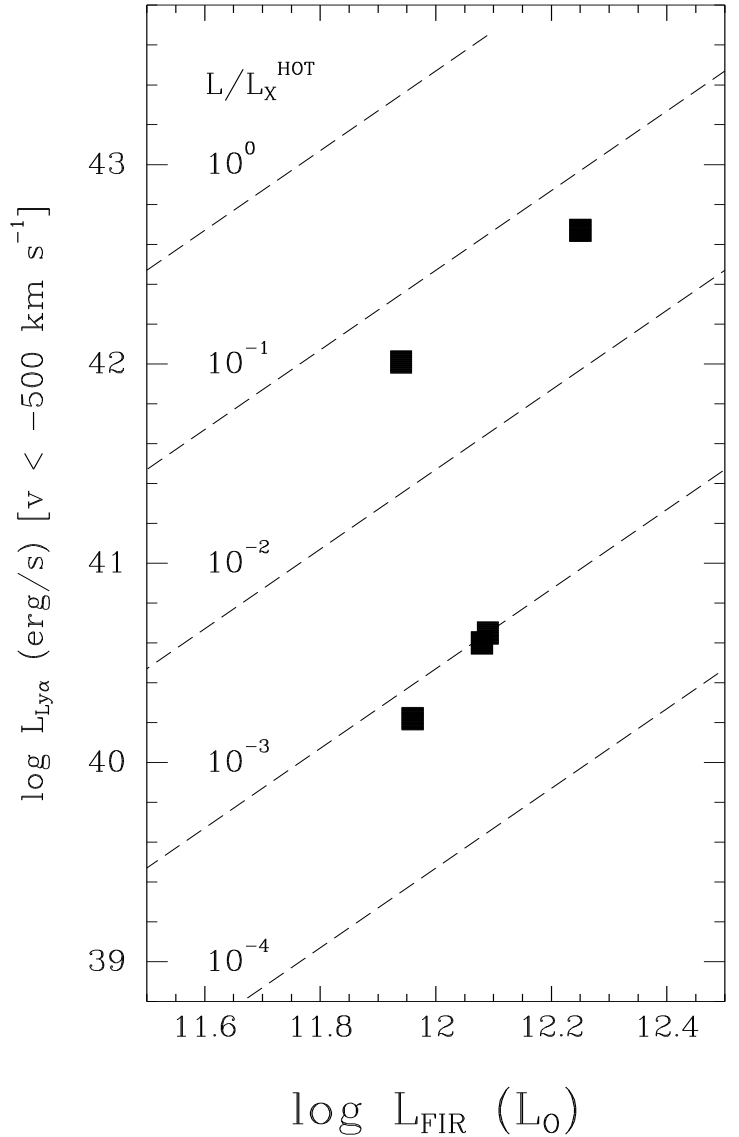}
  \hfill}
   \caption{\footnotesize The \lya\ luminosity in the blue wing, defined here
     as $v < -500$\kms\ increases with the far-infrared luminosity in a manner
     similar to the total \lya\ luminosity in Figure~\ref{fig:lya_lir}.
   }
   \label{fig:ll500} \end{figure}

We compare just the \lya\ luminosity
emitted blueward of $-500$\kms\  with the far-infrared 
luminosity in Figure~\ref{fig:ll500}, 
The luminosity of the highly blueshifted \lya\ emission  
increases with the far-infrared luminosity. 
Among the HII-dominated ULIRGs, 
the highly blueshifted \lya\ emission is about 0.1\% of the
far-infared luminosity.
The blueshifted \lya\ emission is a larger
fraction of $L_{FIR}$, roughly 5\%, in the AGN-dominated ULIRGs.
The steeper than quadratic 
rise in  $L_{\lya}$ with $L_{FIR}$ between the HII-dominated
AGN-dominated ULIRGs shows that the emergence of the AGN increases
the \lya\ emission from ULIRGs.

%Since the AGN contribution to the bolometric luminosity 
%approaches 75\% in Seyfert~1 ULIRGS, compared to less than
%35\% in cool HII and LINER ULIRGs, 

%FIGURE 14

\subsection{Fermi Acceleration of \lya\ Photons}

%In particular, resonance scattering probably shapes the ULIRG \lya\ 
%profiles that show significantly broader wings than the optical lines. 

The framework for the fast outflows should explain the shock 
speeds measured from optical emission-line ratios in addition 
to the lower Doppler shifts of resonance absorption lines.
The scattering of \lya\ photons across shock fronts could be
one such mechanism. 
\cite{Neufeld:1988} have shown that scattering of resonance photons
across shock fronts systematically blueshifts the photons. Since
the photons pick up a Doppler shift of $0.5 v_{sh}$ on each
scattering, the blueshift exceeds the shock velocity after only 
a few reflections across the shock front. 

We considered the shocks identified by \cite{Soto:2012a} \i00262+4251 
as a preliminary test of this idea. A shock speed of $v_{sh} = 225 \pm 50$\kms\
fit the  optical emission-line ratios well. In the Fermi acceleration
model, only three such scattering encounters are sufficient to produce the
observed line wing in \i00262+4251 which has $V_{B20} = -326$\kms.
The Doppler shift of the peak scattered emission
scales as $V_{b,pk} = 270 (N_{20} v_{s7})^{1/3}$ in their model.  

The scaling of this model predicts the velocities of the blue
wings increase with shock speed and the column density of the neutral 
gas, $N_{20}$,
Comparison of the shock velocities from \cite{Soto:2012b}, reproduced
here for convenience in Table~\ref{tab:esi_measure}, against the
velocity of the blue wings from Table~\ref{tab:cos_measure} reveals
no correlation between the two velocities. It follow that the column
density of neutral gas  would have to be  correlated with the velocity 
of the blue wing for the data to support Fermi acceleration acceleration
as the primary explanation for the \lya\ line wings. 
We have argued, however, that the column
densities are lowest in the objects with the bluest line wings because
we measure the largest \fesclya\ in those objects. Hence, we
find no compelling evidence that Fermi acceleration shapes the blue
wings on the ULIRG \lya\ profiles. We emphasize, however, that
the correspondence between the shapes of the \lya\ and optical
line wings varies considerably among the ULIRGs.  And Fermi
acceleration seems like an important physical process that
could play a role in some objects although it awaits compelling
empirical support.

\subsection{Relation between Outflow Velocity and  Galaxy Size}

%The fastest outflows discovered previously  provide general support for this
%scaling. 

Some wind models predict the fastest outflows from the most
compact starbursts. For example, \cite{Murray:2011p66} argue that 
the cluster luminosity cannot grow much beyond its Eddington 
limited value due to outflows driven by radiation-pressure which
disrupt giant molecular clouds faster than the cloud dynamical timescale. 
Momentum conservation requires terminal velocities scale as 
\begin{eqnarray}
V_{term} \propto R_0^{-1/2},
\end{eqnarray}
where $R_0$ again describes the radius of the region where the wind is
launched. In general support of this scenario, the galaxies with 
the most blueshifted resonance {\it absorption} turn out to indeed
be very compact with radii of roughly 100~pc 
(\cite{Tremonti:2007}; \cite{Heckman:2011}; \cite{Diamond-Stanic:2012}).
The discovery of fast outflows from ULIRGs therefore begs the question
of the starburst size in ULIRGs and the SFR surface density.

The starbursts in ULIRGs have long been considered concentrated; and
one of the main differences between ULIRGs and the sub-mm-selected galaxies
at high redshift is the smaller size of the starburst region in the former.
One estimates the effective size of the starburst region in a ULIRG
from a black body argument. This approach assumes the galaxy is optically thick.
For example,
\cite{Scoville:2013} give
\begin{eqnarray} \label{eqn:t_dust}
R_{FIR} = 1.09 {\rm ~kpc}
\left (
 \frac{L_{FIR}}{10^{12} \lsun}
  \right )^{1/2}
\left (
 \frac{35 {\rm ~K}}{T_d}
  \right )^{2},
\end{eqnarray}
where the fiducial dust temperature, $T_d$, was chosen to represent 
starburst dominated ULIRGs. If the source is optically thin, then  
this equation provides a lower limit on the radius. 

Substituting the far-infrared luminosities from Table~\ref{tab:cos_galaxies}
into Eqn.~\ref{eqn:t_dust}, we estimate the starburst radii range from 
0.85 to 1.2~kpc and greatly exceed those of the DCO and the massive, post-starburst 
samples.  This result is clearly sensitive, however, to the assumed dust
temperature. If $T_d \approx 100$~K in the AGN-dominated ULIRGs \i11598-0112,
then Eqn.~\ref{eqn:t_dust} gives a size for the emitting region of
just $R_{FIR} \approx 0.18 {\rm ~kpc~} (100 {\rm ~K}/T_d)^2$.

%This revision gives  $R_{FIR} \approx 0.15 {\rm ~kpc~} (100 {\rm ~K}/T_d)^2$
%for \i12071-0444 and 
%
% Get 0.134 kpc * (100 K / T)^2 = R_FIR in general

From Eqn.~\ref{eqn:t_dust}, we can see
that far-IR sizes for starburst-dominated ULIRGs will increase as the 
square-root of their luminosity, $R_{FIR} \propto L_{FIR}^{1/2}$. 
Hence, any correlations between starburst size and fast outflows
in the COS ULIRGs sample reduces to the strength of the correlation
between the blue wings and the far-infrared luminosity.
Regardless of whether we consider \lya\ or optical emission lines,
the line wings extend to higher velocities in more luminous systems.\footnote{
     Formally, the correlation coefficient between $V_{B20,sys}(\lya)$
     and $L(\lya)$ is $R_S = -0.62$ at only $1.74\sigma$ significance.
     The \lya\ luminosity correlates more strongly with the velocity of the 
     blue wing on the [\ion{O}{3}] line, $V_{B20,sys}([OIII])$; and we
     measure a correlation  $R_S = -0.80$ at $2.26\sigma$ significance.
     The correlation of the \lya\ luminosities with the far-infared
     luminosities is not one-to-one; and it is interesting that the
     correlations weaken when we replace $L(\lya)$  with $L_{FIR}$.
     The correlation coefficient between $V_{B20,sys}(\lya)$
   and $L_{FIR}$ is $R_S = -0.60$ at $1.79\sigma$ significance.
   The far-infared luminosity correlates less strongly with the velocity of the 
   blue wing on the [\ion{O}{3}] line, $V_{B20,sys}([OIII])$; and we
   measure a correlation  $R_S = -0.57$ at $1.61\sigma$ significance.}
We conclude that among the starburst ULIRGs, those with larger far-IR sizes tend
to show more extended blue wings in \lya. Only when the starburst dominated
ULIRGs are compared to the AGN dominated ULIRGs as a class would we recover
the inverse scaling of faster outflows from smaller launch regions that
the escape velocity argument would favor.

%, i.e., a trend opposite to
%the expectation that outflow speeds will be faster from smaller, more
%concentrated regions of star formation.

%While a constant temperture makes the underlying correlation between $V_{B20}$ 
%and luminosity indicate faster outflows from larger sources, variations in
%the dust temperature can easily wipe out this trend. 

%Modeling dust temperatures for these objects is clearly critical for establishing
%the relation between their sizes and outflow speeds. 

For completeness, we also estimate NUV sizes using the WFC3/F225W 
images shown in the middle panel of Figure~\ref{fig:lya_luminosity}. 
Specifically, we measured the area of 
the aperture enclosing all the pixels with surface brightness
greater than a fiducial value of $\Sigma_{\nu}^{Cut} (\lambda 2366) =
7.1 \times 10^{26}$~ergs~s$^{-1}$~cm$^{-2}$~Hz$^{-1}$~kpc$^{-2}$. This
threshold represents a SFR of 0.1 \msunyr kpc$^{-2}$ for continuous star 
formation (of duration greater than 100~Myr with a Salpeter initial mass function)
and  is of interest because it is widely believed to be the minimum value required 
to drive galactic winds \cite{Heckman:2003}. We find values of this $R_{NUV} \equiv (A/\pi)^{1/2}$ 
from 0.5 to 8.2~kpc. 
We find no correlation between these NUV sizes and the FIR sizes;
the assumed dust temperature is irrelevant to this comparison because 
our imaging sample excludes the ULIRGs with the strongest AGN. We find
that \i12071-0444  has a large size $R_{NUV} = 5.3$~kpc which
supports our argument (Section~\ref{sec:sample_agn}) that the starburst 
is at least as luminous as the AGN in this object.
These NUV sizes show no correlation with $V_{B20}$.

 The notable point here is that for a constant dust temperture, the
larger starbursts show higher outflow velocities, and this trend
is opposite to both the theoretical prediction for clouds accelerated 
by radiation pressure and the Doppler shifts of  resonance absorption 
lines in spectra of extremely compact starbursts. In our proposed
picture, however, the broad wings on emission lines come from condensation
of the hot wind whereas the resonance absorption lines come from clouds 
accelerated by some combination of  radiation pressure, cosmic rays,
and/or the hot wind. The difference between the scaling relation for 
the broad line wings and the resonance absorption lines therefore poses no
obvious contradiction but should be explored further via numerical simulations.
We also emphasize the need for resolved measurements of the starburst
sizes in ULIRGs and/or SED constrained dust temperatures; short of these,
the size estimates too uncertain to consider them compelling evidence
for the condensation scenario.

\subsection{Evolution of \lya\ with Merger Progression}

ULIRGs are examples of $L^*$ galaxies caught during
their transition to the red sequence. As we do not yet understand
what makes, and keeps, galaxies 'red and dead,' direct measurements of
feedback in these objects are of particular interest. What we
know already is that towards the final merger stage a larger fraction 
of the bolometric luminosity in ULIRGs appears to be powered by an 
accreting, supermassive black hole \citep{Veilleux:2009}, and the
accompanying increase in bolometric luminosity correlates with
an increase in the speed of the molecular outflows \citep{Veilleux:2013}.
{\rev
It is interesting to explore how the broad line wings fit into this picture.  
}

{\rev
To make this comparison, we refer back to the AGN fractions introduced
in Section~\ref{sec:sample_agn} and supplement those classifications with
measurements of the far-infrared luminosities. 
In Figure~\ref{fig:lya_lir}, for example, we saw that the ULIRG with the dominant AGN is 
25\% more luminous than any of the other objects in the sample; this result
is consistent with the trend that more luminous ULIRGs have a higher AGN 
fraction on average. What Figure~\ref{fig:lya_lir} shows that had not
been seen before is a high ratio of \lya\ to far-infrared luminosity in \i11598-0112 and 
in the ULIRG with the next highest AGN fraction, \i01003-2238. We note that the enormous increase
in the strength of the \lya\ emission (relative to $L_{FIR}$ in these two ULIRGs) and attribute it to the 
emergence of the AGN. This increase in \lya\ production may be directly related
to the AGN spectrum or (indirectly) to changes in the ISM induced by AGN feedback
which enhance the escape fraction of \lya\ photons.
The take-away point is that the ULIRGs with a dominant AGN show evidence for 
additional production of \lya\ emission.}

{\rev
Starbursts, in contrast to AGN, produce the majority of the bolometric luminosity
in the typical COS ULIRG, so an important question is whether the blueshifted wing 
on the \lya\ profile in a starburst-dominated spectrum  could be directly 
related to an AGN, albeit a less luminous one. In our opinion, the properties of 
the starburst ULIRGs measured to date suggest the answer is {\it no}. First, we have
demonstrated that the kinematics of the \lya\ wings indicate they come from the
same parcels of gas emitting the weaker, but equally broad, wings on the optical
emission lines. Among the starburst-dominated ULIRGs, the region of broad optical
emission subtends an angle along the ESI slit that corresponds to up to 20 kpc with
an average radius of 3 kpc. Due to the large size of the excited region, 
it is more natural to attribute the {\it shock-like} line ratios of the broad component 
to outflows rather than to the narrow-line region of an AGN \citep{Soto:2012a,Soto:2012b}.
Second, we found that the ratio $L(\lya)/L_{FIR}$ is over an order of magnitude higher
in the AGN-dominated ULIRGs than in the starburst-dominated ULIRGs. This abrupt jump
indicates strong evolution in either the production mechanism or escape fraction of
\lya\ emission as the AGN luminosity grows. We would have expected a continuous
increase in $L(\lya)/L_{FIR}$ with $L_{FIR}$ if the AGN produced the \lya\ emission
in the starburst-dominated ULIRGs.
}

%Since $L(\lya)/L_{FIR}$ is over ten times larger than the average ratio among the 
%subset of starburst-dominated ULIRGs with \lya\ emission, 

\begin{figure}[t]
 \hbox{\hfill \includegraphics[height=10cm,angle=-90,trim=0 0 0 0]{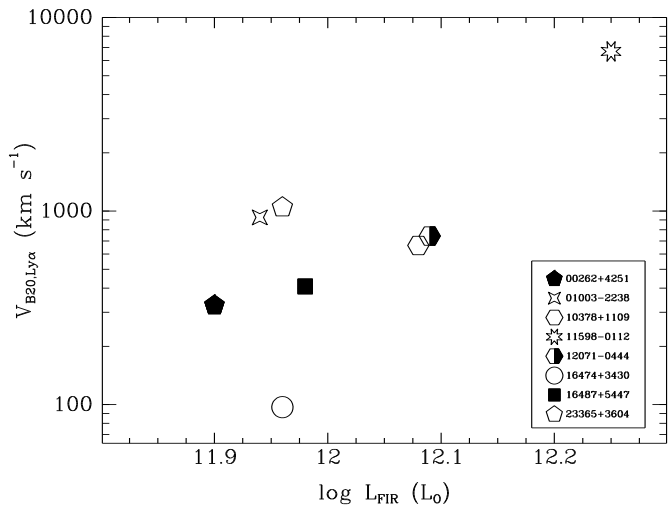}
  \hfill}
   \caption{\footnotesize {\rev
Doppler shift of \lya\ blue wing increases with 
ULIRG luminosity.  The velocity measurement for \i11598-0112 is
an upper limit because the wings of the \lya\ profile are blended with metal
lines; comparison to the unblended wings on the optical emission lines suggests,
nonetheless, that the wings on the deblended \lya\ are the broadest 
in the COS ULIRGs sample. When the ULIRGs with dominant AGN (\i11598-0112,
\i01003-2238, and possibly \i12071-0444) are removed, the tendency for the velocities 
to increase with $L_{FIR}$ is much less pronounced.}
   }
   \label{fig:vb20_lfir} \end{figure}

{\rev
In our initial discussion of Figure~\ref{fig:lya_luminosity}, we noted that the \lya\ 
wings extend to larger blueshifts as the \lya\ luminosity increases. Since the
AGN fraction is known to grow with increasing bolometric luminosity, we plot the velocity 
of the blue wing against far-infrared luminosity in Figure~\ref{fig:vb20_lfir}. In this
plane, the correlation between the Doppler shift and luminosity is less pronounced (than
$V_{B20}$ vs. $L(\lya)$) because \i01003-2238 has the second highest \lya\ luminosity
but an average $L_{FIR}$. It is the broad \lya\ and optical emission-line wings on 
the mini-BAL, \i11598-0112, which give the impression of increasing Doppler shift
with increasing $L_{FIR}$ in Figure~\ref{fig:vb20_lfir}. 
}

{\rev
\cite{Veilleux:2013} found that the velocities of the molecular outflows in 
ULIRGs increased gradually with AGN fraction but showed no significant dependence on the SFR.
While the overall correlation seen in Figure~\ref{fig:vb20_lfir} is not inconsistent with
the trend seen among the molecular outflows, neither do the data strongly support it when
one looks closely at the AGN fraction among individual systems. 
For example, the Seyfert~2 galaxies \i01003-2238 and \i12071-0444  have AGN which
produce up to 50\% of $L_{Bol}$, but their  $V_{B20}$ measurement are nearly the same as that 
of \i10378+1109 and lower than that of \i23365+3604, two ULIRGs with small AGN fractions.
}

{\rev
Detailed mapping of the \lya, optical, and molecular line wings is needed to clarify the
relationships among these outflow diagnostics. As a prelude to such work, we simply
note that \cite{Veilleux:2013} provide measurements of the molecular outflow speeds
for two of the COS ULIRGs: \i12071-0444, $V_{84}(OH) = -321$\kms, and \i23365+3604,
$V_{84}(OH) = -604$\kms. The blueshifts of the broad \lya\ wings ($V_{B20}(\lya) = $ 
-743\kms\ and -1046\kms, respectively) indicate faster outflows than do the OH measurements. 
Whether these discrepancies reveal different components of the bulk flow or merely variations
in measurement technique remain to be determined. 
}

%Figure~\ref{fig:vb20_lfir} also suggests a tendency for
%$|V_{B20}(\lya)|$ to increase with  $L_{FIR}$ 
%among the eight HII-dominated ULIRGs studied here, whereas \cite{Veilleux:2013} 
%found no increase in molecular outflow speed among the starburst-dominated objects.

Further exploration of the variation in outflow diagnostics along the merger sequence 
promises to provide an empirical description of how feedback disrupts
the ISM of a galaxy. We will show in Section~\ref{sec:rad_transfer}
that modeling the \lya\ wings places significant limitations
on the column density of neutral hydrogen along the ULIRG sightlines. In
the most extreme cases, which are the AGN, the feedback has generated holes
in the ISM which have become optically thin to Lyman continuum radiation.

%FIGURE NEW

%-- KEEP ----------------------------------------------------------
%
%While feedback from AGN has been shown
%to be accompanied by faster outflows, AGN
%do not explain the blue wings on ULIRGs like
%\i10378+1109 and \i23365+3604 which are
%clearly starburst dominated and perhaps
%\i1207-0444 although this latter case is more
%ambiguous. 
%
%We do find, however, that while
%the neutral gas column seems to be declining
%along the merger sequence, it is only in 
%the \i01003-2238 that the ULIRG is leaking
%Lyman continuum in our direction.
%
%
%------------------------------------------------------------

%(and $V_{B20}({\rm [OIII]})$) 
 
% ------ SAVE -------------------------------------------------
%The blueshift of the \naI\ absorption 
%from ULIRGs shows little difference between ULIRGs with luminosities
%dominated by HII region emission, LINERs, and Seyfert 2 AGN;
%however, the outflow velocities grow significantly larger among
%ULIRGs with Seyfert 1 characteristics \citep{Rupke:2005c}.

\section{Radiative Transfer Modeling}  \label{sec:rad_transfer}

To better determine the physical conditions along
the UV-selected sightlines, we performed Monte-Carlo radiative transfer
simulations for the \lya\ emission. We first tried a simple model consisting of a thin spherical 
shell moving radially at velocity $V_s$ \citep{Ahn:2003,Verhamme:2006}.
The shell has a column density $N(HI)$ in neutral hydrogen and dust optical depth 
$\tau_d$. {\it We found it difficult to produce emission at the systemic velocity
and the broad blue wings with this model, so we added a single parameter, 
$f_{C}$, to the shell model to describe the covering factor of neutral gas;
in this fiducial model, a fraction of sightlines $1 - f_C\equiv \Omega_{\rm hole}/4\pi$ intersect a `hole' which contains no neutral gas and dust  \citep{Behrens:2014}.} 

{\rev The spectrum of Ly$\alpha$ photons emerging from a spherical shell that contains a hole depends on direction (see Behrens et al. 2014 for a detailed analysis). Here, we capture this effect in a simplified (but faster) way, which is based on the fact that sightlines through the holes contain a superposition of the intrinsic (unscattered) and the scattered spectrum. We denote the (normalized) spectrum Ð as a function of velocity offset $v$ (of Ly$\alpha$ photons that scatter at least once with $S_{\rm scat}(v)$) while we denote the intrinsic spectrum with $S_{\rm int}(v)$.  Under the assumption that scattered radiation emerges isotropically\footnote{Under this assumption the scattered component can be viewed as a single isotropically emitting source with a total flux $f_{\rm scat}F_{\rm tot}$ (where $F_{\rm tot}$ denotes the total emitted flux). We only have an unobscured view of the central source along sight lines passing through holes. This unobscured source simply appears as an isotropic source with a total flux $F_{\rm tot}$. Therefore, the total flux along sight lines with holes is $(f_{\rm scat}+1)F_{\rm tot}$.} 
the normalized spectrum of Ly$\alpha$ photons along sight lines passing through holes is
\begin{equation}
S_{\rm holes}(v) \propto f_{\rm scat}S_{\rm scat}(v)+S_{\rm int}(v),
\end{equation} in which $f_{\rm scat}$ denotes the fraction of photons that scatter, which {\it depends} on the hole covering factor $1-f_C$.}

{\rev We carried out the calculations  using the code described previously
in \cite{Dijkstra:2006a,Dijkstra:2006b}.}  We used $10^5$ \lya\ photons initially distributed in 
frequency according to a model line profile matched to the \Hb\ profile.
This intrinsic \lya\ spectrum implicitly associates the source of the
\lya\ photons with recombination in \ion{H}{2} regions. We start with a
Doppler parameter $b = 12.9$\kms\ but find it necessary to increase it
to 40-50\kms\  to fit \i01003-2238, \i10378+1109, and \i09584+4714-B
Since \cite{Behrens:2014} showed that the spectrum which emerges from
a shell containing holes depends on direction, we also explored the
directional dependence of the spectrum. We studied \lya\ transfer through
shells that contain holes and predicted spectra for those sightlines
passing through the holes. We found the best fitting  models were 
consistent (within a factor of roughly two) with those obtained from the 
fiducial model. We use our fiducial model, however, because it requires 
significantly fewer photons to construct spectra (it assumes spherical
symmetry), and is therefore better suited to exploring the parameter
space describing the shell.

%We also explored the directional dependence of the spectrum. We predicted
%spectra for sightlines passing through the holes and 

\begin{figure*}[t]
% .../hst12533/mark_dijkstra/results_2014aug/v2000.plot
% \hbox{\hfill \includegraphics[height=20cm,angle=-90,trim=0 0 0 -100]{figures/mark_fit.ps} 
 \hbox{\hfill \includegraphics[height=20cm,angle=-90,trim=0 0 0 -100]{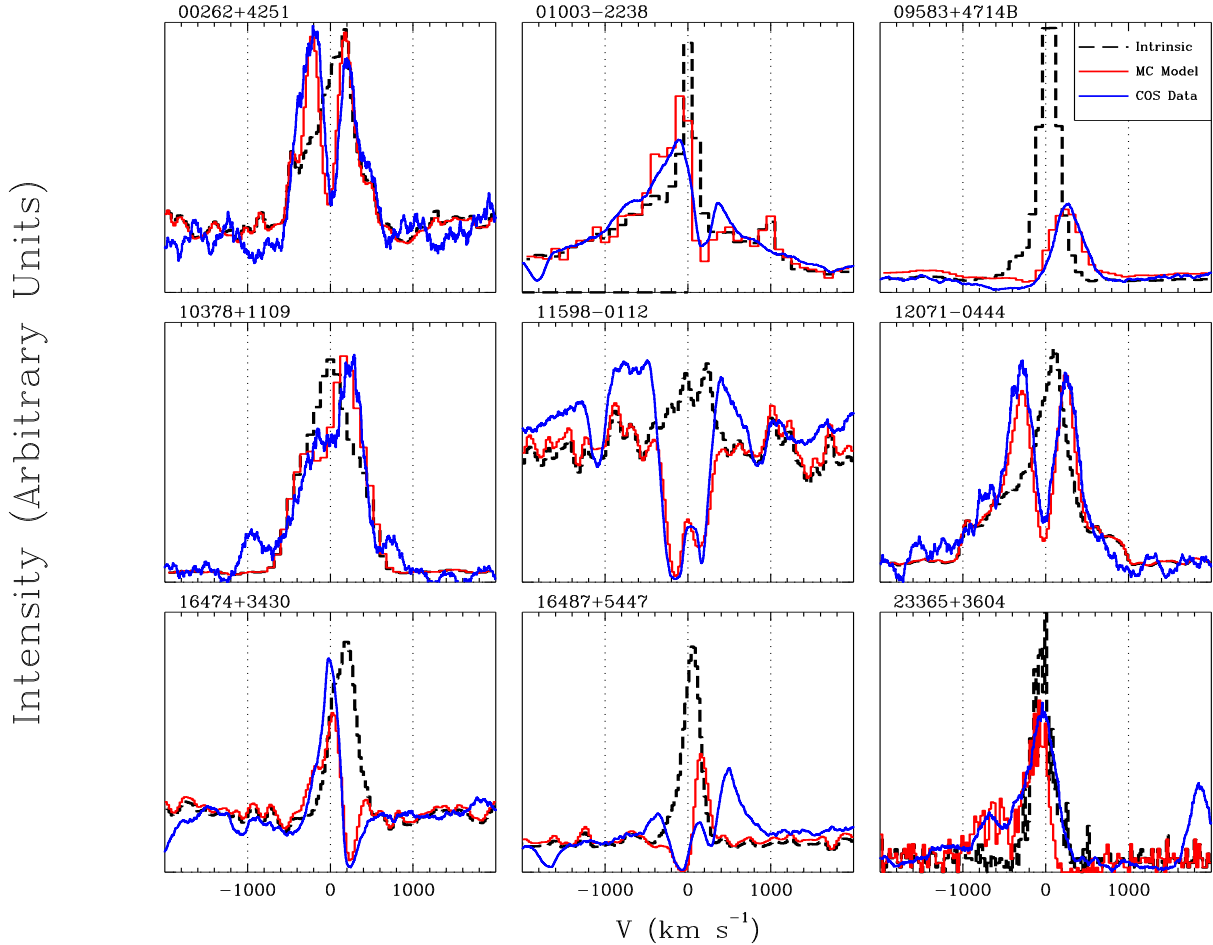} 
                \hfill}
   \caption{\footnotesize Radiative transfer models for \lya\ profiles. Starting from 
an intrinsic line profile (dashed line), we model the transfer of \lya\ photons through
a shell to obtain the emergent spectrum (red curve). These models provide a rough  
description of the observed \lya\ line profiles (blue line), and Table~\ref{tab:rad_transfer} 
lists the parameters of the model shown for each target. The sign convention used here 
for purposes of display defines redshifts as positive velocities.
   }
   \label{fig:mark_fit} \end{figure*}

Our approach was to find shell models that reproduce the basic properties of
the COS \lya\ profiles and then gain physical insight from the model parameters.
The detailed \lya\ shape will clearly depend on  factors in addition to the
model parameters  including 
viewing angle, source size, and shell thickness which are not directly constrained.
Figure~\ref{fig:mark_fit} shows examples of models that describe the main features
of each of the 9 COS \lya\ profiles.  The parameters, which are shown on each plot, span
a substantial range but generally indicate low column densities of neutral gas 
($17.0 \le\ \log N_{HI} ({\rm ~cm^{-2}}) \le 20.3 $), median shell velocity of 80\kms, 
covering fractions from 0.3 to 1, and relatively low dust optical depths ($\tau_d \le\ 2$).
The model parameters, however, are likely representative of the true values (along
the least obscured sightlines) to within a factor of a few as supported, for example,
by our finding similar parameters for models in which we accounted for the directional
dependence of the spectra that is naturally associated with shells that contain holes.

%FIGURE 15
%l b r t

The absorption troughs in the COS spectra determine the velocity of the neutral gas
in the shell model. In our description of \i09583+4714-B, for example, an expanding shell absorbs the 
\lya\ photons emitted slightly blueward of line center due to the Doppler shift of the source in the 
frame of the shell. Figure~\ref{fig:mark_fit} shows the fit with $\log N_{HI} ({\rm cm}^{-2}) = 20.0$, 
$V_s = -80$\kms\ (blueshift), and near unity covering fraction, $f_C = 0.9$. The flux minimum 
in the \i23365+3604 \lya\ profile is also well modeled with a shell moving, but the strong 
transmission at the systemic velocity precludes neutral gas at the systemic velocity and
requires a shell velocity near 400\kms. Similarly, a shell moving  at$\pm 150$\kms\ 
allows transmission near the systemic velocity in \i11598-0112. The redshifted 
absorption in the \i16474+3430 COS spectrum, which exhibits an inverted P~Cygni 
profile with blueshifted emission, is well described by an shell falling towards the
galaxy at roughly 250\kms. We produced the redshifted  dip in the  \i01003-2238 \lya\ 
profile with  infall at 200\kms. In contrast to these neutral gas flows,
a static shell or simply the galactic ISM  absorbs \lya\ photons near the systemic velocity in the spectra of
\i00262+4151 and \i12071-0444; and the low shell velocities in \i10378+1109 (30 \kms) and 
\i16487+5447 (50 \kms) could plausibly arise from wavelength calibration errors rather than
bulk motion.

Transmission at the shell velocity can be recognized by emission lines within saturated 
absorption troughs or absorption troughs that are not black and  requires either partial covering,
and therefore some leakage of the intrinsic spectrum, or a low column $N(HI) < 10^{17}$\col\
\citep{Behrens:2014}. 
The model for the \lya\ spectrum of \i01003-2238 requires columns of  neutral hydrogen 
which are optically thin to the Lyman continuum.
The transmision at zero velocity in the \i11598-0112 COS spectrum required a model
with a low \ion{H}{1} column to $N(HI) < 10^{18.8}$\col.  It is interesting that while \i23365+3604 
has substantial emission at the systemic redshift, Lyman continuum radiation will be absorbed
by the shell at 400\kms\ and will not escape the galaxy. These results are consistent
with a decline in the neutral gas column as the merger progresses since \i23365+3604,
\i11598-0112, and \i01003-2238 are classified, respectively, as a close binary/pre-merger, 
compact merger, and old merger.

% (slightly larger for Calzetti and slightly lower for Fitzpatrick)

The dust optical depth in the models can be directly compared to the measured
nebular reddening. The dust optical depth near \lya\ is $\tau_d = 10.0 E(B-V)$
for the \cite{Cardelli:1989} reddening curve, so it is trivial to convert
the color excess listed in Table~\ref{tab:esi_measure} to a meaured dust optical depth.
Comparison to the model $\tau_d$ in Table~\ref{tab:rad_transfer} shows that the dust
optical depth inferred along the UV sightlines is generally lower than that 
measured towards the HII regions. A number of physical interpretations are possible,
and we note only that this result suggests the UV and optical sightlines do not
sample exactly the same regions of these complex systems.

%{\bf How consistent are these results with the stellar vs gas EB-V results?}

{\rev
This modeling exercise also provides insight into the origin of the broad wings 
of the COS  \lya\ profiles. The blueshifted emission in the
\lya\ line can be attributed primarily to the blue wing of the intrinsic profile.}

\subsection{Comparison to Previous Modeling}

\cite{Verhamme:2014} and \cite{Behrens:2014}  have also recently explored the \lya\ 
profiles of shell models with low neutral columns and partial covering. Our results
strongly support the primary conclusion of those studies, namely that the emergence
of \lya\ emission near the systemic velocity boosts the probability that ionizing
photons may leak through.

The major difference is that we start with intrinsic profiles which have blue wings.
The intrinsic \lya\ profile may have a red wing in addition to the blue wing,
but we start with the simplest profile with which we can model the data.

The \lya\ signatures that betray the escape of Lyman continuum photons may,
however, depend quite a lot on the model. For example, \cite{Verhamme:2014} 
suggest that the peak \lya\ emission from density-bounded HII regions 
\citep{Osterbrock:2006} shifts by less than 150\kms\ from the systemic velocity.  
Close inspection of Figure~\ref{fig:mark_fit}
shows that the shift between the absorption feature and the maximum emission ranges
from 150 to 300\kms\ in \i00262+4251, \i10378+1109, \i16474+3430, and \i12071-0444;
our models for these profiles all have $\log N_{HI} ({\rm cm}^{-2} > 18.0)$, so our
results are consistent with the hypothesis that no Lyman continuum escapes. Our model 
for \i01003-2238, however, requires a low enough column density that we expect 
a significant Lyman continuum escape fraction, yet the emission peaks are still shifted by
more than 150\kms, specifically  $+200$\kms\ and $-300$\kms from the absorption feature 
visible  at 200\kms\ in Figure~\ref{fig:mark_fit}. Adding an anisotropic gas
distribution to the fiducial model could produce velocity shifts as large as we 
measure (even for low neutral gas columns)
when viewed from specific directions \citep{Behrens:2014}.

The \cite{Verhamme:2014} study concludes that peak separation 
will be less than 300\kms\ when Lyman continuum escapes. The peak
separation in our radiative transfer simulation of \i01003-2238 is 
larger, close to 500\kms.  
In our model, we must have the shell moving at 200 km/s to get the absorption right. 
We cannot increase the neutral hydrogen column density of the shell much above 
$10^{17}$\col, or we run into trouble with the observed damping wing. The photons 
that are absorbed in this shell and then backscatter pick up a redshift 
of 1 times the outflow velocity (because they appear practically at resonance in 
the frame of the gas). In this model, the peak-trough separation is twice the
shell velocity.

%This is probably why our simulated spectrum peaks closer to resonance than the observed profile. 
%Thus our model does not get the peak - trough separation quite right. This may be a short coming 
%of the simplicity of the shell-model. I think that is all we can say now. 

Figure~3 of \cite{Verhamme:2014} indicates that the blue peaks go away when the 
shell velocity exceeds $100 $\kms. Our model for \i01003-2238 (the thin case) as well as 
those for \i16474+3430, \i11598-0112, and \i23365+3604 all have blue peaks even though 
the shell velocity in the model is larger than 100\kms. This difference further 
emphasizes our result that the {\it intrinsic} \lya\ profiles of the COS ULIRGs, 
prior to processing by radiative transfer, must have high velocity blue wings 
and suggests the above quantitative result might change with assumptions about
the shape of the intrinsic line profile.

%EW not directly related to covering fraction; can bounce off clouds

%As mentioned above, radiative transfer calculations with an expanding shell have 
%previously been shown to provide an excellent description of \lya\ profiles like
%the one we observe for \i09583+4714-B.  

%The redshifted emission comes from a single scattering off the back side 
%of the shell, and the approaching side of the 

\subsection{Other Evidence for the Escape of Ionizing Radiation}

While radiative transfer models clearly provide extremely useful insight
into the physical conditions in ULIRGs which shape their unusual 
emergent \lya\ profiles, the simple geometry inherent to these models probably
precludes detailed quantitative assessment of Lyman continuum escape from
the separation of profile peaks alone {\it (see \S~\ref{sec:lyasec} for a discussion)}. What they confirm, however, is
that the shapes of the \lya\ profiles is the COS spectra make ULIRGs
leak some Lyman continuum radiation.
A quantitative measurement of this leakage requires direct
detection of the Lyman continuum. Short of that, however,
we can only consider other indirect tracers of the optical
depth. 

In particular, the ESI spectra cover the  $\lambda
\lambda 3726,29$ doublet as well as the [\ion{O}{3}] $\lambda
\lambda 4959, 5007$ doublet. Extreme flux ratios, 
$F[OIII] / F[OII] \sgreat\ 9$ can be an indication of density
bounded HII regions which have undersized $O^+$ zones
\citep{Jaskot:2013}. Hence, our radiative transfer modeling
results predict a very high flux ratio for \i01003-2238.
In support of this prediction, we confirm that this AGN has 
the highest flux ratio in the COS ULIRG sample, $F[OIII] / F[OII] 
\approx\ 6.0$. The only other COS ULIRG with a flux ratio
greater than 2 is \i12071-0444, a possible AGN, with 
$F[OIII] / F[OII] \approx\ 4.5$ and a much higher fitted
neutral gas column in Table~\ref{tab:rad_transfer}.  These 
line ratios would hardly be compelling evidence for Lyman 
continuum escape on their own because the high ratios could also 
be produced by large ionization parameters in these nebulae,
but they support our conclusions that (1) \i01003-2238 almost
certainly has the lowest neutral gas column in the COS
ULIRGs sample, and (2) it is plausibly optically thin
to the Lyman continuum.

Another approach would be to measure the relative residual 
intensity in the cores of the strongest low-ionization-state absorption lines 
\citep{Heckman:2001,Grimes:2009,Heckman:2011,Jones:2013}.
Partial covering of low ionization state gas can be measured
from the depth of saturated absorption lines \citep{Martin:2009}.
Deeper COS spectra, such as one might obtain for purposes of
directly measuring the Lyman continuum, could be used to
measure the covering fraction via lines of \siII\ or \cII.

\subsection{Other Explanations for Ly$\alpha$ Escape at Line Centre}
\label{sec:lyasec}

{\rev Shell models are frequently used to model Ly$\alpha$ lines because of the relatively small number of parameters \citep{Verhamme:2008p89}. We found that ordinary shell models could not fit these lines. We therefore introduced a hole, which represents a simple way to include anistropic escape. These models allow for more Ly$\alpha$ photons to escape without scattering, and at line centre. We stress that however, that the escape of Ly$\alpha$ photons at line center is not a unique signature of clear sight lines to the nebular regions. Two alternative explanations exist in the literature: 
\begin{enumerate}
\item \cite{Garavito:2014} have shown that Ly$\alpha$ transfer through rotating optically thick clouds can give rise to Ly$\alpha$ photons escaping at line center.

\item The `Neufeld mechanism' \citep{Neufeld:1991} allows photons emitted at line center to escape at line center from an optically thick medium if they only scatter off the surfaces of clumps in a multiphase medium. In order for the Neufeld mechanism to transmit photons at line center, we need {\it stationary} clumps (line center photons would not appear at line center in the frame of out-or-inflowing clumps). We can replace our `empty' holes with holes that are fully covered by stationary clumps with $N_{\rm HI}  \sim 10^{17}$ cm$^{-2}$. This would not affect the RT much in models in which shells contain $N_{\rm HI} \gg 10^{17}$ cm$^{-2}$, but it would modify the radiative transfer in e.g. object \i01003-2238 in which we inferred $N_{\rm HI} = 10^{17}$ cm$^{-2}$.
\end{enumerate}

Our models for the observed Ly$\alpha$ lines are therefore likely not unique: for example, it remains possible that some observed line profiles allow our `holes' to contain enough HI to be opaque to ionising radiation. However, in some models (e.g. \i01003-2238) there is so little scattering at all frequencies that this is likely not possible. Addressing both the effects of rotation and clumps in a systematic way is well beyond the scope of this paper, but we expect to explore this in more detail in a separate paper, in which we will also study the additional constraining power of observing the Ly$\beta$ regime (repeated surface scattering of Ly$\beta$ photons is limited by the conversion probability of Ly$\beta$ into H$\alpha+2\gamma$ continuum). }

%{\bf Also mention emission filling?}

% CII 1036.5 and 1334.5

%Even though these
%galaxies are optically thick to much of the optical light, the 
%unambiguous evidence for outflowing gas provides a clear mechanism
%for clearing much of the gas and dust in the direction of blowout.
%Invoking holes or cavities in the ISM therefore seems reasonable.

\section{Summary and Implications}  \label{sec:summary}

Feedback from star-formation on interstellar gas is central
to theories for galaxy formation and evolution, and 
the most plausible mechanism for powering the removal of
gas and metals from galaxies remains a galactic wind.
The challenge at present revolves around translating
the observable signatures of these winds into physical
properties. One promising path forward would be to
connect measurements of outflows to simple, analytic
descriptions of the wind physics.
Our confidence in this mapping would be enhanced if
we could understand different observational diagnostics
in the context of a simple physical framework.

%obtained previously at Keck~II with ESI 

In this paper, we used new far-UV spectroscopy from 
HST/COS in concert with high-resolution optical spectroscopy 
to compare signatures of galactic outflows. The sample
includes the most luminous and dusty starbursts found at low 
redshift, yet we detected \lya\ emission from 9 of the 11 galaxies. 
Even more surprising, however, only the spectrum of the 
galaxy that is not a ULIRG (\i09583+4714-B) shows
the classic P~Cygni profile produced by resonance scattering
off an expanding, neutral medium. We do identify signatures
of resonant scattering in the \lya\ profiles; these
include absorption near the systemic velocity, broad
line cores compared to non-resonant lines, and
a redshifted component absent in the optical lines.

%-------------------------------------------------------
%
%The role of photon diffusion in frequency space
%in creating the wings on the \lya\ profile varies
%widely across the sample. In many systems,
%the black absorption troughs near the systemic 
%redshift draw attention to this frequency 
%redistribution, and we show (using line shape
%parameters) that the \lya\ profiles have
%a larger fraction of photons in the wings
%relative to the optical emission lines. Yet,
%in a few spectra, \lya\ emission at the
%systemic redshift or an emission component superposed on 
%the absorption trough indicates lower neutral gas columns.
%We also demonstrated the relative impact of 
%resonance scattering on the line profiles by
%comparing the attenuation of the line to the continuum
%and identified four ULIRGs for the \lya\ emission
%was not significantly more attenuation than the UV
%continuum. It follows that in the majority of ULIRGs
%the resonance photons must travel a much larger
%distance to escape than do the continuum photons;
%and the exceptions include some early  pre-mergers
%and the oldest merger in our sample.
%
%-------------------------------------------------------

%in \naI, \siII, etc., and also a factor of 2-3 larger
%than the shock velocities  previously inferred 
%from the optical emission-line ratios
%\citep{Soto:2012a,Soto:2012b}. 

%The wings discovered on many of the \lya\ profiles are
%the most interesting feature and include emission 

{\rev
The unanticipated feature of the \lya\ line profiles
is their broad, blue wings. The blueshift of this emission
exceeds  500\kms\ in five of the COS ULIRGs. It seems
plausible, perhaps likely, that these wings arise from
AGN activity in \i11598-0112 and \i01003-2238, the two
objects where an active nucleus provides $\sgreat\ 50\%$
of the bolometric luminosity.  In contrast, the total
\lya\ luminosity is over an order of magnitude lower
(when measured relative to the far-infrared luminosity)
in the ULIRGs with starburst-dominated spectra and cool
far-infrared colors. We find it more natural to consider
origins for the broad wings related to the starburst winds in 
these systems.}

We attribute the large Doppler shifts 
(of order 1000\kms) to bulk outflow based on (1) the close correspondence 
between the shapes of the blue wing on the \lya\ and optical emission-line profiles,
(2) the similar attenuation of the \lya\ line and the underlying
continuum, and (3) the need to include these wings on the intrinsic
\lya\ profile in radiative transfer modeling of the emergent
line profile. This blueshifted \lya\ emission 
must come from the near side of the galaxy and hence outflowing; 
otherwise the resonance photons would be scattered by the 
neutral hydrogen column. The velocity of the outflowing gas
in the line wings is two to three times the escape velocity 
from the starburst.

The high velocity of the \lya\ emission compared to the lower blueshift of
resonance absorption lines poses a paradox. The low-ionization-state gas
traced by \naI\ and  \siII\ absorption is moving outwards more slowly
than the denser gas seen in emission. The Doppler shifts of 
the emission-line wings also exceed the shock velocities 
fit to emission-line ratios by a factor of two to three.
We do detect gas in absorption at these high velocities via 
the \ovi\ 1032, 1038 doublet. This tantalizing evidence for an association 
with a higher-ionization-state component of the outflow is currently
limited to a few ULIRGs, however, because the GO 12533 observations 
were designed to study \lya\ emission rather than UV continuum.

%The difference in outflow speed 
%
%This contradiction is
%compounded by an absence of 
%
%the P~Cygni line profiles observed
%in OH 119.233~$\mu m$ from the molecular outflows from two
%of these ULIRGs \citep{Veilleux:2013}

%We showed that larger \lya\ line widths were a 
%good indication of higher \lya\ escape fraction,
%so the gas kinematics plays a role in determining
%the \lya\ escape fraction. 

%
%The presence of \lya\ flux blueward of the systemic redshift
%can, for example, betray infalling gas \citep{Dijkstra:2006a,Dijkstra:2006b},
%viewing angles marked by holes in the neutral gas
%\citep{Verhamme:2014,Behrens:2014}, or shocks
%\citep{Neufeld:1988}. The optical spectral lines provide 
%information which helps distinguish among these 
%possibilties because photons in both Balmer recombination 
%lines and  collisionally excited forbidden lines escape 
%from galaxies directly without scattering; hence the
%optical emission-line profiles directly 
%reflect the kinematics of the ionized regions
%that emit them modulo attenuation by dust. 

Motivated by our initial goal of tying attributes 
of the line profiles to simple, physical models
for galactic outflows, we advance the conjecture
that this {\it cool, fast} gas is a direct
product of gas condensing out of the  hot phase 
of the wind \citep{CC85,Wang:1995}.
 In support of this scenario, we
demonstrated that the luminosity of this
component increases steeply with the SFR, 
following predictions for the total power radiated
by the hot wind. While most of that energy
would be radiated in X-rays, we belive we have identified
the few times  0.1\% of that cooling radiation emitted
in \lya\ and optical lines (by its large Doppler shift).
Since the cooling timescale will
be decreasing rapidly as the wind temperatures falls 
to where the gas emits strongly in these lines, we
expect it to be unstable and forms dense clumps or filaments. On
qualitative grounds, some association with fast
shocks seems likely; however it remains to be 
shown (using numerical simulations for example) 
why the shock speeds are lower than the fastest 
outflowing gas. Indeed, the low-ionization-state absorption is 
widely believed to mark outflowing clouds accelerated
by the hot wind and/or radiation; and
the shocks could be generated by the interaction
of this low-ionization-state component with the cooling hot wind.

If the broad emission-line wings do in fact mark
the cooling hot wind, then the absence of this
feature in the spectra of many starburst galaxies
requires explanation. While fully exploring this topic 
is beyond the scope of this paper (see Thompson \et, in prep), 
the winds from the galaxies with the highest SFR surface 
densities would appear to be good candidates for small cooling radii.
Denser winds cool faster and reach the critical inflection
in the cooling curve closer to the starburust region where
the emission is brighter due to the higher density (prior
to geomtrical dilution by the expansion).

% When comparing galaxies of similar mass,
%if mass-loss rates increase with SFR, then the 
%extreme star formation in the ULIRGs would produce
%high density winds with shorter cooling times and
%higher emission measures.

%{\bf Transition below is too abrupt.}
%
%---------------------------------------------

The relationship between the various outflow diagnostics
in ULIRGs should be applicable to a wide range of 
intensely star-forming galaxies over cosmic time.
These ULIRGs share a common property with high-redshift galaxies, 
namely high gas accretion rates, and evolve towards
a quiescent state on the merger timescale.
When ordered in a temporal sequence based on the 
merger progression, we find the fastest outflows and the
lowest columns of neutral gas in the most advanced merger stages.
These late stages show a larger contribution to $L_{FIR}$ from AGN,
but the outflows develop long before the AGN dominates $L_{FIR}$.
The combination appears to drill holes through the neutral gas 
which allow a small percentage of the Lyman continuum to
escape from even these extremely dusty galaxies.

One of the important outcomes of this study is that the prominent
blue wings on the \lya\ profiles also exist on the optical emission
lines; the optical wings are less prominent but cover the same 
velocity range as the \lya\ wings. It should be possible to 
resolve and study these fast outflows over much of cosmic time 
using highly-mutiplexed, infrared spectroscopy from the James 
Webb Space Telescope. Deeper observations with COS could 
provide detections of resonance absorption lines across a 
broad range of ionization states further defining the 
relationship between the signatures of outflows in absorption
and emission lines.

%as we demonstrated for \i16487+5447. 

Facilities: \facility{Keck:II}, \facility{HST}

\acknowledgements
This research was supported by the National Science Foundation under AST-1109288 (CLM)
and was partially carried out at the Aspen Center for Physics which is
supported by the National Science Foundation under Grant No. NSF PHYS-1066293
and the Kavli Institute for Theoretical Physics under support from
the National Science Foundation under Grant No. NSF PHY11-25915.
We thank Tim Heckman, Norman Murray, and Todd Thompson for stimulating
discussions about this work. 
We also wish to recognize and acknowledge the highly significant
cultural role that the summit of Mauna Kea
has always had within the indigenous Hawaiian community. It is a
privilege to be given the opportunity to conduct observations from
this mountain.

%%%%%%%% REFERENCES
\bibliography{Papers_cos}

\clearpage

% TABLES

%TABLE 1
%\label{tab:cos_galaxies}
\begin{turnpage}
 \begin{deluxetable}{lllllllll}
\tablecaption{ULIRG Properties}
\tabletypesize{\normalsize}
%\rotate
\tablewidth{0pt}
\tablehead{
\colhead{Name} &
\colhead{z} &
\colhead{$\log L_{FIR}$} &
\colhead{$F_{\nu}(25\mu m) / $} &
\colhead{$E_{B-V}^{MW}$}   &
\colhead{$m_{NUV}$} &
\colhead{$m_{FUV}$} &
\colhead{Merger} &
\colhead{Separation} 
\\
\colhead{}      &
\colhead{}  &
\colhead{($L_{\odot}$)} &
\colhead{$F_{\nu}(60\mu m)$}  &
\colhead{}  &
\colhead{ (AB mag)}  &
\colhead{ (AB mag)}  &
\colhead{Class} &
\colhead{(kpc)} 
\\
\colhead{(1)} &
\colhead{(2)} &
\colhead{(3)} &
\colhead{(4)} &
\colhead{(5)} &
\colhead{(6)} &
\colhead{(7)} &
\colhead{(8)} &
\colhead{(9)}
}
\startdata
IRAS 00262+4251  &   0.097240\tablenotemark{a}   & 11.90 & 0.11 & 0.066  & 18.73  & 19.22 & ${\rm IVa}^\dagger$    & $< 1.4$  \\
IRAS 01003-2238  &   0.117701\tablenotemark{b}   & 11.94 & 0.25 & 0.015  & 18.49  & 18.51 & ${\rm  V}^\dagger$     & $< 3.4$  \\
IRAS 08030+5243  &   0.083500\tablenotemark{a}   & 11.82 & 0.07 & 0.037  & 21.78  & 21.17 & ${\rm  IVb}^\dagger$   & $<1.3$  \\
IRAS 09583+4714  &   0.085887\tablenotemark{c}   & 11.78 & 0.18 & 0.012  & 18.46  & 19.09 & ${\rm  IIIa}^\dagger$  & 14.5  \\
IRAS 10378+1109  &   0.136238\tablenotemark{b}   & 12.08 & $\le 0.15$ & 0.028  & 21.04  & 21.22 & ${\rm  IVb}$           & $<1.2$  \\
IRAS 11598-0112  &   0.150694\tablenotemark{b}   & 12.25 & $\le 0.22$ & 0.023  & 17.54  & 17.87 & ${\rm  IVb}$           & $<1.6$  \\
IRAS 12071-0444  &   0.128360\tablenotemark{b}   & 12.09 & 0.21 & 0.039  & 19.46  & 20.82 & ${\rm  IVb}$           & 2.25  \\
IRAS 16474+3430  &   0.111471\tablenotemark{b}   & 11.96 & 0.11 & 0.002  & 18.39  & 18.93 & ${\rm  IIIb}$          & 3.73  \\
IRAS 16487+5447  &   0.103600\tablenotemark{a}   & 11.98 & 0.063 & 0.013  & 18.16  & 18.47 & ${\rm  IIIb}$          & 5.16  \\
IRAS 17028+5817  &   0.106090\tablenotemark{b}   & 11.99 & 0.032 & 0.020  & 20.26  & 20.82 & ${\rm  IIIa}$          & 25.3  \\
IRAS 23365+3604  &   0.064480\tablenotemark{a}   & 11.96 & 0.11 & 0.097  & 18.91  & 19.88 & ${\rm  IIIb}$          & $<1.1$  \\
\enddata
\tablenotetext{a}{Redshift from CO in \cite{Solomon:1997}}
\tablenotetext{b}{Redshift from optical emission lines.}
\tablenotetext{c}{Redshift from infrared \Pa\ emission line.}
\tablecomments{
  {\it Col 1:} Name of COS target.
  {\it Col 2:} Redshift.
  {\it Col 3:} Far-infrared luminosity defined as
               $ L_{FIR} = 3.86 \times 10^5 d_{Mpc}^2 [2.58 F_{\nu} (60\mu m) + F_{\nu}(100 \mu m)] L_{\odot} $.
               The fluxes (in Janskys) come from \cite{Murphy:1996}, and the luminosity
               distance assumes a universe with $\Omega_{\Lambda} = 0.7, \Omega_m = 0.3,
               h = 0.7$. In the limit that stars, rather than an AGN, heat the dust which emits the far-IR
               radiation, the inferred star formation rate is $SFR = L_{FIR} / 5.8 \times 10^9$\msunyr
               \citep{Kennicutt:1998araa}. The bolometric luminosities, $L_{IR}$, are a factor of two larger than $L_{FIR}$.
  {\it Col 4:} Infrared color. Ratio computed from the flux densities at 25\um and 60\um listed in Table~1 of \cite{Murphy:1996}.
               \cite{Sanders:1988b} used this ratio to identify ULIRGs with unusually high dust temperature; 
               The
               {\it warm  ULIRGs} have $0.2 < F_{\nu}(25\mu m) / F_{\nu}(60\mu m) < 2.0$.
  {\it Col 5:} Foreground extinction \citep{Schlafly:2011} recalibrated from
               the infrared-based dust map of \cite{Schlegel:1998}. The recalibration 
               assumes a \cite{Fitzpatrick:1999} reddening law with $R_V$ = 3.1.
  {\it Col 6:} Near-UV magnitude from GALEX (not corrected for foreground extinction).
  {\it Col 7:} Far-UV magnitude from GALEX (not corrected for foreground extinction).
  {\it Col 8:} Interaction classification as described in Section~3.4 of \cite{Veilleux:2002}:
               I - First approach, II - First contact, III(a/b) - Pre-merger (Wide binary/Close binary), 
               IV(a/b) - Merger (Diffuse/Compact), V - Old Merger. Dagger indicates object classified by
               \cite{Soto:2012b}.
  {\it Col 9:} Separation of nuclei. Upper limits from \cite{Murphy:1996}; all others made from the 
  observations presented here.
}
\label{tab:cos_galaxies}  
\end{deluxetable}
\normalsize

  \end{turnpage}

\clearpage

% TABLE 2
\begin{turnpage}
 \begin{deluxetable}{llllllll}
 \tablecaption{COS Imaging Observations of ULIRGs}
  \tabletypesize{\small}
%\rotate
\tablewidth{0pt}
\tablehead{
\colhead{Program} &
\colhead{Visit} &
\colhead{HST ID} &
\colhead{Config} &
\colhead{Filter} &
\colhead{Target} &
\colhead{Date} &
\colhead{$\tau$} 
\\
\colhead{GO HST}  &
\colhead{} &
\colhead{} &
\colhead{}  &
\colhead{}  &
\colhead{}  &
\colhead{}&
\colhead{ (s)}  
\\
\colhead{(1)}  &
\colhead{(2)} &
\colhead{(3)} &
\colhead{(4)}  &
\colhead{(5)}  &
\colhead{(6)}  &
\colhead{(7)}&
\colhead{(8)}  
}
\startdata
13407  &  1 & JC8T01010 & ACS/SBC & F125LP & IRAS 00262+4251 & 2013-12-28  & 892 \\          
13407  &  1 & IC8TA1010 & WFC3/UVIS & F225W & IRAS 00262+4251 & 2013-12-28 & 892 \\          
13407  &  1 & IC8TA1YMQ & WFC3/UVIS & F625W & IRAS 00262+4251 & 2013-12-28 & 71 \\          
%
%5982   & n/a & U2U20103T  & WFPC2     & F814W & IRAS 01003-2238 &1995-10-21& 1100 \\ %(600) \\ 
% ABOVE, 04T, 02T add 400 and 100 s, respectively
%5982&n/a & U202010[1,5,6,7,8]T & WFPC2&F439W&IRAS 01003-2238&1995-10-21   & 3030 \\ %(2400) \\ 
%5982 & n/a & U202010[1,5,6,7,8]T & WFPC2&F439W&IRAS 01003-2238&1995-10-21   & 3030 \\ %(2400) \\ 
5982 & n/a & U2020106T & WFPC2&F439W&IRAS 01003-2238&1995-10-21  & 800 \\ 
12533  & 21 & JBRM21010 & ACS/SBC & F125LP  &  IRAS 09583+4714 & 2012-06-08& 934 \\
12533  & 21 & IBRMA1010 & WFC3/UVIS & F225W &  IRAS 09583+4714 & 2012-06-08& 704 \\
12533  & 21 & IBRM1AM0Q & WFC3/UVIS & F625W &  IRAS 09583+4714 & 2012-06-08& 75  \\
12533  &  9 & JBRM09010 & ACS/SBC & F125LP &  IRAS 12071-0444 & 2012-05-17 & 934 \\
12533  &  9 & IBRMA9010 & WFC3/UVIS & F225W &  IRAS 12071-0444 & 2012-05-17& 704 \\
12533  &  9 & IBRMA9F2Q &  WFC3/UVIS& F625W &  IRAS 12071-0444 & 2012-05-17& 75  \\
12533  & 22 & JBRM22010 & ACS/SBC & F125LP &  IRAS 16474+3430 & 2012-12-14 & 934 \\
12533  & 22 & IBRMA2010 & WFC3/UVIS & F225W &  IRAS 16474+3430 & 2012-12-14& 710 \\
12533  & 22 & IBRMA2F2Q & WFC3/UVIS & F625W &  IRAS 16474+3430 & 2012-12-14& 75  \\
13407  &  2 & JC8T02010 & ACS/SBC & F125LP & IRAS 16487+5447 & 2014-01-12  & 932 \\
13407  &  2 & IC8TA2010 & WFC3/UVIS & F225W & IRAS 16487+5447 & 2014-01-12 & 932 \\  
13407  &  2 & IC8TA2NDQ & WFC3/UVIS & F625W & IRAS 16487+5447 & 2014-01-12 & 91 \\ 

12533  & 23 & JBRM23010 & ACS/SBC & F125LP &  IRAS 23365+3604 & 2012-09-23 & 934 \\
12533  & 23 & IBRMA3010 & WFC3/UVIS & F225W &  IRAS 23365+3604 & 2012-09-23& 704 \\
12533  & 23 & IBRMA3PEQ & WFC3/UVIS & F625W &  IRAS 23365+3604 & 2012-09-23& 76  \\
\enddata
\tablecomments{
  {\it Col 1:} HST Guest Observer program number.
  {\it Col 2:} Visit number.
  {\it Col 3:} Program identifier.
  {\it Col 4:} Instrument configuration.
  {\it Col 5:}  Imaging filter.
  {\it Col 6:} Target name.
  {\it Col 7:} Date
  {\it Col 8:} Exposure time.
}
\label{tab:cos_imaging}  
\end{deluxetable}
\normalsize

  \end{turnpage}

\clearpage

% TABLE 3
%\label{tab:cos_observations}
\begin{turnpage}
 \begin{deluxetable}{lrlllcllllr}
 \tablecaption{COS G130M Observations of ULIRGs}
  \tabletypesize{\footnotesize}
%\rotate
\tablewidth{0pt}
\tablehead{
\colhead{Program} &
\colhead{Visit} &
\colhead{HST ID} &
\colhead{Target} &
\colhead{Date} &
\colhead{Acquisition} &
\colhead{RA} &
\colhead{DEC} &
\colhead{CWL} &
\colhead{Band (A/B)} &
\colhead{Duration} 
\\
\colhead{GO HST}  &
\colhead{}  &
\colhead{}  &
\colhead{}  &
\colhead{}  &
\colhead{} &
\colhead{}  &
\colhead{}  &
\colhead{(\AA)}  &
\colhead{(\AA)} & 
\colhead{(s)}  
\\
\colhead{(1)}  &
\colhead{(2)}  &
\colhead{(3)}  &
\colhead{(4)}  &
\colhead{(5)}  &
\colhead{(6)} &
\colhead{(7)}  &
\colhead{(8)}  &
\colhead{(9)}  &
\colhead{(10)} & 
\colhead{(11)}  
}
\startdata
12533  & 1  & LBRM01010  &  IRAS 00262+4251 & 2011-11-11 & peak & 00 28 54.1526 & +43 08 15.82 & 1291 & 1137-1274/1292-1432 & 592    \\
12533  & 1  & LBRM01020  &  IRAS 00262+4251 & 2011-11-11 & peak & 00 28 54.1526 & +43 08 15.82 & 1309 & 1154-1293/1310-1448 & 592    \\
12533  & 2  & LBRM02010  &  IRAS 01003-2238 & 2011-12-03 & peak & 01 02 49.9631  & -22 21 57.02 & 1291 & 1137-1274/1292-1432 & 1716    \\
12533  & 4  & LBRM04010  &  IRAS 08030+5243 & 2011-10-30 & peak & 08 06 50.8856 & +52 35 7.70  & 1291 & 1137-1274/1292-1432 & 932    \\
12533  & 18 & LBRM18010  &IRAS 09583+4714-B & 2012-11-08 & peak & 10 01 30.4490 & +46 59 51.74 & 1291 & 1137-1274/1292-1432 & 1832   \\
12533  & 6  & LBRM06010  &  IRAS 10378+1109 & 2011-11-11 & peak & 10 40 29.1743 & +10 53 18.17 & 1291 & 1137-1274/1292-1432 & 420     \\
12533  & 6  & LBRM06020  &  IRAS 10378+1109 & 2011-11-11 & peak & 10 40 29.1743 & +10 53 18.17 & 1309 & 1154-1293/1310-1448  & 420     \\
12533  & 8  & LBRM08010   &  IRAS 11598-0112 & 2011-11-18 & peak & 12 02 26.7505 & -01 29 15.49 & 1309 & 1154-1293/1310-1448  & 652    \\
12533  & 8  & LBRM08020  &  IRAS 11598-0112 & 2011-11-18 & peak & 12 02 26.7505 & -01 29 15.49 & 1327 & 1173-1312/1328-1468 & 652    \\
12533  & 17 & LBRM17010  &  IRAS 12071-0444 & 2013-01-24 & offset  & 12 09 45.1000 & -05 01 13.20 & 1291 & 1137-1274/1292-1432 & 588    \\
12533  & 17 & LBRM17020  &  IRAS 12071-0444 & 2013-01-24 & offset  & 12 09 45.1000 & -05 01 13.20 & 1309 & 1154-1293/1310-1448  & 588    \\
12533  & 24 & LBRM24010  &  IRAS 16474+3430 & 2013-04-14 & offset & 16 49 14.2380 & +34 25 8.68 & 1291 & 1137-1274/1292-1432 & 2204   \\
12533  & 13 & LBRM13010  &  IRAS 16487+5447 & 2011-11-06 & peak & 16 49 47.0306 & +54 42 35.41 & 1291 & 1137-1274/1292-1432 & 1888    \\
12533  & 14 & LBRM14010  &  IRAS 17028+5817 & 2011-10-26 & peak & 17 03 42.0084 & +58 13 44.39 & 1291 & 1137-1274/1292-1432 & 1456    \\
12533  & 16 & LBRM16010  &  IRAS 23365+3604 & 2011-11-07 & peak & 23 39 1.2726  & +36 21 8.45  & 1327 & 1173-1312/1328-1468  & 1340    \\
12533  & 20 & LBRM20010  &  IRAS 23365+3604 & 2012-11-24 & offset  & 23 39 1.2290  & +36 21 8.03  & 1327 & 1173-1312/1328-1468  & 1876    \\
\enddata
\tablecomments{
  {\it Col 1:}  HST Program ID.
  {\it Col 2:}  Visit number.
  {\it Col 3:} HST ID.
  {\it Col 4:} Target name. 
  {\it Col 5:} Date of observation
  {\it Col 6:} Type of acquisition: blind offset or peak up.
  {\it Col 7,8:} Coordinates. For blind offset acquisitions, these coordinates are the exact position of the PSA. 
  For peak ups, the telescope was pointed near these coordinates and then moved in a spiral pattern to cover a square grid 
(ACQ/SEARCH  with SCAN-SIZE = 2 and STEP-SIZE = 1\arcsec) recording a spectrum at each stop. 
The telescope was then moved to the flux-weighted mean position of the target. We then 
improved the centering in the cross-dispersion direction by taking a spectrum, collapsing 
it in the dispersion direction, finding the offset in the cross-dispersion direction to 
the mean of the collapsed spectrum, and finally slewing the telescope by this offset. After
this ACQ/PEAKXD, we moved the telescope along the dispersion direction to obtain                                                                                                              
several ACQ/PEAKD exposures  (STEP-SIZE=1.118; CENTER FLUX WEIGHTING) and found
the center of the knot along the dispersion direction. The science spectra were then                                                                                                          
obtained through the PSA.                              
  {\it Col 9:} Central wavelength.
  {\it Col 10:} Wavelength coverage.
Each spectrum has a gap between detector Segments B and A. For some targets,
we simply chose a CWL that avoid loosing important lines to
this gap, but for others we needed continuous coverage and requested
exposures at two CWL's as indicated.
  {\it Col 11:} Exposure time.
}
\label{tab:cos_observations}  
\end{deluxetable}
\normalsize

  \end{turnpage}

\clearpage

% TABLE 4
%\label{tab:cos_measure}
\begin{turnpage}
 \begin{deluxetable}{llllllllll}
\tablecaption{Measured Properties from COS Spectra}
\tabletypesize{\small}
%\rotate
\tablewidth{0pt}
\tablehead{
\colhead{Object} &
\colhead{$F(\lya) \times 10^{15}$} &
\colhead{$\log L(\lya)$} &
\colhead{\fesclya} &
\colhead{$W_r(\lya)$} &
\colhead{E(B-V)$^*$}    & 
\colhead{$V_{B20}$} &
\colhead{$V_{R20}$} &
\colhead{$W_r(\siII)$} &
\colhead{$W_r(\siIII)$} 
\\
\colhead{}  &
\colhead{(\flux)}  &
\colhead{(ergs s$^{-1}$)}  &
\colhead{}  &
\colhead{(\AA)}  &
\colhead{(mag)}  &
\colhead{(\kms)}  &
\colhead{(\kms)}  &
\colhead{(\AA)}  &
\colhead{(\AA)}  
\\
\colhead{(1)}  &
\colhead{(2)}  &
\colhead{(3)}  &
\colhead{(4)}  &
\colhead{(5)}  &
\colhead{(6)}  &
\colhead{(7)}  &
\colhead{(8)}  &
\colhead{(9)}  &
\colhead{(10)}  
}
\startdata
  IRAS 00262+4251 & $3.52^{0.53}_{-0.61}$ &40.92 &$2.05^{+0.30}_{-0.34} \times 10^{-2}$   & $-10.8 ^{+3.9 }_{-5.4 }$  & $0.32 ^{+0.03 }_{-0.03 }$ & $-326_{-4}^{+63}$       & $449_{-18}^{+36}$  & 1.04 & 0.93  \\
  IRAS 01003-2238 & $79.1^{16.6}_{-4.00}$ &42.53 &$7.2^{+0.2}_{-0.4} \times 10^{-2}$      & $-32.4 ^{+15 }_{-2.0 }$& $0.02 ^{+0.02 }_{-0.02 }$    & $-1002_{-198}^{+75}$   & $692_{-104}^{+180}$ & \nodata     & \nodata     \\
  IRAS 08030+5243 & $ \le\ 0.072$         &$< 39.09$   &$<1.2 \times 10^{-4}$                        & \nodata                  & \nodata                    & \nodata                & \nodata             & \nodata     & \nodata  \\
IRAS 09583+4714-B & $2.91^{0.48}_{-1.32}$ &40.73 &$1.3^{+0.2}_{-0.6} \times 10^{-2}$    & $-7.3 ^{+3.6 }_{-4.0 }$  & \nodata                    & $212_{-27}^{+48}$      & $500_{-1}^{+2}$     & \nodata     & \nodata  \\
  IRAS 10378+1109 & $5.33^{0.12}_{-0.92}$ &41.42 &$2.44^{+0.006E}_{-0.21} \times 10^{-2}$   & $ \le -13$               & \nodata                    & $-664_{-4}^{+275}$      & $440_{-72}^{+2}$   & \nodata     & \nodata      \\
  IRAS 11598-0112 & $151^{121}_{-33.0}$   &$< 43.00$   &$5.5^{+4.4}_{-1.2}$               & $-30.1 ^{+15.5 }_{-20.7}$ & $0 ^{+ }_{- }$            & $-6679_{-2408}^{+2137}$ &$4291_{-203}^{+955}$& \nodata     & 0.05  \\
  IRAS 12071-0444 & $5.89^{0.88}_{-1.03}$ &41.41 &$0.24^{+0.04}_{-0.04}$            & $ \le -17$               & \nodata                    & $-743_{-120}^{+153}$    & $531_{-28}^{+42}$  & \nodata     & \nodata      \\
  IRAS 16474+3430 & $2.31^{1.01}_{-1.11}$ &40.87 &$1.33^{+0.60}_{-0.60} \times 10^{-2}$   & $-1.2 ^{+1.7 }_{-3.4 }$  & $0.16 ^{+0.02 }_{-0.13 }$  & $-97_{-101}^{+97}$      & $27_{-27}^{+18}$   & 1.32 & 3.42  \\
  IRAS 16487+5447 & $2.76^{1.86}_{-0.87}$ &40.88 &$1.31^{+0.85}_{-0.39} \times 10^{-2}$   & $-2.2 ^{+1.5 }_{-2.1 }$  & $0.12 ^{+0.03 }_{-0.02 }$  & $-408_{-127}^{+16}$     & $744_{-42}^{+4}$   & 0.91 & 1.01  \\
  IRAS 17028+5817 & $ \le\ 0.059$         &$< 39.23$   &$<4.6 \times 10^{-4}$                        & \nodata                  & \nodata                    & \nodata                & \nodata             & \nodata     & \nodata  \\
  IRAS 23365+3604 & $5.69^{0.07}_{-1.75}$ &40.78 &$2.10^{+0.03}_{-0.64} \times 10^{-2}$   & $ \le\ -29$              & \nodata                    & $-1046_{-40}^{+217}$    & $146_{-44}^{+20}$  & \nodata     &  \nodata     \\
%
%20 &  IRAS 23365+3604 & & \nodata                  & \nodata                      \\
\enddata
\tablecomments{
  {\it Col 1:} Name
  {\it Col 2:} \lya\ flux. The uncertainties
  include a term for the continuum placement. We adopted the linear continuum fit
  to bandpasses on both sides of the line and compared this to the maximum and
  minimum values of the continuum one could fit by adopting different bandpasses,
  e.g., such as our global fit to the continuum.
  {\it Col 3:} \lya escape fraction.
  {\it Col 4:} \lya luminosity 
  {\it Col 5:} Equivalent width of \lya. 
  Where the continuum was not detected, we use the $3\sigma$ upper limit on the
  continuum, as measured in a 20\AA\ wide window, to estimate the minimum 
  equivalent width (which appears as an upper limit since negative $W_r$ indicates 
  net emission). 
  {\it Col 6:} Stellar reddening. Values assume an intrinsic far-UV continuum slope of
  $\beta = -2.3 \pm 0.1$, where $F_{\lambda} \propto \lambda^{\beta}$ and the
starburst reddening curve of \cite{Calzetti:2000}.
The continuum bandpasses regions were chosen to avoid telluric emission, 
foreground Galactic absorption lines,
and interstellar absorption from resonance transitions.
  {\it Col 7:} Interpercentile velocity at $p = 20\%$ for the blue wing (relative
to the systemic velocity.
  {\it Col 8:} Interpercentile velocity at $p = 20\%$ for the red wing (relative
to the systemic velocity. The line width is $IPV_{20} = V_{B20} + V_{R20}$.
  {\it Col 9:} Absorption equivalent width of \siII\ 1190. The stronger \siII\
  1260 line is often blended with \sII\ 1259.
  {\it Col 10:} Absorption equivalent width of \siIII\ 1206.
}
\label{tab:cos_measure}
\end{deluxetable}
\normalsize

  \end{turnpage}

\clearpage

% TABLE 5
%\label{tab:esi_measure}
\begin{turnpage}
 \begin{deluxetable}{llllllll}
\tablecaption{Measured Properties from Optical Spectra}
\tabletypesize{\small}
%\rotate
\tablewidth{0pt}
\tablehead{
\colhead{Object} &
\colhead{E(B-V)} &
\colhead{$F(\Ha) \times 10^{15}$} &
\colhead{Aper.} &
\colhead{$W_r(\naI)$} &
\colhead{$V_{B20}(5007)$} &
\colhead{$V_{R20}(5007)$} &
\colhead{$V_{SH}$}     
\\
\colhead{}  &
\colhead{(mag)}   &
\colhead{(\flux)} &
\colhead{Corr.} &
\colhead{(\AA)}  &
\colhead{(\kms)}  &
\colhead{(\kms)}  &
\colhead{(\kms)}  
\\
\colhead{(1)}  &
\colhead{(2)}   &
\colhead{(3)} &
\colhead{(4)} &
\colhead{(5)}  &
\colhead{(6)}  &
\colhead{(7)}  &
\colhead{(8)}  
}
\startdata
IRAS 09583+4714-B & $0.138 \pm 0.003 $ & $4.63  \pm 0.01 $  & 1.81  & 0                         &  $-48_{-0}^{+11}$  & $72_{-0}^{+0}$         &   $\sles\ 150$              \\
  IRAS 16487+5447 & $0.115 \pm 0.001 $ & $13.75 \pm 0.01 $  & 1.31  & $ < 0.5^{\dagger}$       &  $-217_{-11}^{+0}$  & $238_{-0}^{+22}$      & $220 \pm 10 $     \\
  IRAS 16474+3430 & $0.199 \pm 0.001 $ & $7.849 \pm 0.005$  & 1.51  & 2.53                    &  $-80_{-10}^{+0}$  & $351_{-0}^{+10}$       & $300 \pm 25$       \\
  IRAS 00262+4251 & $0.242 \pm 0.004 $ & $6.466 \pm 0.007$  & 1.51  & $1.0 \pm 0.3^{\dagger}$  &  $-503_{-32}^{+64}$  & $172_{-21}^{+0}$      & $225 \pm 50$      \\
  IRAS 23365+3604 & $0.318 \pm 0.006 $ & $7.91  \pm 0.02 $  & 1.55  & $4.5 \pm 0.4^{\dagger}$  &  $-687_{-65}^{+172}$  & $150_{-0}^{+0}$     & $150 \pm 50$      \\
  IRAS 10378+1109 & $0.339 \pm 0.002 $ & $6.600 \pm 0.004$  & 1.64  & 6.7                      &  $-671_{-339}^{+31}$  & $254_{-0}^{+0}$     & $320 \pm 10$      \\
  IRAS 12071-0444 & $0.098 \pm 0.001 $ & $1.578 \pm 0.001$  & 1.33  & 1.45                      &  $-723_{-11}^{+0}$  & $164_{-0}^{+11}$      & $350 \pm 20$     \\
  IRAS 01003-2238 & $0.047 \pm 0.004 $ & $6.73  \pm 0.02 $  & 1.64  & 0.77                       &  $-1650_{-0}^{+10}$  & $68_{-0}^{+0}$     & $260 \pm 10$      \\
  IRAS 11598-0112 & \nodata            & $1.76  \pm 0.02 $  & 1.71  & 0.53                    &  $-1826_{-252}^{+516}$  & $652_{-76}^{+55}$  & $300 \pm 15$      \\
  IRAS 08030+5243 & $0.513 \pm 0.002 $ & $8.700 \pm 0.005$  & 2.27  & $7.4 \pm 1.0^{\dagger}$  &  \nodata               & \nodata            & $150 \pm 100$     \\
  IRAS 17028+5817 & $0.143 \pm 0.007 $ & $6.485 \pm 0.008$  & 1.56  & \nodata                  &  \nodata               & \nodata            & $225 \pm 15$      \\
%
%20 &  IRAS 23365+3604 & & \nodata                  & \nodata                      \\
\enddata
\tablecomments{
  {\it Col 1:} Target name.
  {\it Col 2:}  Nebular reddening inferred from Balmer decrement measured in ESI spectrum. 
  We adopted
  an intrinsic flux ratio $F(\Ha)/F(\Hb) = 2.86$ appropriate to Case~B recombination
  at an electron temperarture of $T_e = 10^4$~K \citep{Osterbrock:2006} and used
  the  \cite{Cardelli:1989} reddening curve with a ratio of total-to-selective
  extinction $R_V = 3.1$. We extracted the ESI spectrum from
  2\farcs5 long aperture centered on the COS G130M pointing. The optical emission
  lines were fit simultaneously using two emission components plus stellar Balmer
  absorption as described in \cite{Soto:2012a}.
  {\it Col 3:} \Ha\ flux in 2\farcs5 wide apeture along the ESI slit.
  {\it Col 4:} Aperture correction required to match \Ha aperture to COS G130M \lya aperture. 
  We used an $r$-band image of each ULIRG as a template for the spatial distribution of the
  \Ha emission \citep{Kennicutt:1989}. We scaled the relative intensities by the COS vignetting
  function described in the STSci Instrument Science Report COS 2010-09(v1); this correction
  lowers the weight of flux at radii greater than approximately 0\farcs4. We define the
  aperture correction by the ratio of the vignetted flux in the COS aperture (2\farcs5 diameter
  circle)   to the ESI aperture (2\farcs5 by 1\farcs0 rectangle).
  {\it Col 5:} Absorption equivalent width of \naI. A dagger denotes measurements from 
  \cite{Martin:2005}; all other measurements were made from the ESI spectra using the method
  described therein.
  {\it Col 6:} Interpercentile $p = 20\%$ velocity (blueshifted relative to systemic).
  {\it Col 7:} Interpercentile $p = 20\%$ redshifted velocity (redshifted relative to systemic).
  {\it Col 8:} Shock velocity fitted to $F([OIII]) / \Hb$ vs. $F([OI]) / \Ha$
  flux ratio using the \cite{Allen:2008} library of fully radiative shock models calculated
  with the MAPPINGS III shock and photoionization code. We list the value from \cite{Soto:2012b}
  aperture that falls within the COS primary science aperture.
}
\label{tab:esi_measure}
\end{deluxetable}
\normalsize

  \end{turnpage}

\clearpage

% TABLE 6
%\label{tab:rad_transfer}
%\input table_rad_transfer.tex
\begin{deluxetable}{lllll}
\tablecaption{Radiative Transfer Models for a Neutral Shell}
\tabletypesize{\normalsize}
%\rotate
\tablewidth{0pt}
\tablehead{
\colhead{Name} &
\colhead{$\log N_{HI}$} &
\colhead{V} &
\colhead{$\tau_d$}   &
\colhead{$f_C$} 
\\
\colhead{}      &
\colhead{(\col)}  &
\colhead{(km~s$^{-1}$)} &
\colhead{}  &
\colhead{}  
}
\startdata
IRAS 00262+4251  & 19.9  & 0     & 0 & 0.3  \\
IRAS 01003-2238  & 17.0  & 200  & 0.01 & 0.5  \\
IRAS 09583+4714B & 20.0  & -80   & 0.4 & 0.9  \\
IRAS 10378+1109  & 19.5  & -30   & 0.01 & 0.5  \\
IRAS 11598-0112  & 18.8  & $\pm 150$& 2.8  & 1.0 \\
IRAS 12071-0444  & 20.3  & 0     & 0 & 0.8  \\
IRAS 16474+3430  & 18.5  & +250  & 2.0 & 1.0  \\
IRAS 16487+5447  & 19.5  & -50  & 0.6 & 1.0  \\
IRAS 23365+3604  & 19.6  & -400  & 0.05 & 1.0  \\
\enddata
\tablecomments{
  {\it Col 1:} Name.
  {\it Col 2:} Column density of neutral hydrogen.
  {\it Col 3:} Shell velocity using the convention that outflows have a negative sign.
  {\it Col 4:} Dust optical depth.
  {\it Col 5:} Covering fraction of neutral gas.
}
\label{tab:rad_transfer}
\end{deluxetable}
\normalsize

\clearpage
%FIGURES
%\clearpage

%%%%%%%%%%%%%%%%%%%%%%%%%%%%%%%%%%%%%%%%%%%%%%%%%%%%%%%%%%%%%%%%%%%%%%%%%%%%%%%%%%%%%%%%%%%%%%%%%%
\appendix
\section{Velocity Measurements} \label{sec:appendix}

In order to compare \lya\ profile to  the optical emission lines in this work,
we require quantitative measurements of the shapes of the emission-line profiles.
Below, we follow the method introduced by \cite{Whittle:1985a} to describe the [\ion{O}{3}] 
profiles of active galaxies. This quantitative descriptions should also facilitate
comparison of the highly unusual \lya\ profiles of ULIRGs to those of other 
samples of \lya\ emitting galaxies. 

The strategy rests on measuring line widths at a specified fraction of the total
profile area. We take the profile area to be the net area in emission where the integration
is over the entire line profile; any other definition would be sensitive to spectral resolution.
For purposes of measuring the line width, asymmetry, etc., one  chooses a value of the fractional 
area that isolates the desired profile feature such as the line core, wings, etc.

We call the width at a specific fractional profile area the interpercentile velocity (IPV).
In summary, $IPV_{10}$, $IPV_{20}$, $IPV_{30}$, ...., and $IPV_{90}$
mark the bandpasses that include, respectively, 90\%, 80\%, 70\%, ..., and 10\% of the profile
area. To measure $IPV_{20}$, for example, we integrate the velocity profile up to a velocity
$V_B$ that includes 10\% of the total area and then continue integrating up to a velocity, $V_R$,
that includes 90\% of the total area. The interpercentile velocity width $IPV_p$ 
includes the central $1-p$ percent of the line flux, where $p $ denotes the fraction
the line emission detected at velocities larger than the interpercentile velocity width.
Following \cite{Whittle:1985a}, we define
\begin{eqnarray}
IPV_p \equiv V_{Bp} + V_{Rp}
\end{eqnarray}
and an asymmetry parameter
\begin{eqnarray}
A_p \equiv \frac{V_{Bp} - V_{Rp}}{V_{Bp} + V_{Rp}}
\end{eqnarray}
When applying this scheme to \lya, or any line that have absorption as well
as emission components, we have chosen to define the area of the line profile 
as the total area such that absorption components subtract from the total area.
We estimate the uncertainty in each marker by varying the continuum placement
as described in Section~\ref{sec:observations}.

In Figures~a-l, we have marked velocities of the blue wing, $V_{Bp}$, and red 
wing, $V_{Rp}$, where 20\%, 50\%, and 80\% of the total emission in the line 
lies at higher velocity. These two markers are defined for any percentile
$p$ of the total profile area, so we show how the line width grows as
$p$ decreases to include more of the total profile area. The Doppler shifts
of the blue and red wings are generally not symmetric about either the
systemic redshift or the median velocity the profile area. When the
profile is entirely in absorption blueward of the systemic
velocity, as for example in  Figure~c, then
the velocity $V_{B20}$ will be redward of the systemic
velocity because it is defined in terms of the net profile area
{\it in emission}. 

Since we have defined profile asymmetry relative to the median velocity of 
each emission-line profile, a profile with a net Doppler shift can have
$A \approx 0$ (no asymmetry) if the two sides of that profile are symmetric.
Returning to the P~Cygni profile in Figure~c as an example, again, we
see the emisssion is only mildly asymmetric about the median velocity of
300\kms; and the \lya\ asymmetry parameter is driven slightly redward by 
the presence of the blueshifted absorption. This gross asymmetry with 
respect to the systemic redshift is recognized instead by the comparison
of $V_{B,p}$ and $V_{R,p}$ which have the same sign instead of opposite
signs.

As a guide towards interpreting these measurements, we provide a few brief 
comments on the individual objects.

\paragraph{\i00262+4251} Black \lya\ absorption near the center of the [\ion{O}{3}]
emission profile. Scattered Resonance photons broaden \lya. Despite the disimilar
line shapes, the blue marker velocities have similar values in both lines. The
[\ion{O}{3}] profile shows a distinct blue asymmetry in the line wings whereas
the \lya\ profile has a red asymmetry at percentile markers.

\paragraph{\i01003-2238} This object has a substantial AGN luminosity.
The \lya\ profile has a blueshited (-200\kms)  median velocity about which it is
fairly symmetric. The [\ion{O}{3}] emission has a net blue asymmetry in the line
wings.

\paragraph{\i09583+4714-B} This non-ULIRG spectrum shows the classical
P~Cygni profile as described in the text. The asymmetry parameter of the
\lya\ profile is not very strong because it is measured relative to the
median velocity which is redshifted 300 \kms.

\paragraph{\i10378+1109} The blue wings of the \lya\ and [\ion{O}{3}]
profiles share the identical shape blueward of -200\kms. The \lya\
profile is missing photons in its core relative to the [\ion{O}{3}]
profile but shows more redshifted emission.

\paragraph{\i11598-0112} We show this narrow-line Seyfert 1 whose 
UV spectrum has mini-BAL characteristics for completeness. The
[\ion{O}{3}] profile blends with \feII\ lines, so we compare to
\Hb\ instead. The \Hb \ profile shows the blue asymmetry that
we typically find on optical lines from ULIRGs. The large
\lya\ velocities are probably misleading because \lya\ is likely
blended with metal lines. In the absence of blending, this spectrum
would be the only one with \lya\ emission extending to larger
blueshifts than the optical lines.

\paragraph{\i12071-0444} Resonance photons scatterd out of the
deep absorption trough 200\kms\ emerge at bluer and redder wavelengths.
The blueshifted component has a prominet wing which also shows optical
emission. The \lya\ and optical line shapes are very similar blueward
of -500 \kms.

\paragraph{\i16474+3430} The COS aperture includes a second nucleus offset 1\farcs84
from the target. In the ESI spectrum, this offset nucleus is redshifted about
300\kms relative to the target. Although the COS spectrum contains no spatial
information, the redshifted \lya\ absorption trough may be associated with the offset
nucleus. The fast orbital motion of these nuclei explains why the optical emission
lines are broad and double peaked. The  lack of \lya\  emission from the offset nucleus 
explains why the \lya\ profile is narrower than the optical lines.

\paragraph{\i16487+5447} Much of the \lya\ profile shows absorption. The emission
is largely redshifted. With a median velocity of roughly 500 \kms\ for the net
emission, the weak, the blueshifted  emission at -300 to -500 \kms\ produces 
a strong blue asymmetry as shown by the $IPV_{50}$ and $IPV_{20}$ markers.

\paragraph{\i23365+3604} The \lya\ profile is not delineated by narrow
absorption components, yet the larger amplitude of the peak [\ion{O}{3}]
emission suggests some scattering out of the line core. The blue
wings of the \lya\ and [\ion{O}{3}] lines cover the same velocity.
The extra wiggles in the blue wing of the \lya\ result from absorption
by an expanding shell in the radiative transfer model.

%     The interpercentile velocity markers, illustrated at  at $p = 20\%$ (magenta), $p = 50\%$ (yellow), 
%     and $p = 80\%$ (cyan), pick out the emission component because they are defined by the net
%     positive area. 

%% CHECK YOUR SIGN CONVENTION

%%%%%% APPENDIX FIGURES %%%%%%%%%%%%%%%%%%%%%%%%%%
\setcounter{figure}{0}

\clearpage
%a
\begin{figure}[t]
 \subfigure[\i00262+4251]{\label{fig:ipv_00262+4251}
%    \hbox{\hfill \includegraphics[height=15cm,angle=-90,trim=0 0 0 0]{figures/ipv_v1.ps} \hfill} }
    \hbox{\hfill \includegraphics[height=15cm,angle=-90,trim=0 0 0 0]{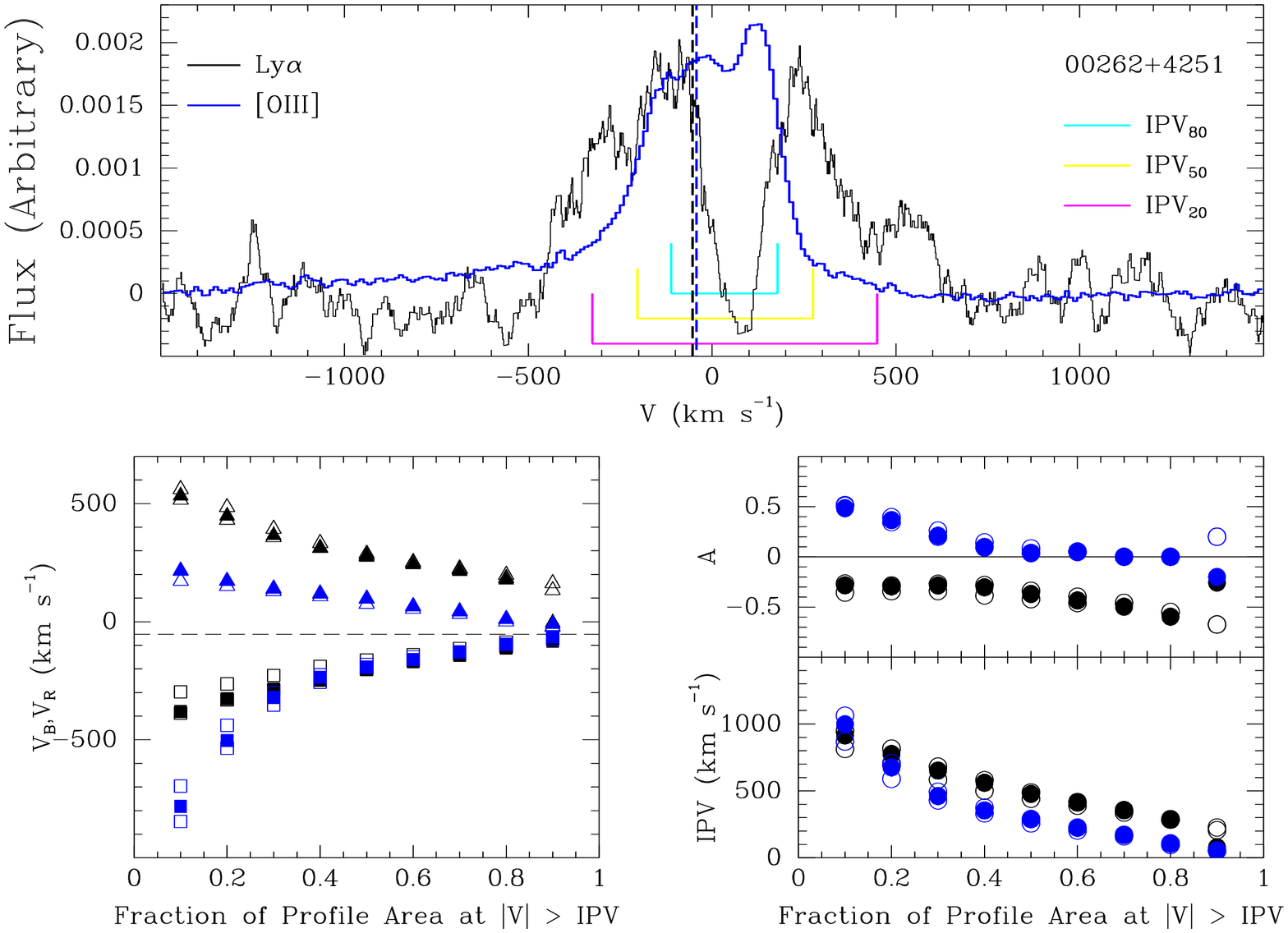} \hfill} }
   \caption{\footnotesize 
     {\it Top Row:}  The \lya\ and [\ion{O}{3}] line profiles 
     normalized by the net flux in emission.  For the \lya\ line, we have marked the 
     interpercentile velocities at $p = 20\%$ (magenta), $p = 50\%$ (yellow), and $p = 80\%$ 
     (cyan) to illustrate the meaning of $V_{B,p}$ and $V_{R,p}$. 
     {\it Bottom Row Left:} 
     Interpercentile velocity width and asymmetry parameter vs. fractional 
     area of the profile for \lya\ and [\ion{O}{3}]. Note that the asymmetry
     is measured relative to the median velocity which is marked by the dashed
     line in the top and right panels. 
     {\it Bottom Row, Right:} Comparison of the velocity markers $V_{B,p}$ and $V_{R,p}$ 
     for \lya\ and [\ion{O}{3}]. The blue wings (squares) are more correlated than the
     red wings (triangles). Blue and white points denote [\ion{O}{3}] and \lya, respectively;
     and the open symbols denote upper and lower limits on the solid point.
}
   \label{fig:ipv} \end{figure}

\clearpage
%b
\addtocounter{figure}{-1}
 \addtocounter{subfigure}{1}
  \begin{figure}[t]
   \subfigure[\i01003-2238]{\label{fig:ipv_01003-2238}
%    \hbox{\hfill \includegraphics[height=15cm,angle=-90,trim=0 0 0 0]{figures/ipv_v2.ps}\hfill}  }
    \hbox{\hfill \includegraphics[height=15cm,angle=-90,trim=0 0 0 0]{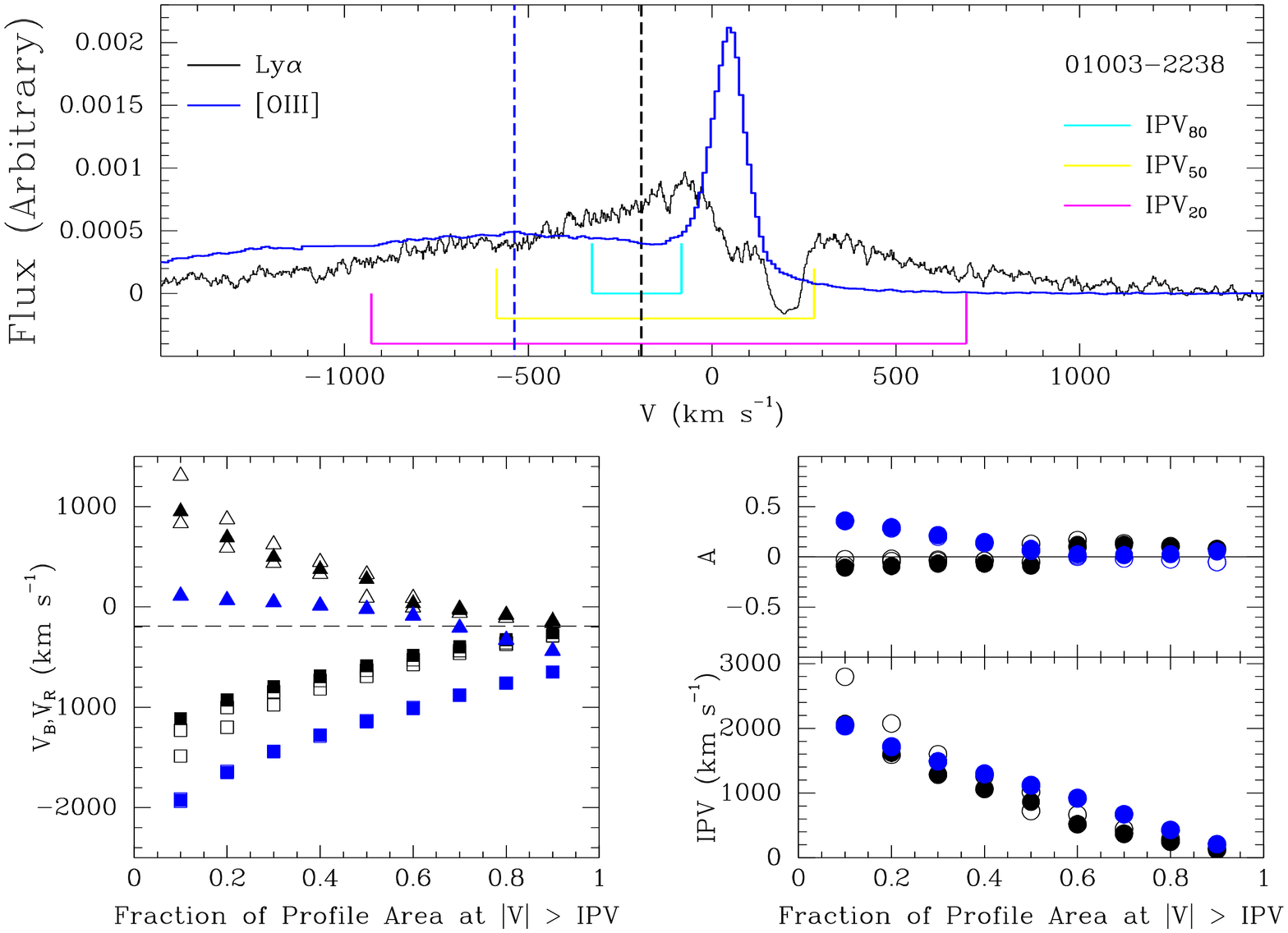}\hfill}  }
     \end{figure}

\clearpage
%c
\addtocounter{figure}{-1}
 \addtocounter{subfigure}{0}
  \begin{figure}[t]
   \subfigure[     \i09583+4714-B]{\label{fig:ipv_09583+4714B} 
%    \hbox{\hfill \includegraphics[height=15cm,angle=-90,trim=0 0 0 0]{figures/ipv_v18.ps} \hfill} }
    \hbox{\hfill \includegraphics[height=15cm,angle=-90,trim=0 0 0 0]{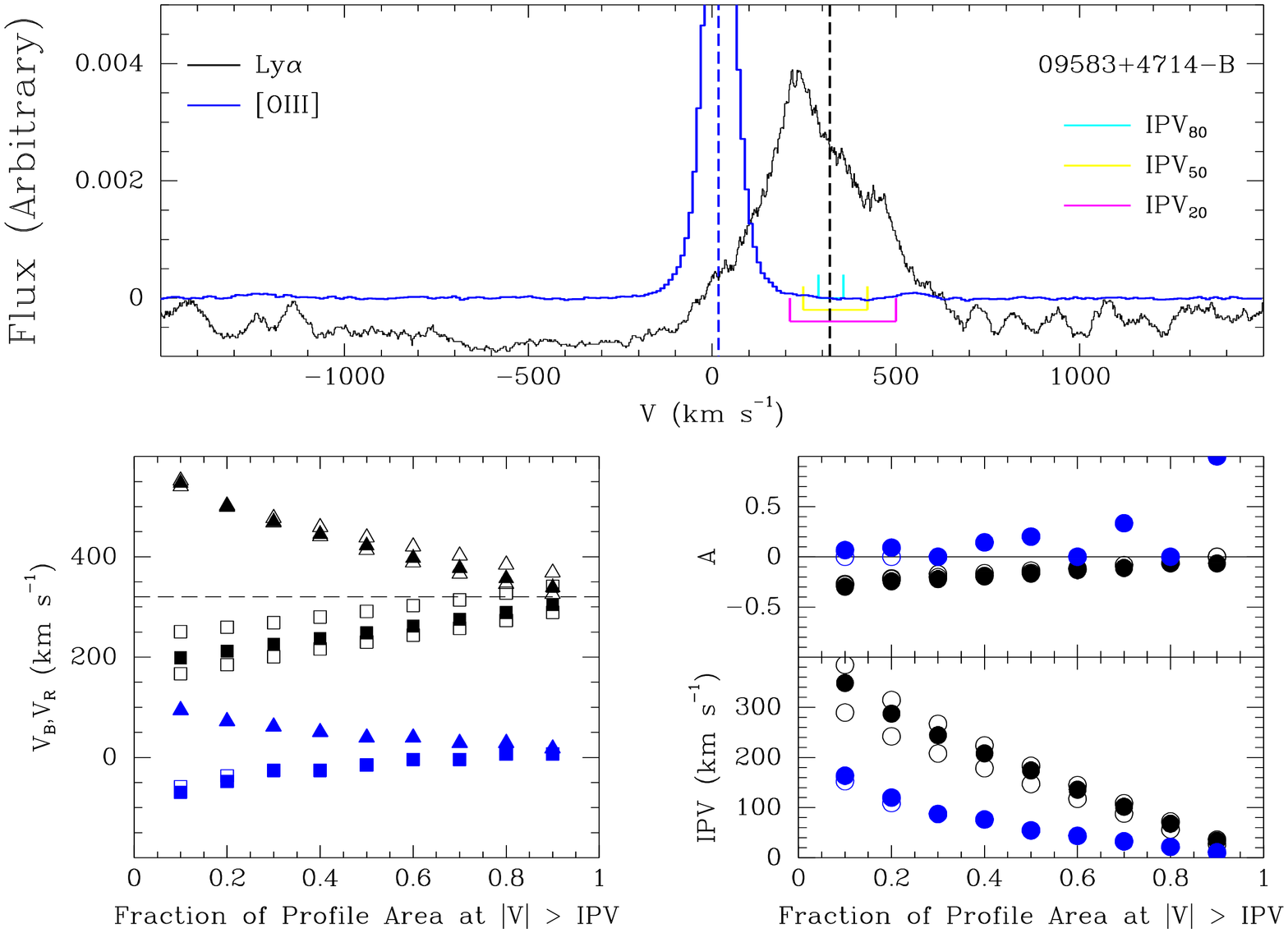} \hfill} }
     \end{figure}

\clearpage
%d
\addtocounter{figure}{-1}
 \addtocounter{subfigure}{0}
  \begin{figure}[t]
   \subfigure[\i10378+1109]{\label{fig:ipv_10378+1109}
%    \hbox{\hfill \includegraphics[height=15cm,angle=-90,trim=0 0 0 0]{figures/ipv_v6.ps} \hfill} }
    \hbox{\hfill \includegraphics[height=15cm,angle=-90,trim=0 0 0 0]{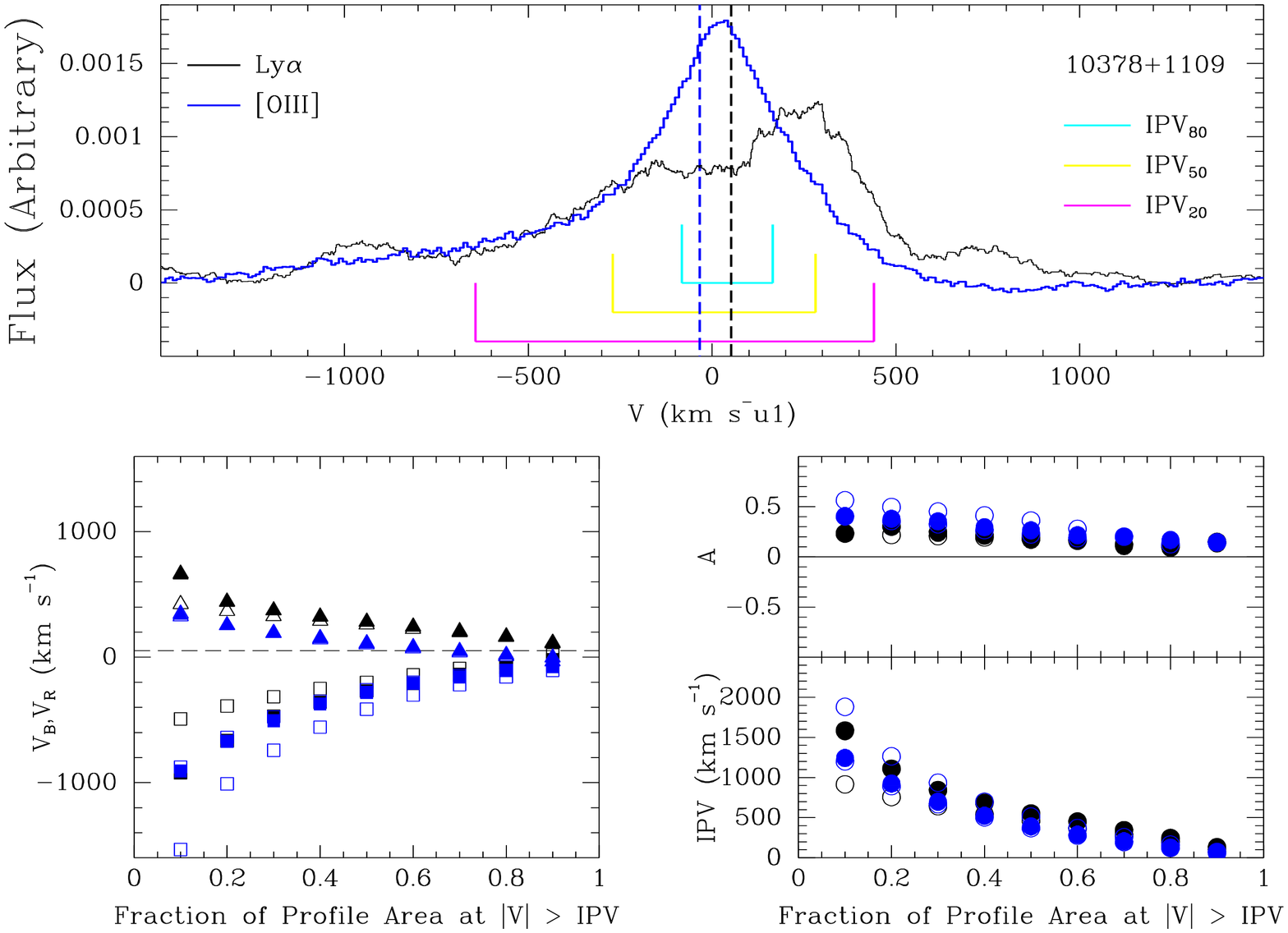} \hfill} }
     \end{figure}

\clearpage
%e
\addtocounter{figure}{-1}
 \addtocounter{subfigure}{0}
  \begin{figure}[t]
   \subfigure[\i11598-0112]{\label{fig:ipv_11598-0112}
%    \hbox{\hfill \includegraphics[height=15cm,angle=-90,trim=0 0 0 0]{figures/ipv_v8.ps} \hfill} }
    \hbox{\hfill \includegraphics[height=15cm,angle=-90,trim=0 0 0 0]{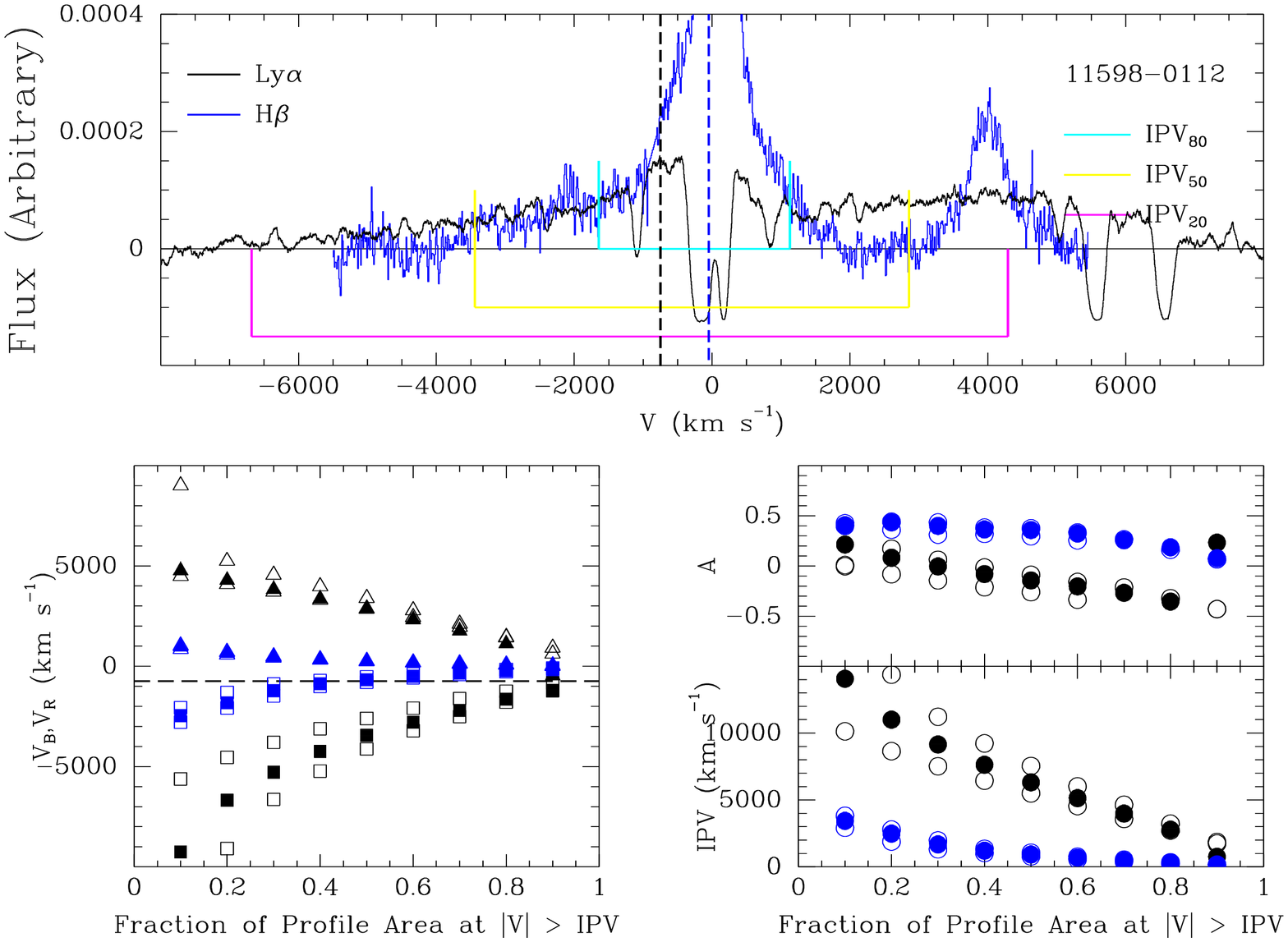} \hfill} }
     \end{figure}

\clearpage
%f
\addtocounter{figure}{-1}
 \addtocounter{subfigure}{0}
  \begin{figure}[t]
   \subfigure[\i12071-0444]{\label{fig:ipv_12071-0444}
%    \hbox{\hfill \includegraphics[height=15cm,angle=-90,trim=0 0 0 0]{figures/ipv_v17.ps}\hfill}  }
    \hbox{\hfill \includegraphics[height=15cm,angle=-90,trim=0 0 0 0]{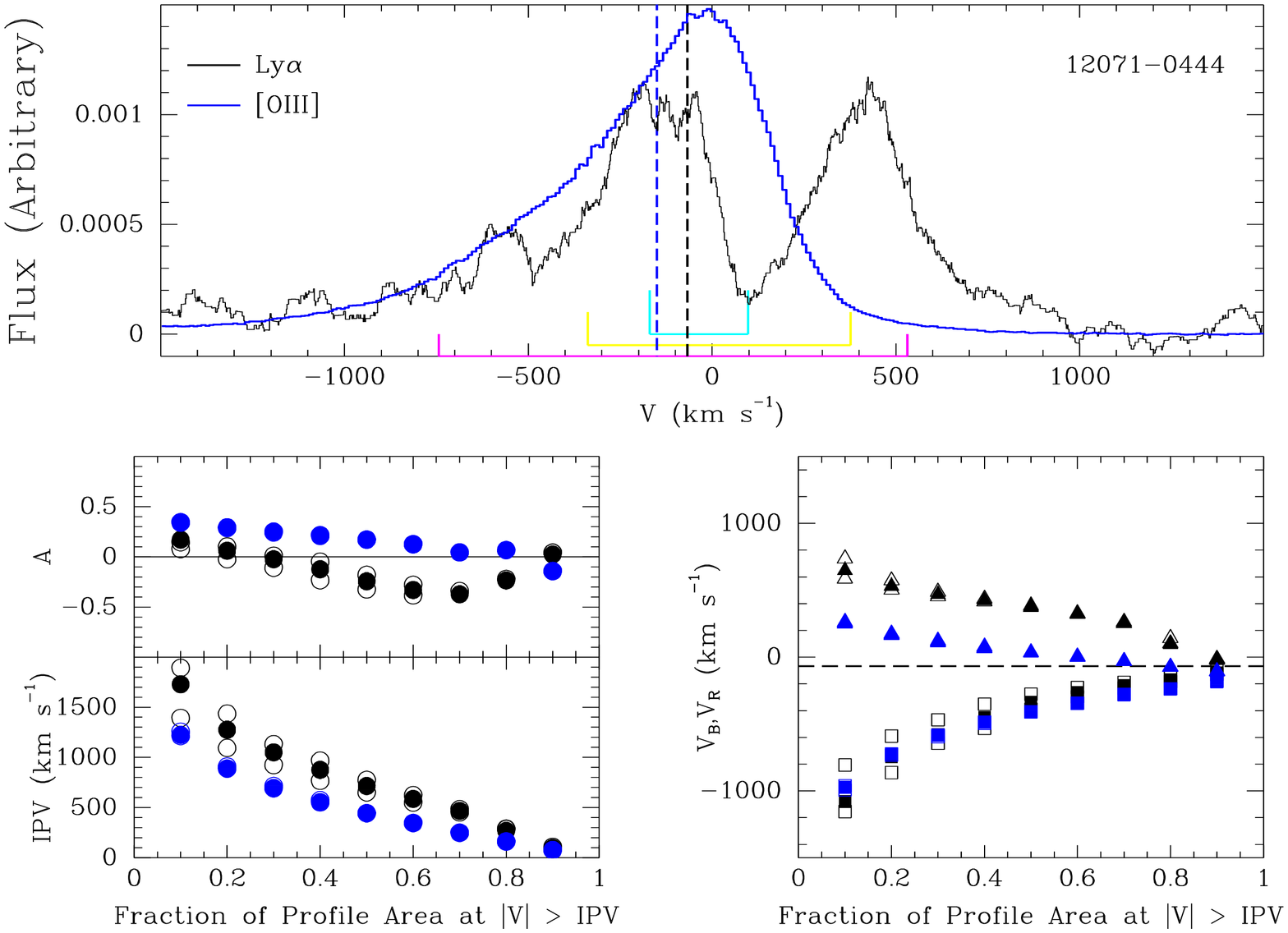}\hfill}  }
     \end{figure}

\clearpage
%g
\addtocounter{figure}{-1}
 \addtocounter{subfigure}{0}
  \begin{figure}[t]
   \subfigure[\i16474+3430]{\label{fig:ipv_16474+3430}
%    \hbox{\hfill \includegraphics[height=15cm,angle=-90,trim=0 0 0 0]{figures/ipv_v24.ps} \hfill}  }
    \hbox{\hfill \includegraphics[height=15cm,angle=-90,trim=0 0 0 0]{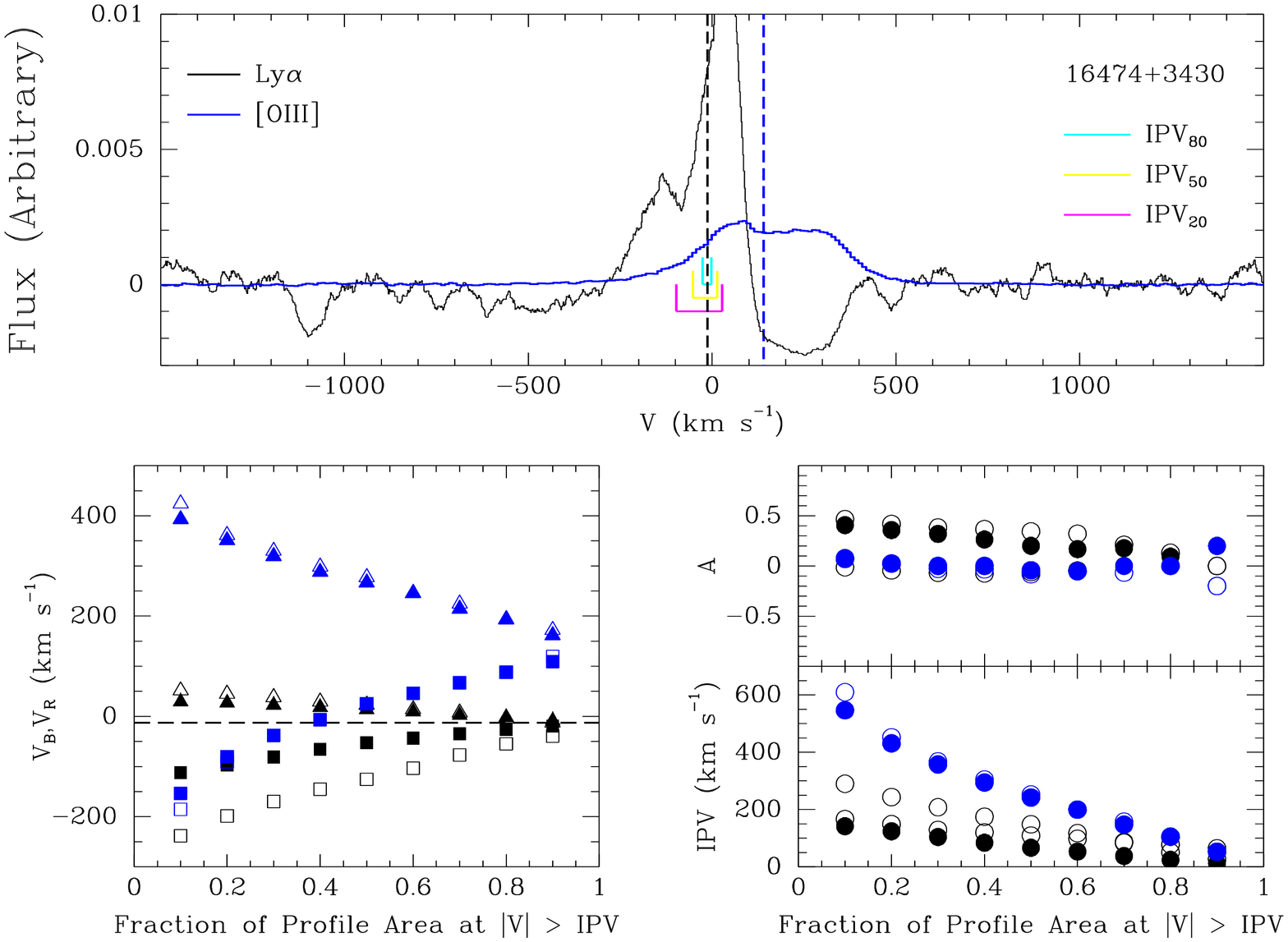} \hfill}  }
     \end{figure}

\clearpage
%h
\addtocounter{figure}{-1}
 \addtocounter{subfigure}{0}
  \begin{figure}[t]
   \subfigure[\i16487+5447]{\label{fig:ipv_16487+5447}   
%    \hbox{\hfill \includegraphics[height=15cm,angle=-90,trim=0 0 0 0]{figures/ipv_v13.ps} \hfill} }
    \hbox{\hfill \includegraphics[height=15cm,angle=-90,trim=0 0 0 0]{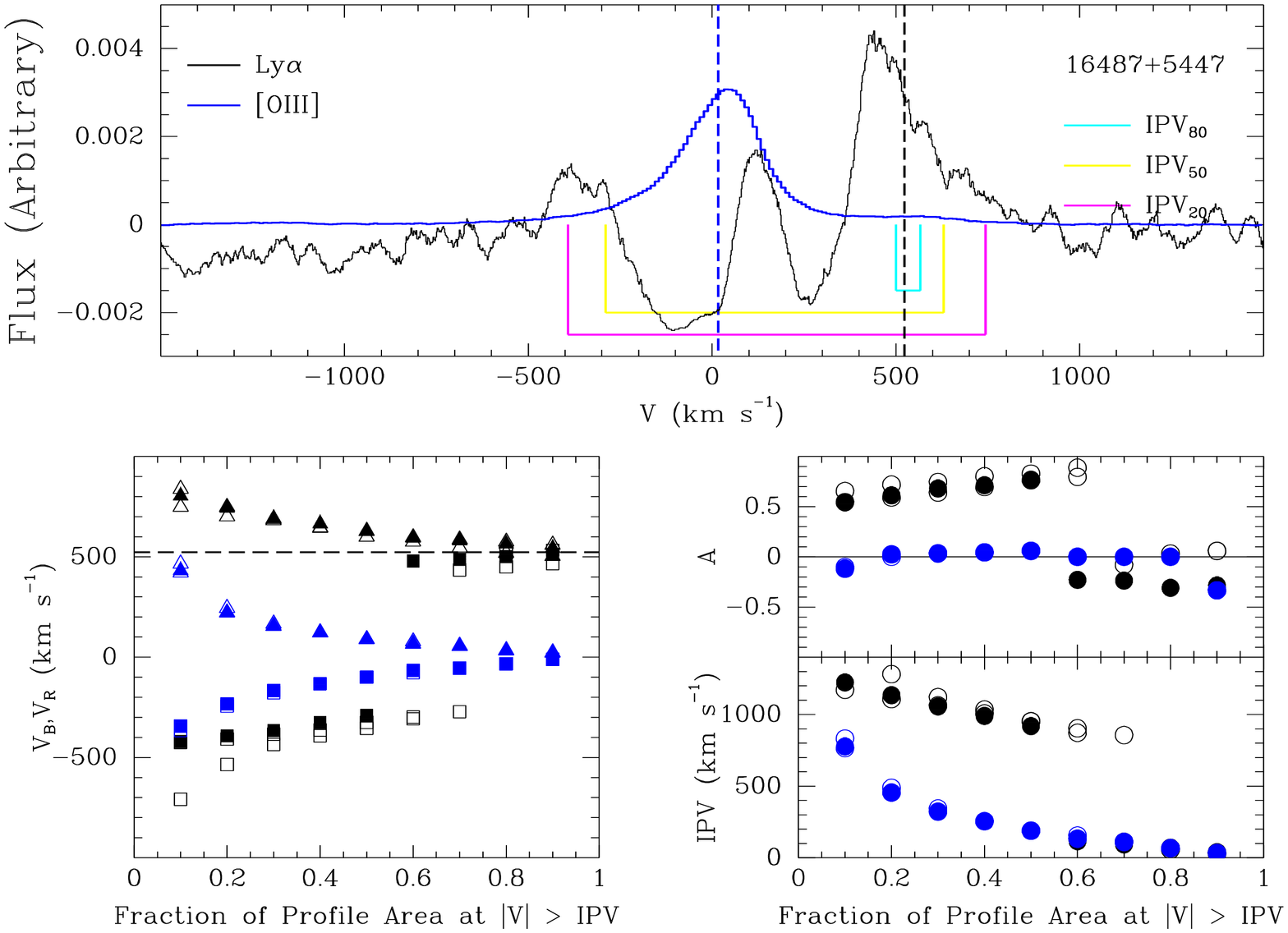} \hfill} }
     \end{figure}

\clearpage
%i
\addtocounter{figure}{-1}
 \addtocounter{subfigure}{0}
  \begin{figure}[t]
   \subfigure[\i23365+3604]{\label{fig:ipv_23365+3604}
%    \hbox{\hfill \includegraphics[height=15cm,angle=-90,trim=0 0 0 0]{figures/ipv_v16.ps}  \hfill} }
    \hbox{\hfill \includegraphics[height=15cm,angle=-90,trim=0 0 0 0]{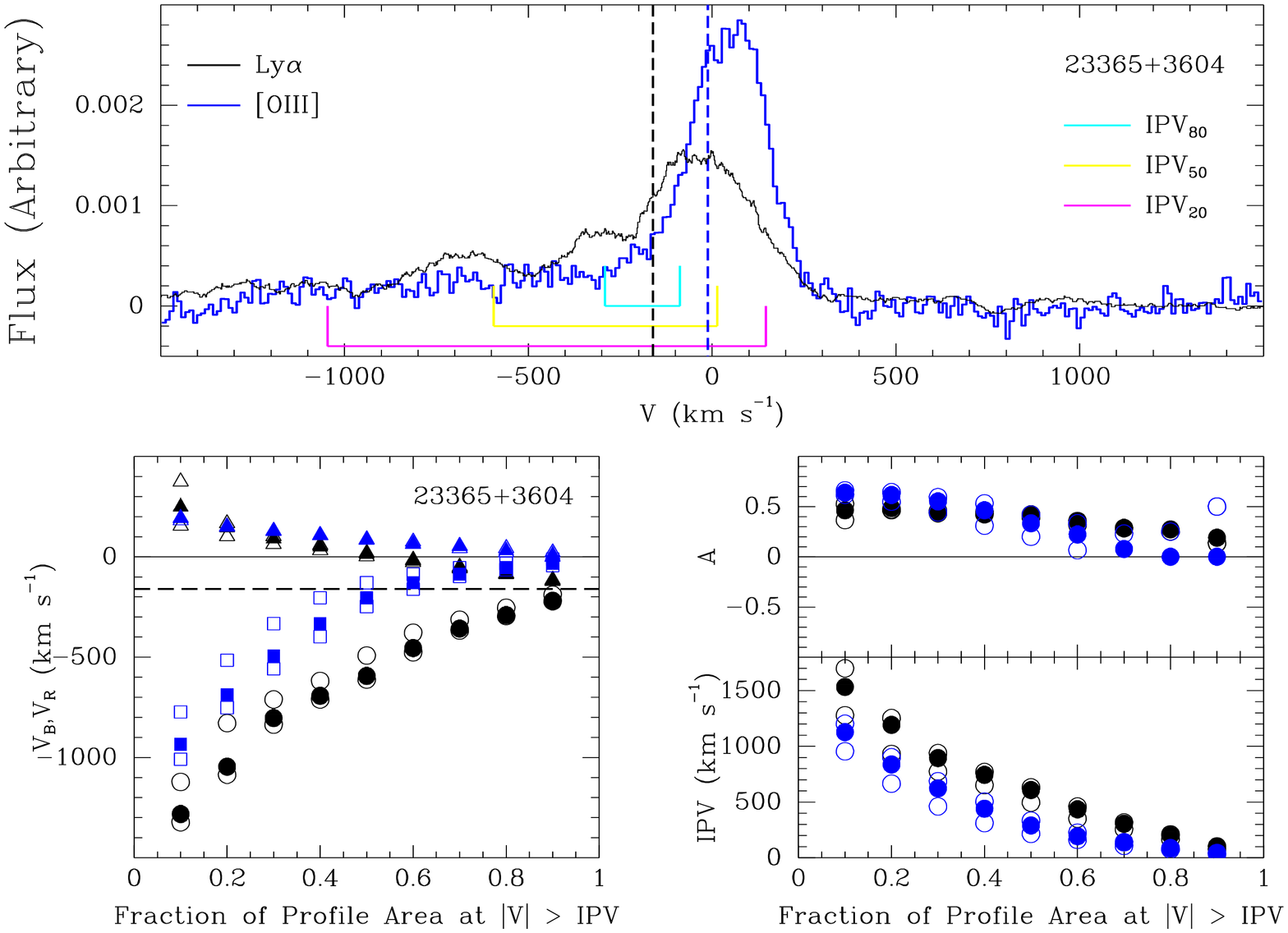}  \hfill} }
     \end{figure}

\end{document}